%% file: Thesis.tex
\documentclass[11pt,dvips,openright,a4paper,twoside,english,dutch]{book}
\usepackage{babel,varioref}
\usepackage[normal,small,bf]{caption}% give figures hanging indent captions 

\usepackage{epsfig}
\usepackage{cite}         % to sort citations like [2-6]
\usepackage[sectionbib]{chapterbib}
\renewcommand{\bibname}{References}

\begin{document}
\frontmatter
\selectlanguage{dutch}
\include{title}

\selectlanguage{english}
\renewcommand{\bibname}{References}
\tableofcontents
\mainmatter
\include{introduction}

\include{experiments}

\include{potentials}

\include{ch4scattering}

\include{isotope}

\include{excitations}

\include{trajectory}

\include{conclusion}

\backmatter
\include{summary}

\selectlanguage{dutch}
\include{samenvatting}

\selectlanguage{english}
\include{listofpub}
\selectlanguage{dutch}
\include{dank}

\include{cv}

\end{document}

%% file: title.tex
\thispagestyle{empty}

%%%%%%%%%%%%%%%%%%%%%%%%%%%%%%%%%%%%%%%%%
%\begin{center}
%\vspace*{1cm}
%\noindent\rule{12cm}{0.5mm}
%\vspace*{1cm}
%{\Huge
%Monte Carlo Simulations of \\[5mm] Catalytic Surface Reactions
%}
%\vspace*{1cm}
%\noindent\rule{12cm}{0.5mm}
%\vspace*{2.5cm}
%{\LARGE Ronald Gelten}
%\vfill\noindent
%\end{center}
%\newpage
%%%%%%%%%%%%%%%%%%%%%%%%%%%%%%%%%%%%%%%%%%%%

\begin{center}
%\vspace*{1cm}

{\Huge  Quantum and Classical Dynamics  \\[3mm]

        of Methane Scattering  \\[3mm]  }

\vspace*{3.5cm}
{\sc
{\large PROEFSCHRIFT}

\vspace*{10mm}

\begin{center}
ter verkrijging van de graad van doctor \\
aan de Technische Universiteit Eindhoven, \\
op gezag van de Rector Magnificus, \\
prof.dr.~M.~Rem, \\
voor een commissie aangewezen \\
door het College voor Promoties \\
in het openbaar te verdedigen \\
op donderdag 14 juni 2001 om 16.00 uur
\end{center}

\vspace*{12mm}

door

\vspace*{12mm}

{\Large Robin Milot}

\vspace*{5mm}

geboren te Utrecht
}
\end{center}
\newpage
%%%%%%%%%%%%%%%%%%%%%%%%%%%%%%%%%%%%%%%%%%%%%%%%%%%%%%%%%%%%%%%%%%%%%%%%%%%%%%%%\begin{center}
\thispagestyle{empty}

\noindent
Dit proefschrift is goedgekeurd door de promotoren:
\vskip 3mm
\par\noindent prof.dr.~R.A.~van~Santen
\par\noindent en
\par\noindent prof.dr.~A.W.~Kleyn
\vskip 3mm
\noindent Copromotor: dr.~A.P.J.~Jansen

\vfill

%%\noindent
%%Copyright \copyright\ 2001 by Robin Milot, Geldrop, The Netherlands.\\
%%\vskip 5mm
\noindent
CIP-DATA LIBRARY TECHNISCHE UNIVERSITEIT EINDHOVEN
\vskip 3mm
\noindent
Milot, Robin.
\vskip 3mm
\noindent
Quantum and classical dynamics of methane scattering / \\
by Robin Milot. - Eindhoven: Technische Universiteit Eindhoven,
2001. - Proefschrift. - ISBN 90-386-2782-3 
\vskip 3mm
\noindent
NUGI 813 
\vskip 3mm
\noindent
Trefwoorden: moleculaire dynamica / rotationele en vibrationele
energie-overdracht / overgangsmetaal-katalysatoren / dissociatieve
chemisorptie; methaan \\
Subject headings: molecular dynamics / rotational and vibrational energy
transfer / transition metal catalysts / dissociative chemisorption; methane 
\\[5mm]
Printed at {\sl Universiteitsdrukkerij}, Eindhoven University of Technology
\\[5mm]
\noindent
{\sl This research has been financially supported by the Council for
    Chemical Sciences of the Netherlands Organization for Scientific
    Research (CW-NWO). }

\noindent
{\sl The work described in this thesis has been carried out at the
  Schuit Institute of Catalysis (part of NIOK, the Netherlands School
  for Catalysis Research), Eindhoven University of Technology, The
  Netherlands.}

%%%%%%%%%%%%%%%%%%%%%%%%%%%%%%%%%%%%%%%%%%%%%%%%%%%%%%%%%%%%%
%%\newpage
%%\thispagestyle{empty}
%%
%%\vfill
%%~\vskip 50mm
%%\par\noindent{\sl \hfill Like molecules in constant motion ...} \\[5mm]
%%\par\noindent\hfill David Byrne
%%\vfill

%\cleardoublepage

%% file: introduction.tex
\chapter{General introduction}
\label{chap:genintro}

%\vfill
%\vspace*{2cm}
%\section*{abstract}
\begin{quote}
  {\it This thesis describes the scattering dynamics of methane from
    transition metals. The dynamics is studied by wave packet (quantum
    dynamics) and classical (Newtonian) trajectory simulations. Although
    the dissociation of methane is not studied itself, I try to deduce
    consequences for the dissociation from the scattering simulations.
    I give a general introduction in this chapter with a description of
    theoretical research in general, and explain the chemical theories
    which are used nowadays. Next I discuss catalysis and how my
    research is related to this. I end with an overview of the next
    chapters.}
\end{quote}

%\newpage

%%%%
\section{Theory in action}

This dissertation describes theoretical chemical research on the
scattering of methane from transition metal surfaces. I try to deduce from
the scattering consequences for the dissociation of methane, which is
important for catalysis. I will explain in this introduction: What is
theoretical research supposed to do? What kind of chemical theories are
in use? What can be the relevance of my research for catalysis?  And
finally: What can be found in the rest of this book?

The core business of theoretical research is a mixture of speculation,
calculation, model building, and approximation.\footnote{I follow here
  Chapter 12 of Ref.~\cite{representing}, which you can read for a more
  extended discussion on theoretical research. } There are different
levels of speculation, which you normally can find back if words like
hypotheses and assumption are used.  Simple speculation can be for
example that there are particles, and these particles have interactions
with each other.  More complex speculation occurs if the interaction is
expressed by a mathematical equation, which is called then a theory. The
theory has some parameters, which can be later on filled in for a
specific experiment. The parameters can be fitted on the experiment,
obtained from other theories, or just guessed for a lucky shot. At that
moment we are already busy with calculations. If the parameters are
filled in or the mathematical equations are combined, we apply the
general theory into a specific model. Sometimes the model is too complex
for drawing simple conclusions from it, or it is very hard or even
impossible to solve all the equations (calculation) within a reasonable
amount of time. Approximation is then a way to overcome such a problem.
Approximation contains also a bit of speculation itself. We assume that
the subset of the model will be able to describe the experiment (at less
in part) the same as the original model. The subset of the original is
an model itself on which the same tricks can be applied later on.

Most experimentalists do use theory for the interpretations of their
experiments. For some problems the developed theory or the system under
study will become so complex or the model so big, that calculation can
only be performed if one only focus on theory. My research is an example
of this. A lot of approximation have been done to obtain models, which
are small enough to be calculated on a computer. I need to know
something about computer programming and numerical methods, before I can
put them into a computer. I also have to compare the results of my
calculated approximated models with the available experiments. I have to
look for agreement, and speculate how we can overcome the disagreements
between the models and experiments. Of course my models are not build
totally from scratch, but are developed from established chemical
theories.

%%%%%
\section{Chemical theory}

Chemistry studies things we normally denote with substance, matter,
material, and chemicals. It is not only interested in the matter itself,
but also how it can be changed in something else. We all know a lot of
this matter, because we use at lot of it on a daily basis. We cook,
eat, do the dishes, wash our clothes, clue our fingers, and making fire
for heating. Still there are some things we let over to specialists. The
chemists are specialists in (change of) matter. Chemist make use of
theoretical concepts for the description of chemical phenomena. I will
give an small and global overview of the three most important
theoretical concepts; Thermodynamics, kinetics, and molecular dynamics.

\subsection{Thermodynamics and kinetics}

Thermodynamics is a very well established chemical theory. It is
a spin off from the developments on the steam engine at the end of the
18th century, to which its name still refer. Thermodynamics
stands for how heat is transfered into work \cite{thermodyn}. The theory
itself is not about dynamics at all, but deals with thermostatic
phenomena. Thermodynamics describes equilibrium conditions.

In thermodynamics everything is related to different kinds of energy. A
chemical reaction is thermodynamical possible (runs on its own), if the
energy of the products is lower then the energy of the reactants. Such
an reaction is then called \textsl{exothermic}, because the loss of
energy from the reactants to the products is given back as heat. If the
products have a higher energy than the reactants, then the chemical
reaction is called \textsl{endothermic}. An endothermic reaction cannot
run independently, but it needs a side reaction that delivers the heat
necessary for the reaction.

%\subsection{Kinetics}

\begin{figure}[t]
  \begin{center}
    \epsfig{file=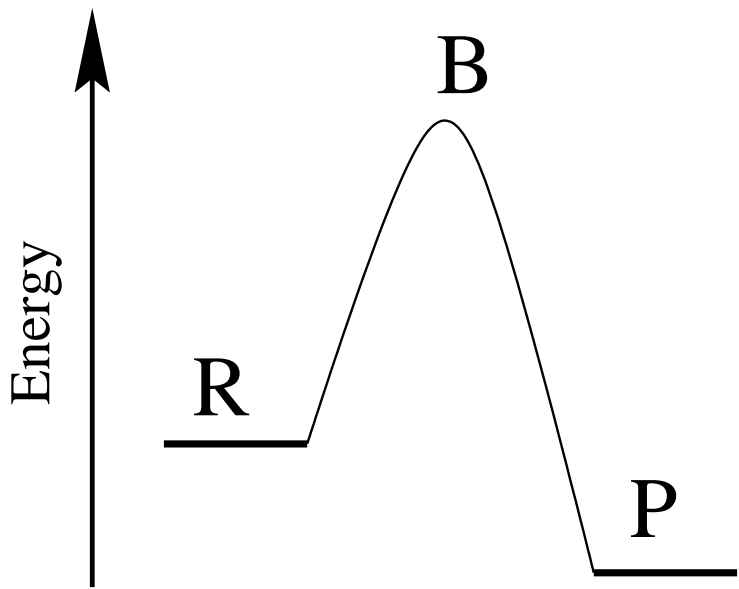}    
    \caption{Sketch of a reaction barrier (B) between reactants (R) and
      products (P) for an exothermic reaction.}
    \label{fig:Eact}
  \end{center}
\end{figure}

If a reaction is thermodynamical possible (exothermic), then a reaction
does not have to occur at all. Thermodynamics does not say anything
about the time-scale of a reaction. The time-scale is left over to
chemical kinetics, which studies the rate of the reaction as a function
of concentration, pressure, and temperature. The reaction rate $r$ can
be linear dependent of the concentration $c$ (or pressure) of the
reactants in a simple case. This can be written as
\begin{equation}
  \label{reactrate}
  r = k \ c ,
\end{equation}
where $k$ is the rate constant. The rate constant can be expressed by
the Arrhenius rate law
\begin{equation}
  \label{arrhenius}
  k = A  \ e^{-E_a/RT},
\end{equation}
where $A$ is the prefactor, $E_a$ is the activation energy, $R$ is the
gas constant, and $T$ the temperature. The activation energy can be seen
as a barrier which needs to be overcome to let the reaction occur. This
barrier is sketched in Figure~\ref{fig:Eact} for an exothermic reaction.
We can see now that we first need energy to overcome the barrier before
the reaction can occur, although the energy of the products is lower
then the energy of the reactants. Thermal energy can be used to overcome
the barrier. Raising the temperature increases the rate of the reaction,
while a higher activation energy decreases it. The backward reaction
from product to reactant is also possible, but the activation energy for the
backward reaction ($E_B-E_P$) is higher than for the forward reaction
($E_B-E_R$). The consequence of this is that the rate constant of the
forward reaction is larger than the rate constant of the backward
reaction, which means that the concentration of the product is higher
than of the reactants at the time the thermodynamics equilibrium is
reached.

A conversion from reactants to products does not always occur in one
step, but can consist out of multiple steps. In such a case we have to
overcome multiple barriers. The reactants are then converted into an
intermediate, and the intermediate into other intermediates or finally
into the products. The reaction step with the highest barrier is the
rate limiting step, because it is the bottle neck for the overall
reaction rate.

\subsection{Molecular dynamics}

Nowadays the world is full of molecules. However, this was not always
the case. Of course there was an atomic world-view suggested by
Democritus, and the corpuscles of Boyle went in the direction of
molecules, but they were not very useful at their time. Thermodynamics
and kinetics works very well without any notion of molecules. The real
victory of the atoms and molecules is established at the beginning of
the twentieth century. Beside that (chemist assume that) there are small
particles like molecule, we also take for granted that they move and do
interact with each other. The kinetic energy of the random motion of
particles is now associated with the temperature.

Chemists think that molecules exist out of positive charged nuclei and
negative charged electrons. If the molecule has only one nucleus then it
is called an atom. Electrons have always an electron charge of $-e$, and
nuclei can have charges of $+e$ to more than $+100e$.  The mass of an
electron is much lower than the mass of a nucleus. The mass of the
nucleus increases with the charge of the nucleus, but there can exist
nuclei with different masses while their charge is the same. These
nuclei are called isotopes.

Negative and positive charged particle have an attractive interaction
with each other.  Two negative or positive charged particles have a
repulsive interaction.  If the nuclei are in a certain position from
each other with in between them some electrons, and the total attraction
is larger than the total repulsion, then the result is a stable
molecule. The spatial arrangement of a stable molecule, which is
normally called the molecular structure, is responsible for the physical
and chemical properties of a substance.  The structure of a stable
molecule has a lower energy than other spatial arrangements
near that structure. We are talking about isomers, if two molecule have
the same number of electrons and types of nuclei, but the nuclei are
arranged different in space.

Within the molecular world-view a chemical reaction is a spatial
rearrangement of nuclei and electrons over a period of time, which gives
molecules with a different structure (the products) than the original
molecules (the reactants). This rearrangement is possible, because the
nuclei have kinetic energy that can be used to overcome a barrier of
potential energy between different molecular structures. We do not only
have to know the height of the barrier, but we are also interested how
we can use the kinetic energy to overcome this. We do so by studying the
dynamics of nuclei during the reaction. We do not have to look at the
rearrangement of the electrons, because they rearrange much faster than
nuclei. Three things are important for the reaction dynamics; The
initial momentums and positions of the nuclei, and the potential energy
for each spatial arrangement of the nuclei. If we know these, then we
can calculate for every time the positions of the nuclei with a
dynamical theory.

There are at least three dynamical theories, which are all useful for a
specific set of natural phenomena. Classical dynamics is very useful for
describing the fall of an apple out of a tree. Quantum dynamics needs to
be used for very small particles like electrons. For spaceships
traveling near the speed of light we apply relativistic dynamics.
Molecules have properties such as translational and rotational motion,
which can be described sometimes as accurate with classical and as with
quantum dynamics. Quantum dynamics is better in treating vibrational
motion, but it has the draw back that is harder to calculate with.  In
this thesis I use both classical and quantum dynamics for simulating the
scattering of methane on transition metal surfaces. (We use normally the
term wave packet dynamics for quantum dynamics applied on molecular
motion.)

%%%%
\section{Catalysis}

A catalyst is a chemical, which can change the rate of a reaction step
without ending up in a reaction product. This means that it changes the
kinetics of a reaction and not the thermodynamics. A catalyst can be
used for changing the \textsl{reactivity} or the \textsl{selectivity} of
a reaction. Reactivity means that the reaction rate of a reaction is
enhanced, and by so that the reactants are converged faster into the
products. However, there is not always only one reaction possible if we
start with a set of reactants. This can result in a bunch of products.
We have not always equal interest in all products. If we are not
interested in a product at all, then we call it even waste. We can try
to avoid this undesired product by increasing the reaction rate of the
reactions to the desired products more than the reaction rate of the
reactions to the waste or less desired products. In this case we make
use of the selectivity of a catalysts.

Catalysis is the name of the science that studies catalysts. Catalysis
has many technological aspects, because the output of the research has
to benefit us. The main goal is to let make catalysts better. The main
question is: What is the optimal catalyst for a given reaction?
Chemistry is not the only science that plays a role in answering this
question. Of course we can search for an optimum between reactivity and
selectivity. The found optimum catalyst is not always the most economic
favourable catalysts. The optimum catalyst in terms of reactivity and
selectivity is unfortunately most of the time also the most expensive,
so a cheaper but less selective or reactive catalyst can give more
profit. Another chemical and economical problem is the
\textsl{stability} of the catalyst. After a certain time a catalyst is
not anymore as reactive and selective as at the time it was placed in
the reactor. This can be caused by thermal instability or poisoning of
the catalyst. The consequence of this is that the original catalyst has
to be replaced in the reactor by a new one. This can be very costly,
because one has to buy a new catalyst and the reactor cannot be used
during the replacement of the catalyst.

Catalysts can be divided into two classes: Homogeneous and heterogeneous
catalysts. Homogeneous catalysts are in the same phase as the reactants,
and has the advantage that they are well-dispersed and well-mixed with
the reactants. They are also in general more complex, and by so more
selective than heterogeneous catalysts. Examples of homogeneous catalyst
are the enzymes in living creatures, and organo-metallic complexes used
for polymerisation reactions. A disadvantage of homogeneous catalyst is
that the products are hard to separate from the catalyst.

Heterogeneous catalysts are in a different phase than the reactants and
products. Normally this means that the catalyst is solid and the
reactants are gaseous or liquid. The reaction takes place at the
interphase between the catalyst and reactant phase. Heterogenous
catalysts are widely used in industry for the conversion of bulk
chemicals. Examples of industrial heterogeneous catalysts are zeolites
and transition metal clusters deposited on micro-porous carrier materials. 

The last example is important for steam (H${}_2$O) reforming of methane
(CH${}_4$) for the production of syngas (CO + H${}_2$). Syngas can be
used as building blocks for the productions of various organic
molecules. The rate limiting step in the steam reforming of methane is
the breaking of an C--H bond \cite{hoo80}. It is very difficult to study
this reaction step on its own in an industrial catalyst. Therefore we
have to look for model systems which make it possible to do so.

Surface science has given us such a model. The metal clusters are
modeled by a metal surface with a well-defined topology. The metal
surface can be placed in a box and deposited on a gas mixture. If the
box is under ultra high vacuum (UHV), then it is also possible to study
reactions at the interphase with spectroscopic techniques. One can also
shoot molecules with a specific velocity at the surface with a gun
called molecular beam. Over the last decades a lot of new phenomena have
been created with these molecular beams, which have never been seen
before.  I did not work with molecular beams, but I have tried to make
some dynamical models for describing the observations with molecular
beams of methane on metal surfaces. So at the end my research might
contribute a very little bit to a better understanding of the working of
a heterogeneous industrial catalyst. If you like to see any relevance of
my research with daily live issues, then you can find it through this
small connection. I leave the judgement to you.

%\section{Surface science}
%
%It is difficult to study chemical reactions within a industrial
%heterogeneous catalyst. 
%\begin{quote}
%``One of the prominent arguments for performing surface science studies
%have for many years been to improve and design new and better
%catalysts. Although surface science has provided the fundamental
%framework and tools for understanding heterogeneous catalysis until now
%there have been extremely few examples of actually designing new
%catalysts based solely on surface science studies.''\cite{lar00}
%\end{quote}

\section{Overview of this thesis}

This thesis describes simulations of methane scattering on a nickel
metal surface. Although the dissociation of methane is not studied, I
try to deduce from the scattering simulations implication for the
dissociation and to relate them with the available molecular beam
studies. Therefore, I will give an overview of the experimental and
theoretical studies on methane dissociation on transition-metal
surfaces in Chapter~\ref{chap:exper} first.

Chapter~\ref{chap:wp} gives an introduction and the computational
details of our wave packet simulations discussed in
Chapters~\ref{chap:ch4scat}, \ref{chap:iso}, and \ref{chap:exc}.  It
describes four model potential energy surfaces (PESs) for the scattering
of oriented of CH${}_4$ and CD${}_4$ from a flat surface on which we
performed our wave packet simulations. I also gives a short survey of
the multi-configurational time-dependent Hartree (MCTDH) method, which
we have used for our wave packet propagation, because it can deal with a
large number of degrees of freedom and with large grids \cite{bec00}.

The scattering of CH${}_4$ in the vibrational groundstate on the model
PESs is presented in Chapter~\ref{chap:ch4scat}. At a translational
energy up to 96 kJ/mol we find that the scattering of almost completely
elastic.  Therefore, we used vibrational excitations when the molecule
hits the surface and the corresponding deformation for the analysis of
the scattering. From these we deduce consequences for the dissociation
mechanism, and find that for methane to dissociate the interaction of
the molecule with the surface should lead to an elongated equilibrium
C--H bond length close to the surface.

The isotope effect is described in Chapter~\ref{chap:iso} by comparing
the scattering of CD${}_4$ with CH${}_4$.  First, we look again at the
scattering, excitation probabilities when the molecule hits the surface,
and the corresponding deformations. The scattering is still
predominantly elastic, but less so for CD${}_4$. Second, we do an energy
distribution analysis of the kinetic energy per mode and the PES terms
when the molecule hits the surface. They indicate that the orientation
with three bonds pointing towards the surface is mostly responsible for
the isotope effect in methane dissociation.

Chapter \ref{chap:exc} presents the role of vibrational excitation of a
single mode in the scattering of CH${}_4$ and CD${}_4$. Energy
distribution analysis shows that initial vibrational excitations enhance
the transfer of translational energy towards vibrational energy and
increase the accessibility of the entrance channel for dissociation. The
simulations predict that initial vibrational excitations of the
asymmetrical stretch ($\nu_3$) and especially the symmetrical stretch
($\nu_1$) modes will give the highest enhancement of the dissociation
probability of methane.

In Chapter~\ref{chap:traj} I study the full rotational vibrational
scattering of a non-rigid CH${}_4$ molecule on a Ni(111) surface with
classical (Newtonian) trajectory calculations. Energy dissipation and
scattering angles have been studied as a function of the translational
kinetic energy, the incidence angle, the (rotational) nozzle
temperature, and the surface temperature.  Scattering angles are
somewhat below the incidence angles of 30${}^{\circ}$, 45${}^{\circ}$,
and 60${}^{\circ}$ at a translational energy of 96 kJ/mol. Energy loss
is primarily from the normal component of the translational energy and
transfered for somewhat more than half to the surface and the rest
mostly to rotational motion.

I will end this thesis with some concluding remarks, and suggestion for
further research in Chapter~\ref{chap:concl}.

%\bibliography{Methane,STS}
%\bibliographystyle{prstyfull}

%% file: experiments.tex
\chapter{The dissociation of methane}
\label{chap:exper}

\begin{quote}
  
  {\it I will give in this chapter an overview of the experimental and
    theoretical studies on methane dissociation on transition-metal
    surfaces. The dissociation of methane on transition metals occurs in
    general through a direct dissociation mechanism. Vibrational energy
    is overall about as efficient in enhancing dissociation as
    translational energy, which suggests a late reaction barrier.}
\end{quote}

%\vspace*{2cm}

\section{General theory of surface catalytic dissociation}

\begin{figure}[p]
  \begin{center}
    \epsfig{file=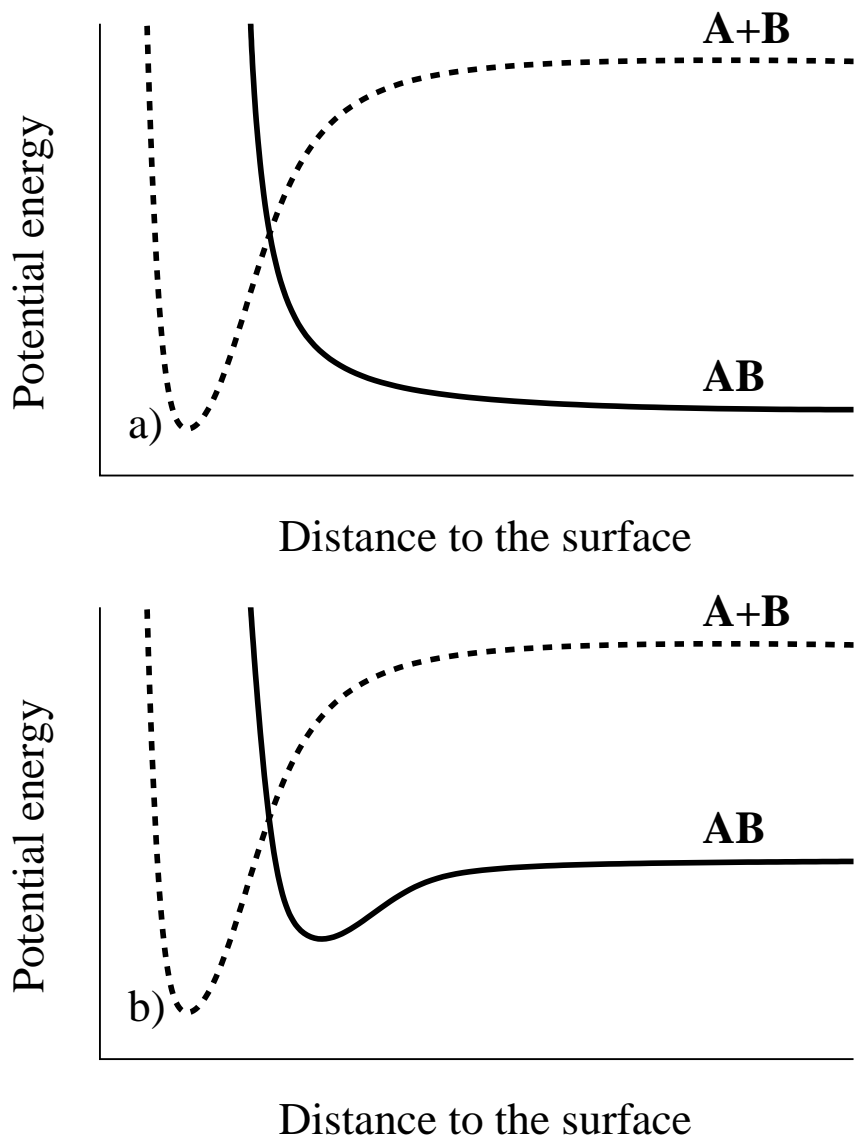, height=15cm}
    \caption{Schematic interaction potentials of a surface with a
      molecule AB and its dissociation products A and B for; a) the {\sl
      direct} activated process, and b) the {\sl indirect} trapping
      mediated process.}
    \label{fig:directpot}
  \end{center}
\end{figure}

I will now give an overview of the experimental and theoretical
literature of methane dissociation on surfaces. Before we enter into
details, I will discuss the catalytic activity of metal surfaces in
general qualitative terms. Let us assume therefore a diatomic molecule
A--B, which we want to dissociate by an catalytic metal surface. The
overall reaction will be:
\begin{equation}
  \label{eq:ABdis}
  \hbox{A--B (gas)} \rightarrow \hbox{ A (ads.) + B (ads.)}
\end{equation}
In this reaction equation we only see the initial reactant (molecule
A--B) and the final products (A and B atoms adsorbed on the metal
surface). We are interested in the mechanism of this catalytic reaction:
What is the nature of the interaction of the reactant with the surface
during the reaction?  Of course there are a lot of mechanisms
imaginable, but they can all be classified in two types of mechanism.
One of these mechanisms is the so called {\sl direct} activated process,
where the molecule will dissociate directly at the moment that the
molecule approaches the surface. The A--B bond will be activated by
energy transfer at the collision with the surface. The direct reaction
can be written as:
\begin{equation}
  \label{eq:ABdirect}
  \hbox{A--B (gas)} \rightarrow \hbox{A --- B} \rightarrow \hbox{ A
  (ads.) + B (ads.)} 
\end{equation}
The other mechanism is an {\sl indirect} process, which is also called
a trapping mediated process. The molecule has to adsorb at the surface
first, before it can dissociate. The indirect reaction can be written as: 
\begin{equation}
  \label{eq:ABindirect}
  \hbox{A--B (gas)} \rightarrow \hbox{A--B (ads.)} \rightarrow \hbox{ A
  (ads.) + B (ads.)} 
\end{equation}
The differences between both mechanisms can be explained theoretically
on the basis of the interaction potentials of the surface with the
molecule and its dissociation products. Figure \ref{fig:directpot}~a)
shows the potential energy of the molecule AB and its dissociation
products A and B as a function of the distant to the surface. The
molecule AB has far away from the surface a much lower energy than the
dissociation product A and B. It is therefore hard to dissociate for the
molecule AB in the gas phase. If the molecule comes close to the
surface, then its energy will go up. The energy of the dissociation
product A and B drops towards the surface. The dissociate product A and
B are even lower in energy at the point were the line of AB crosses the
line of A + B.\footnote{The actual location and height of the crossing
  point cannot be taken from the one dimensional schematic plots of
  Figure \ref{fig:directpot}, because it also dependent of other
  coordinates like the bond length.} The energy difference between the
gas phase energy of the molecule and the energy of the crossing point is
smaller than the difference between the molecule and dissociation
products in the gas phase.  The surface is now a catalyst, because it
makes the dissociation easier.  The molecule needs translational energy
to run up against the surface repulsion and reach the crossing point.
This is the picture of a direct dissociation mechanism, because the
dissociation occurs directly at the collision with the surface.

Figure \ref{fig:directpot}~b) shows again the potential energy of the
molecule AB and its dissociation products A and B as a function of the
distant to the surface. The energy difference in the gas phase is still
large, so dissociation is hard. The main difference with Figure
\ref{fig:directpot}~a) is that the interaction of the molecule AB with
the surface is changed. There is a well in the potential near the
surface. Because of this it is possible for the molecule AB to adsorb at
the surface. Translational energy is not needed to do so. The molecule
can now reach the crossing point with use of the available kinetic
energy at the surface, when it is adsorbed at the surface. The
dissociation occurs then through a indirect trapping mediated process.
It is still possible to see a direct dissociation mechanism within the
potential energy scheme of Figure \ref{fig:directpot}~b), if the
translational energy of the molecule AB is high and it will be able to
reach the crossing point without absorbing first.

It is now clear how we can experimentally distinguish a direct and
indirect dissociation mechanism. The dissociation probability of a
direct mechanism is primary dependent on the translational energy of the
molecules in the gas phase, and by so on the gas temperature. An indirect
mechanism is on the other hand primary dependent on the surface
temperature.

\begin{figure}[p]
  \begin{center}
    \epsfig{file=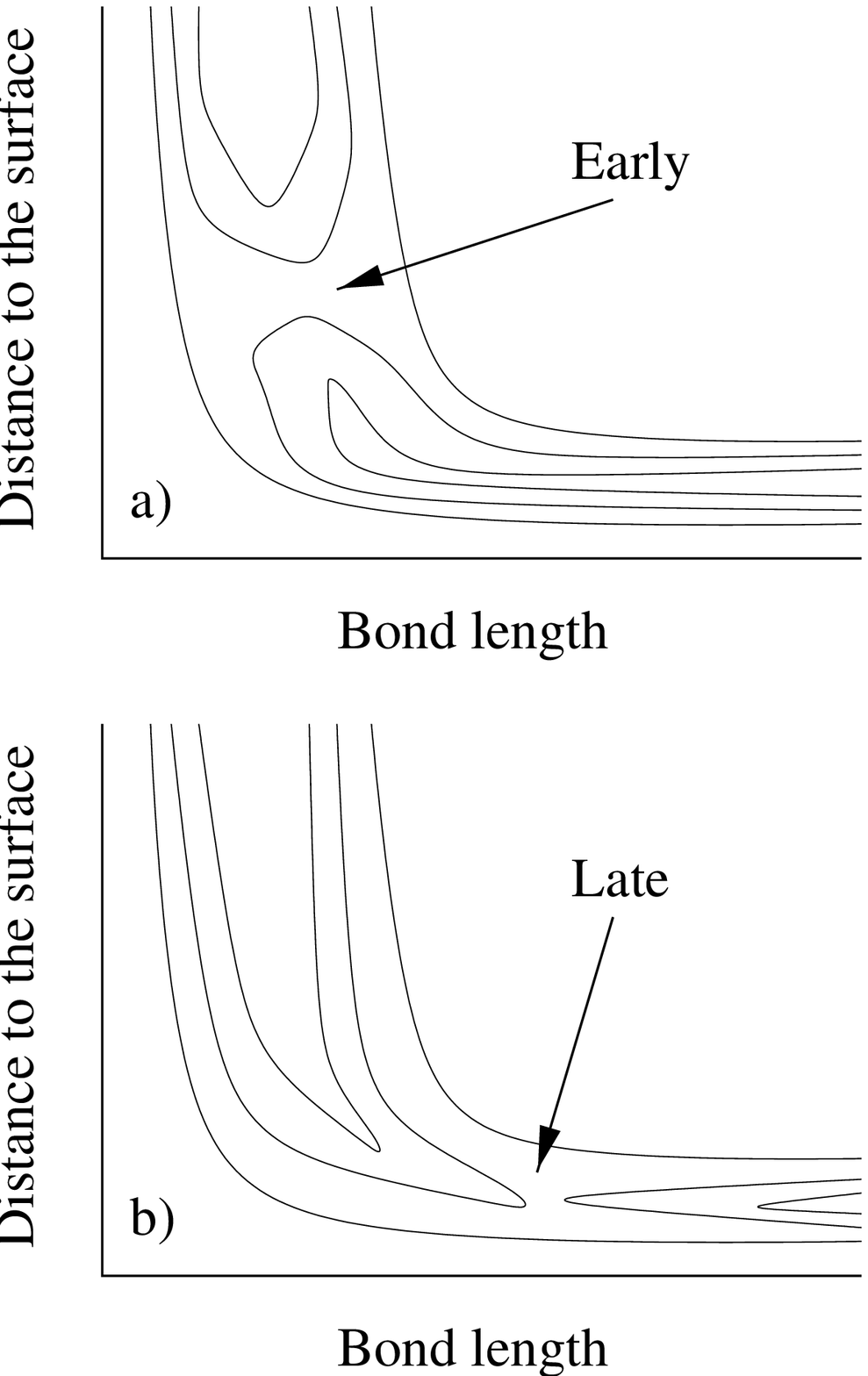, height=17cm}
    \caption{Schematic contour plot of the potential energy surface as a
      function of the distance to the surface and the bond length of the
      molecule for; a) an early, and  b) late dissociation barrier.}
    \label{fig:contpot}
  \end{center}
\end{figure}

Another point of interest is the role of the internal vibrations.
Sometimes the dissociation probability is enhanced by both vibrational
and translational energy. We need a potential, which is dependent of two
dimensions to explain how this is possible. Figure \ref{fig:contpot}
shows two possible contour plots of potentials which are dependent of
the distance between the molecule and the surface, and the bond length
within the molecule. The difference between the two plots is the location
of the reaction barrier. Figure \ref{fig:contpot}~a) shows a potential
with a so called \textsl{early} barrier. The reaction barrier can be
accessed by moving the molecule towards the surface. For a potential
with a \textsl{late} barrier as sketched in Figure \ref{fig:contpot}~b)
a movement of the molecule towards the surface is not enough. The bond
need also be lengthened to before the reaction barrier can be reached.
For a potential with an early barrier only translational energy is
needed to overcome the barrier. For a potential with a late barrier also
vibrational energy is needed. Vibrational energy can be available
initial by vibrational excitations, or generated during the reaction by
transferring translational energy to vibrational energy.

The real interaction potential is even more complex than I sketched
here. It is also dependent of the translational coordinates parallel to
the surface and the orientation of the molecule. For a polyatomic
molecule such as methane it becomes even more complicates, because the
potential is then dependent of multiple vibrational modes.   

\section{Experimental studies}

The dissociation mechanism on transition metal have been studied by
either bulb gas or molecular beam experiments. A bulb gas experiment is
performed by placing a metal surface in reactor and exposing it to a
methane gas (mixture) with a certain pressure and temperature. The
surface is placed in a ultra high vacuum chamber at molecular beam
studies. The velocity and rovibrational temperature of the molecules in
the beam can be varied by changing the gas mixture and temperature of
the nozzle. Table \ref{tab:overview} shows that one can get more
direct detailed information from molecular beam than from bulb gas
experiments. It is possible with the use of lasers to study the role of
rotational en vibrational excitation in more detail, which can be later
on compared with the outcome of molecular dynamics simulations.  
I will give here a now small survey of all experimental studies. A more
extended review can be read in Ref.~\cite{lar00}.

\begin{table}[t]
  \caption{Overview of relating observables in bulb gas and molecular
    beam experiments, and molecular dynamics simulations.}
  \label{tab:overview}
  \begin{tabular}[h]{|| c | c || c ||}
    \hline
    \hline
      Bulb gas & Molecular beam & Simulations \\
    \hline
    \hline
      & Translational energy  &  Translational energy  \\ 
      \cline{2-3}
      Gas temperature   & Nozzle   & Rotational energy \\ 
      \cline{3-3}
                     &  temperature   & Vibrational energy \\ 
    \hline
      Surface temperature & Surface temperature  & Surface motion \\ 
    \hline
    \hline
  \end{tabular}
\end{table}

\subsection{Bulb gas}

One of the first surface science studies of the methane dissociation was
under bulb gas condition on a tungsten surface \cite{win75}. A large
kinetic isotope effect was observed for CH${}_4$ in comparison with
CD${}_4$. In a second article also the dissociative chemisorption of
CH${}_3$D, CH${}_2$D${}_2$, and CHD${}_3$ were reported \cite{win76}.
The apparent activation energies were discussed in terms of tunneling of
hydrogen through a potential barrier, vibrational excitation, and the
lifetime of undissociated methane on the surface. Around the same time,
it was reported that the dissociation on rhodium surfaces can be
enhanced by raising only gas temperature of methane \cite{stew75}. A
large kinetic isotope effect was observed again, and this was attributed
to the dominant role of vibrational activation.  Nevertheless two
independent bulb gas experiment with laser excitation of the $\nu_3$
asymmetrical stretch and $2\nu_4$ umbrella modes on the Rh(111)
surface,\cite{yates79} and laser excitation of the $\nu_3$ and $2\nu_3$
modes on thin films of rhodium\cite{brass79} have not revealed any
noticeable enhancement in the reactivity of CH${}_4$.

Many years later, new thermal activation studies on rhodium by raising
independently the gas temperature and the surface temperature
observed that the dissociation increases in all cases with the surface
temperature \cite{brass87a,brass87b}. It was suggested therefore that
both a direct and an (indirect) trapping mediated mechanism plays a
role.  The differences in activation energies between CH${}_4$ and
CD${}_4$ were again explained in terms of internal vibrational
excitations. A measurement of the kinetic isotope effect on W(211) \cite{lo87}
initiated a debate, whether this experiment could be \cite{kay88a,kay88b}
or could {\sl not} be \cite{lo88a,lo88b} described by a tunneling mechanism.
An article about thermal activation experiment under
isothermal conditions reported a kinetic isotope effect of a factor 20
on Ni(100), whereas none was observed on Ni(110) \cite{bee87}. However,
Ref.~\cite{cho90,olg95} found under the same conditions an activation energy
twice as high for CH${}_4$ on Ni(100) as Ref~\cite{bee87}.
A direct reaction mechanism has been found on Ni(111)\cite{han91} and
Pt(110)\cite{lun94}, which contrasts reported trapping mediated mechanism on
Ni(100)\cite{cam93}. On Ir(111) both mechanisms have been reported for
${}^{13}$CH${}_4$ and CD${}_4$ \cite{jachim97}.  

\subsection{Molecular beam}

A large number of molecular beam experiments in which the dissociation
energy was measured as a function of translational energy have already
been done on many different surfaces. The first experiments were
performed on W(110) and show that the sticking probability increases
exponentially with the normal component of the incident translational
energy \cite{ret85,ret86}. This is in agreement with a direct
dissociation mechanism, and is observed by all later molecular beam
experiments (at the higher incident translational energies). It was also
observed that vibrationally hot ${\rm CH}_4$ dissociates more readily
than cold ${\rm CH}_4$, with the energy in the internal vibrations being
about as effective as the translational energy in inducing dissociation,
which have been confirmed by studies on other surfaces:
Ni(111)\cite{lee87,hol96}, Ni(100)\cite{xxx13,hol95},
Pt(111)\cite{lun89}, Pt(110)\cite{walker99,walker00}, and
Ru(0001)\cite{lar99}. There has been also measured a substantial kinetic
isotope effect for CD${}_4$ compared with CH${}_4$ on W(110), which have
been reported also on Ir(110)\cite{ver93,ver94,ver95},
Ni(100)\cite{hol95}, and Pt(111)\cite{lun89}. This indicates again that
the internal vibrational do play an important role in the dissociation
mechanism.  A more detailed assessment of the importance of the internal
vibrations could not be made, because of the large number of internal
vibrations.  However, two models about the role of the internal
coordinates have been suggested in relation with the experiments on
Ni(111). By making use of EELS vibrational spectroscopy it is
demonstrated that the initial step in the dissociation is the breaking
of a single CH bond and the formation of adsorbed CH${}_3$ and H
fragments \cite{lee86,lee87,cey87}. One of the suggestions is the direct
activation of the C--H bond, and another the so-called splats model
\cite{lee87}. The splats model suggests that the critical requirement
for methane dissociation is angular deformation of methane which allows
a Ni--C bond to form before a C--H bond breaks. In the splats model the
increased transfer of translational kinetic energy to the bending and
umbrella vibrational energy is responsible for the strong dependency on
normal incident translational energy. This has been tested by inert gas
atoms induced chemisorption \cite{bec87,bec89,ceyer90}. Later studies on
Ni(100) have related the enhancement of dissociation probability by
nozzle temperature to the excitation of a general vibrational stretch
mode \cite{hol95}.

A molecular beam experiment with  laser excitation of the
$\nu_3$ mode succeeded in measuring a dramatical enhancement of the
dissociation on a Ni(100), which was still, however, much too low to
account for the vibrational activation observed in previous studies.
This indicates that other vibrationally excited modes contribute
significantly to the reactivity of thermal samples \cite{juur99,mcc00}.
It has been shown also that the laser excitation of the $\nu_3$ mode can
be used to deposit carbon on a sub-micrometer scale \cite{juur00}.
Later on, the effect of rotational excitation in combination with the
$\nu_3$ vibrational mode was studied, which indicated that rotational
excitation does not alter reactivity by more than a factor two
\cite{juur00b}. Very recently the effect of a 2$\nu_3$ excitation on the
methane dissociation probability on Pt(111) has been measured
\cite{higgins01}. They find that 72 kJ/mol of vibrational energy in the
excited CH${}_4$ is approximately equivalent to at least 30 kJ/mol
normal translational energy.

Although one does not expect strong surface temperature effects on the
sticking probability for a direct dissociation mechanism, it has been
reported anyway for some metal surfaces. It has been noticed for the
first time by Ref.~\cite{lun89} on Pt(111), but it was not observed by
Ref.~\cite{xxx11}. Later on surface temperature effects were observed on
Ir(110)\cite{see97,see97c}, Ir(111)\cite{see97b}, and
Pt(110)\cite{walker99,walker00} at the lower incident translational energies.

There are further more some other molecular beam experiments, which
studies the effect of adsorbate coverage on the dissociation
probability; oxygen on Pt(111)\cite{val96a}, Pd(110)\cite{val96b}, and
Pt(110)\cite{walker00b}; and the effect of surface alloying coverage; Au
on Ni(111)\cite{hol96}, Co on Cu(111)\cite{lar98,lar98b}, and K on
Ni(100) and Ni(111)\cite{beng99}.  The dissociation probability of
methane was also compared with higher alkanes
\cite{xxx13,mcmaster93,kel97,wea97,wea00}.

Summarizing, we can say that the current opinion about methane
dissociation on transition metals based on the beam molecular studies is
as follow: The dissociation is dominated by a direct dissociation
mechanism, because it is strongly dependent on translational energy. A
trapping mediated mechanism can also play a role on some metals at low
translational energies, because the dissociation probability is
sometimes enhanced by the surface temperature. Vibrational excitation
also enhances strongly the dissociation probability, which suggest a
late barrier for dissociation. State resolved information has only been
reported for the $\nu_3$ asymmetrical stretch mode. This mode enhances
the dissociation probability, but another mode is possibly more
reactive.

The scattering of methane has also been studied with molecular beams on
Ag(111) \cite{asada81,asada82}, Pt(111)
\cite{yagyu99,yagyu99b,yagyu00,hiraoka00}, and Cu(111)
\cite{Andersson00} surfaces. It was reported in
Refs.~\cite{yagyu99,yagyu99b} that the scattering angles are in some
cases anomalous with the outcome of the classical Hard cube model (HCM)
described in Ref.~\cite{logan66}. We will show in Chapter
\ref{chap:traj} that the assumption of the HCM model are too crude for
describing the processes obtained from our simulation. The time of
flight experiments show that there is almost no vibrational excitation
during the scattering \cite{yagyu00,hiraoka00}.

\section{Theoretical studies}

There are multiple ways to study a chemical reaction. One can simulate a
model of the dynamics, or use transition state theory. For both methods
it is necessary to obtain an idea of the interaction between particles
in the system. Information of an interaction is theoretically expressed
as the potential. The potential can be calculated empirical from fitting
experimental data (dissociation energies, spectra, etc.) or from
electronic structure calculation. If we want information about the
thermodynamic properties of the system, then we only have to calculate
the potential energies of the reactant and the products. 

Reaction rates can be obtain with transition state theory. This means
that we calculate the minimum energy path between reactants and
products, which will give us the energy of the transition state.
Transition state theory gives us only thermally average dissociation
probabilities, which is useful if we have to deal with trapping mediated
dissociation mechanism. The main assumption of transition state theory
is that the available thermal energy will be distributed equally over
all modes.  This is not the case for molecular beam studies on a direct
dissociation of methane on transition metal surfaces. The energy can be
distributed unequally over translational, rotational, and vibrational
modes. So it is not in thermal equilibrium. The best way to study
theoretically in detail the contribution of kinetic energy available in
specific modes to the reaction probability from the molecular beam
experiments, is with molecular dynamics simulations. For an accurate
molecular dynamics simulation we do not only need to know the minimum
energy path, but also the potential energy of other possible paths (in
other directions). The reaction path does not necessary have to be the
minimum energy path, but can occur also over a higher barrier which is
better accessible with the available kinetic energy. Non reactive paths
are only important for calculating the reaction probability.  This means
that we need to calculate for more points the potential energy. We can
construct with these points a potential energy surface (PES) in multiple
dimensions.  The number of points we can calculate is limited by the
availability of computer resources. The number of points increases
dramatically with the dimensionality of the dynamical system.  Computer
resources limits also the dimensionality and the time-scale of the
dynamics simulation itself.

% There are multiple ways to study a chemical reaction. One can simulate a
% model of the dynamics, or use transition state theory. For both methods
% it is necessary to obtain an idea of the interaction between particles
% in the system. Information of an interaction is theoretically expressed
% as the potential. The potential can be calculated empirical from fitting
% experimental data (dissociation energies, spectra, etc.) or from
% electronic structure calculation. If we want information about the
% thermodynamic properties of the system, then we only have to calculate
% the potential energies of the reactant and the products. Reaction rates
% can be estimated with transition state theory, if we also calculate the
% energy of the transition state.  Normally this means that we calculate
% the minimum energy path. However, transition state theory gives only
% thermally average dissociation probabilities, and could therefore not be
% compared directly with the detailed molecular beam experiments of a
% direct dissociation mechanism.  For an accurate description we have to
% try to simulate the dynamics, which also means that we need to calculate
% for more point the potential energy. We can construct with these points
% a potential energy surface (PES) in multiple dimensions. The number of
% point we can calculate is limited by the availability of computer
% resource. The number of points increases dramatically with the
% dimensionality of the dynamical system.  Computer resources limits also
% the dimensionality and the time-scale of the dynamics simulation.

\subsection{Electronic structure calculations}

A convenient way of performing electronic structure calculations
involving metal atoms is by using density functional theory (DFT). Full
configurational interaction (CI) are too hard to perform, since the
metal atoms in the surface model consist of too many electrons. The
surface can be either modeled as a cluster of a few metal atoms or as a
periodic slab consisting of a few layers of metal atoms.

%Electronic structure calculations of molecule surface interactions
%cannot be calculated with a full configurational interaction (CI), because
%the metal atoms in the surface model consist of many electrons. Therefore
%an approximation method called density functional theory (DFT) is used
%often to reduce calculation time and memory usage.  The reliability of
%DFT is highly dependent on the used functional. The surface is normally
%modeled as a cluster of a few metal atoms or as a periodic slab
%consisting of a few layers of metal atoms.

The first electronic structure calculations of methane dissociation were
self consisted field (SCF) calculations on Ni(111) and substitutional
Fe/Ni(111) clusters \cite{yan92,yan93}. We performed in our group DFT
calculation on Ni and Co clusters
\cite{bur93a,bur93b,bur93c,bur94,bur95a,burghgraefPhD}.  Others have
done the same on various clusters of Ru, Os, Rh, Ir, Pd, Pt, Cu, Ag, and
Au \cite{au97,liao97,au98,au99}.

DFT calculations on periodic slabs have studied the methane dissociation
reaction on pure and gold-alloyed Ni(111) \cite{kra96}, Ni(100) and
Ni(111) with preadsorbed potassium \cite{beng99,watwe00}, Ru(0001)
\cite{ciob99,ciob00}, and Ir(111) \cite{henkelman01} surfaces.  The
obtained transition state for CH${}_4$ dissociation on Ni(111) involves
considerable internal excitation of the molecule \cite{kra96}.
The breaking of the C--H bond occurs preferentially on top of a Ni atom,
with a dissociation barrier of about 100 kJ/mol. The reaction coordinate
is mainly a C--H stretch.

There is also calculated a PES for the methyl radical and methane in
interaction with nickel surfaces based on the embedded
diatomics-in-molecules (EDIM) formalism, which involves mixing the
semi-empirical diatomics-in-molecules (DIM) valence bond method for the
covalent part of the system with the embedded atom method (EAM) for the
metal \cite{wonchoba98}. 

\subsection{Dynamics simulations}

Wave packet simulations are being used more and more to study the
dynamics of molecule surface reactions. A lot of progress have been made
over the decade especially for the reaction of H${}_2$ on metal surfaces
\cite{darling95,kroes99}. It is very interesting to simulate the
dynamics of the dissociation, because of the direct dissociation
mechanism, and the role of the internal vibrations.  The published wave
packet simulations on the methane dissociation reaction on transition
metals have treated the methane molecule always as a diatomic up to now
\cite{har91,lun91,lun92,lun95,jan95,car98}.  Besides the C--H bond and
molecule surface distance, a combination of other coordinates were
included, like (multiple) rotations and some lattice motion. None of
them have looked at the role of the other internal vibrations.

The first model included the C--H bond, the molecule surface distance,
and a dynamic surface. It has been used for comparison of molecular beam
experiments of the dissociation of methane on Pt(111)
\cite{har91,lun91,lun92}, and Ni(100) \cite{lun95}.  The second included
a rotational degree of freedom, but left out the dynamical surface for a
study on Ni(111) \cite{jan95}. Both models found that the dissociation
occurs via a tunneling mechanism.

The most recent wave packet simulation on the dissociation probability of
CH${}_4$ and CD${}_4$ on Ni(100) used both a dynamic surface and
rotational freedom. It showed a semiquantitative agreement
between the theoretical results and the experiments of Ref.~\cite{hol95},
except for the isotope effect and the extracted vibrational
efficacy \cite{car98}. One of the possible explanation of the incorrect
isotope effect can be the role played by the non-included intramolecular
vibrations.

%\subsection{Classical trajectory simulations}

Finally, I like to mention that there have also been performed some
classical stochastic trajectory simulations \cite{ukr94,sti96}.

%\bibliography{Methane}
%\bibliographystyle{prstyfull}

%% file: potentials.tex
\chapter{Wave packet simulations}
\label{chap:wp}

\begin{quote}
  {\it We have used the multi-configurational time-dependent Hartree
    (MCTDH) method for our wave packet simulations, because it can deal
    with a large number of degrees of freedom and with large grids. We
    have developed four efficient model potentials for this method for
    studying the full vibrational scattering of oriented methane on a
    flat surface. We have described the MCTDH method and the model
    potentials for the simulations discussed in the next three
    chapters.}
\end{quote}

%\vspace*{2cm}

\section{Introduction}

Every wave packet simulation consist of three parts. First we have to
define the potential energy surface (PES) and generate an intial state
(wave packet) of the system. Secondly we have to propagate the wave
packet during a certain time. Finally we have to analyse the final and
intermediate states of the wave packets.

A general problem of computer simulations is that the size and
simulation time of the system under study is limited by the available
computational resources. Wave packet simulations are mainly limited by
the number of dimensions, because the total number of grid point for the
numerical integration scales with the product of the number of grid
points per dimension.  This means that we have to keep the system small
(typically 6 dimensions or less) or we have to use a computational
trick. We did both. We reduced the dimensionality to ten by leaving out
three rotional and two translational coordinates of methane. The
computational trick is that we used an approximation method for our wave
packet propagation for which the total number of grid points scales with
the sum of the number of grid points per dimension.

The published wave packet simulations on the methane dissociation
reaction on transition metals have treated the methane molecule always
as a diatomic up to now \cite{har91,lun91,lun92,lun95,jan95,car98}.
Besides the C--H bond and molecule surface distance, a combination of
other coordinates were included, like (multiple) rotations and some
lattice motion. None of them have looked at the role of the other
internal vibrations.

We are not able yet to simulate the dissociation including all internal
vibrations. Instead we have simulated the scattering of methane, for
which all internal vibrations can be included. We deduce the
consequences for the dissociation at of them. In this chapter I will
give the theoretical background and computational details of our wave
packet simulations of the scattering of fixed oriented methane on a flat
surface described in the Chapters \ref{chap:ch4scat}, \ref{chap:iso},
and \ref{chap:exc}. I describe our wave packet propagation method first.
Then the various potential energy surfaces (PESs) that we have used are
derived. (A harmonic intramolecular PES is adapted to include
anharmonicities in the C--H/D distance, the decrease of C--H/D bond
energy due to interactions with the surface, and the increase of C--H/D
bond length also due to interactions with the surface.) I end with a
description of the initial states.

\section{The MCTDH Method}
\label{sec:mctdh}

The multi-configurational time-dependent Hartree (MCTDH) method is an
approximation method, which has like every approximation method
advantages and disadvantages. Its main advantages are that it can be
several orders of magnitude faster than conventional methods, and it
requires a very small amount of memory. It is especially very usefull if
it has to deal with large number of degrees of freedom and with large
grids.  The main disadvantage is that the MCTDH method requires the
Hamiltonian to be expanded as a sum of products of one-particle
operators. Normally the kinetic energy is already in the right form. The
potential is often only in such a form when considering model problems.
So most of the time it is necessary to transform a potential to MCTDH.
Since the computation time grows linearly with the number of Hamiltonian
terms, the MCTDH method can become slow, if an accurate transformation
requires a lot of potential terms. Another drawback of this is that the
MCTDH method is only fast if the time dependent wavepacket can be
expanded into a small product basis set.  The MCTDH method has been
applied successfully to gas phase reactions and reactions at surfaces,
and further method development is still going on
\cite{man92,jan93,man92b,man93,meyer93,ham94,fan94,fan95,fan95b,fan95c,liu95,cap95,eha96,wor96,jac96,man96,mat96,ger97a,ger97,bec97,jac98a,jac98b,wor98,mil98,meyer98,mat99,jac99,worth99,raab99,raab99b,raab00,raab00b,mil00a,mil00b,worth00,heitz01,burghardt01,worth01,beck01}.
It has been reviewed recently in Ref.~\cite{bec00}.

%\subsection{Formalism}

We give here a short overview of the MCTDH formalism for
completeness. More details can be found in Refs.~\cite{man92} and
\cite{jan93}. The 
exact wave-function of a $D$-dimensional system, is approximated by an
expression of the form
\begin{equation}
\label{eD}
  \Psi_{\rm MCTDH}(q_1,\ldots,q_D;t)
  =\sum_{n_1\ldots n_D}
  c_{n_1\ldots n_D}(t)
  \,\psi_{n_1}^{(1)}(q_1;t)\ldots\psi_{n_D}^{(D)}(q_D;t).
\end{equation}
From this expression, it is possible to obtain
the equations of motion for the one-dimen\-sional functions
$\psi_{n_i}^{(i)}(q_i;t)$ and for the correlation coefficients 
$c_{n_1\ldots n_D}(t)$.  
Without loss of generality we can choose the $\psi_{n_i}^{(i)}$'s
to be natural single-particle states \cite{jan93}.
They insure that we obtain the best approximation to the exact
$\Psi(q_1,\ldots,q_D;t)$ for a fixed number of configurations; i.e.,
they minimize the
expression $\langle\Delta\vert\Delta\rangle$ where 
\begin{equation}
  \Delta=\Psi-\Psi_{\rm MCTDH},
\end{equation}
and where $\Psi$ is the exact wave-function.
The natural single-particle functions are
eigenstates of the reduced density operators.
\begin{eqnarray}
  &&\rho_j(q_j,q_j^\prime)
  =\int\!\!dq_1\ldots dq_{j-1}dq_{j+1}\ldots dq_D\\
  &&\quad\Psi(q_1\ldots q_{j-1}q_jq_{j+1}\ldots q_D)
  \Psi^*(q_1\ldots q_{j-1}q_j^\prime q_{j+1}\ldots q_D).
  \nonumber
\end{eqnarray}
The equations of motions for the natural single-particle states
can be obtained by differentiation of the eigenvalue equation:
\begin{equation}
  \int\!dq_j^\prime\rho_{i}(q_j,q_j^\prime)\psi_{n}^{(i)}(q_j^\prime)
  =\nu_{n}^{(i)}\psi_{n}^{(i)}(q_j).
\label {eq:rho}
\end{equation}
This gives us
\begin{eqnarray}
  i\hbar{\partial\over\partial t}\psi_{n_j}^{(j)}
  &&=h_j\psi_{n_j}^{(j)}+\sum_{m_j}B_{n_jm_j}^{(j)}\psi_{m_j}^{(j)}
  \nonumber\\
  &&+{\langle\tilde\psi_{n_j}^{(j)}\vert V\vert\Psi\rangle
   \over\langle\tilde\psi_{n_j}^{(j)}\vert
   \tilde\psi_{n_j}^{(j)}\rangle}
  -\sum_{m_j}{\langle\psi_{m_j}^{(j)}\tilde\psi_{n_j}^{(j)}
   \vert V\vert\Psi\rangle
   \over\langle\tilde\psi_{n_j}^{(j)}\vert
   \tilde\psi_{n_j}^{(j)}\rangle}
   \psi_{m_j}^{(j)},\label{eB}
\end{eqnarray}
where
\begin{equation}
  B_{n_jm_j}^{(j)}={\langle\psi_{m_j}^{(j)}\tilde\psi_{n_j}^{(j)}
   \vert V\vert\Psi\rangle-\langle\Psi\vert V\vert
   \psi_{n_j}^{(j)}\tilde\psi_{m_j}^{(j)}\rangle\over
   \langle\tilde\psi_{n_j}^{(j)}\vert\tilde\psi_{n_j}^{(j)}\rangle
  -\langle\tilde\psi_{m_j}^{(j)}\vert\tilde\psi_{m_j}^{(j)}\rangle},
  \label{eE}
\end{equation}
and
\begin{equation}
  \tilde\psi_{n_j}^{(j)}=\sum_{n_1\ldots n_{j-1}}
  \sum_{n_{j+1}\ldots n_{D}}c_{n_1\ldots n_D}
  \psi_{n_1}^{(1)}\ldots\psi_{n_{j-1}}^{(j-1)}
  \psi_{n_{j+1}}^{(j+1)}\ldots\psi_{n_D}^{(D)},
\end{equation}
with the Hamiltonian $H$ given by
\begin{equation}
  H=\sum_{j=1}^Dh_j+V.\label{Ham}
\end{equation}
The equations of motion for the coefficients
\begin{equation}
  c_{n_1\ldots n_D}=\langle\psi_{n_1}^{(1)}\ldots\psi_{n_D}^{(D)}
  \vert\Psi\rangle
\end{equation}
are again obtained by differentiation.
\begin{equation}
  i\hbar{d\over dt}c_{n_1\ldots n_D}
  =\langle\psi_{n_1}^{(1)}\ldots\psi_{n_D}^{(D)}\vert
    V\vert\Psi\rangle
  -\sum_{j=1}^Dc_{n_1\ldots n_{j-1}m_jn_{j+1}\ldots n_D}
  B_{m_jn_j}^{(j)}.
  \label{eC}
\end{equation}
Equations~(\ref{eB}) and (\ref{eC}) are a particular form
of the more general equations obtained from a time-dependent variational
principle \cite{man92}.
They conserve the norm of the wave-function and the mean energy
of a time-independent Hamiltonian.
The resulting system of first-order differential equations, 
has to be solved with a general-purpose integrator.
We used the variable-order variable-step Adams method, as implemented in
the NAG library \cite{nag93}. This method gave good convergence for all
described simulations.
The singularities in Eqs.\ (\ref{eB}), (\ref{eE}), and (\ref{eC}) have
been treated numerically by the regularization procedure described
in Ref.~\cite{jan93}.

The natural single-particle states have some small advantages over other
possible choices of single-particle states.
The most important one is that one can directly see from 
$\langle\tilde\psi_{n}^{(i)}\vert\tilde\psi_{n}^{(i)}\rangle$ how well
$\Psi_{\rm MCTDH}$ approximates the exact wave-function.
How much a natural single-particle functions contributes to the wave-function
is given by the eigenvalue $\nu_n^{(i)}$
of the reduced density matrix.
In an approximate MCTDH simulation
$\langle\tilde\psi_{n}^{(i)}\vert\tilde\psi_{n}^{(i)}\rangle$
is an approximation for this exact eigenvalue.
The natural single-particle functions are also convenient for interpreting the
results of a simulations.

%\subsection{Applications}

\section{Potential energy surfaces}
\label{sec:pes}
The PESs we used for our wavepacket simulations (as described in
Chapters~\ref{chap:ch4scat}, \ref{chap:iso}, and \ref{chap:exc}) can all be written as
\begin{equation}
  \label{Vgen}
  V_{\rm total}=V_{\rm intra}+V_{\rm surf},
\end{equation}
where $V_{\rm intra}$ is the intramolecular PES and $V_{\rm surf}$ is the
repulsive interaction with the surface. For the $V_{\rm intra}$
we looked at four different types of PESs. Two of the four different
PESs include changes in the intramolecular potential due to interactions
with the surface in $V_{\rm intra}$.

\subsection{A harmonic potential}

The first one is completely harmonic. We have used normal mode
coordinates for the internal vibrations, because these are coupled only
very weakly.
In the harmonic approximation this coupling is even absent so that we
can write $V_{\rm intra}$ as 
\begin{equation}
  \label{Vharm}
  V_{\rm intra}=V_{\rm harm}= {1\over 2}\sum_{i=2}^{10}k_iX_i^2,
\end{equation}
the summation is over the internal vibrations, $X_i$'s are mass-weighted
displacement coordinates and $k_i$ are mass-weighted force constants. (see
Table \ref{tab:gen} for definitions and values); ($X_1$ is the
mass-weighted overall translation along the surface normal) \cite{wil55}.
The force constants have been obtained by fitting them on the
experimental vibrational frequencies of CH${}_4$ and
CD${}_4$ \cite{gray79,lee95}.

\begin{table}[t]
  \caption{Overview of the relations between the mass-weighted coordinates
  $X_i$; the force constants $k_i$ (in atomic units), the designation,
  and the symmetry in T${}_d$, C${}_{3v}$ and C${}_{2v}$.} 
  \label{tab:gen}
  \begin{tabular}{r l l l c c c}
    \hline $i$ & $k_i$(CH${}_4$) & $k_i$(CD${}_4$) & designation &
    T${}_d$ &
    C${}_{3v}$ & C${}_{2v}$ \\
    \hline
    $1$ &                      & & translation & $t_2$ & $a_1$ & $a_1$ \\
    $2$ & $1.780\cdot 10^{-4}$ & $8.897\cdot 10^{-5}$ & $\nu_1$;
    symmetrical stretch & $a_1$ & $a_1$  & $a_1$ \\
    $3$ & $3.599\cdot 10^{-5}$ &  $2.008\cdot 10^{-5}$ & $\nu_4$; umbrella & $t_2$ & $a_1$ & $a_1$ \\
    $4$ & $1.892\cdot 10^{-4}$ &  $1.060\cdot 10^{-4}$ & $\nu_3$;
    asymmetrical stretch & $t_2$ & $a_1$ & $a_1$ \\
    $5$ & $4.894\cdot 10^{-5}$ & $2.447\cdot 10^{-5}$  & $\nu_2$;
    bending & $e$ & $e$ & $a_1$ \\ 
    $6$ & $4.894\cdot 10^{-5}$ & $2.447\cdot 10^{-5}$  & $\nu_2$;
    bending & $e$ & $e$ & $a_2$ \\ 
    $7$ & $3.599\cdot 10^{-5}$ &  $2.008\cdot 10^{-5}$ & $\nu_4$;
    umbrella & $t_2$ & $e$ & $b_1$ \\ 
    $8$ & $3.599\cdot 10^{-5}$ &  $2.008\cdot 10^{-5}$ & $\nu_4$;
    umbrella & $t_2$ & $e$ & $b_2$ \\ 
    $9$ & $1.892\cdot 10^{-4}$ &  $1.060\cdot 10^{-4}$ & $\nu_3$;
    asymmetrical stretch & $t_2$ &  $e$ & $b_1$ \\
    $10$ & $1.892\cdot 10^{-4}$ &  $1.060\cdot 10^{-4}$ & $\nu_3$;
    asymmetrical stretch & $t_2$  & $e$ & $b_2$ \\
    \hline
  \end{tabular}
\end{table}

We have assumed that the repulsive interaction with the surface is only
through the hydrogen or deuterium atoms that point towards the surface.
We take the $z$-axis as the surface normal. In this case the surface PES
is given by
\begin{equation}
  \label{gen_Vsurf}
  V_{\rm surf}={A \over{N_H}}\sum_{i=1}^{N_H} e^{-\alpha z_i},
\end{equation}
where $N_H$ is the number of hydrogens or deuteriums that points towards
the surface, $\alpha$=1.0726 atomic units and $A$=6.4127 Hartree. These
parameters are chosen to give the same repulsion as the PES that has
been used in an MCTDH wavepacket simulation of CH${}_4$
dissociation \cite{jan95}.

\begin{figure}[t]
  \begin{center}
    \epsfig{file=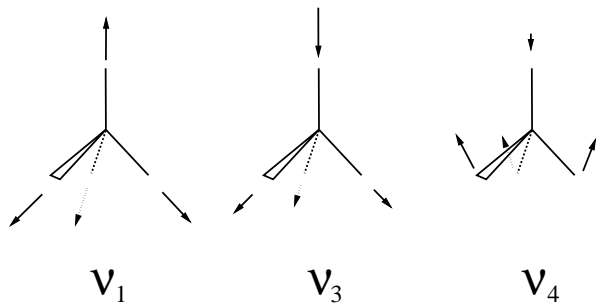, height=4cm}
  \end{center}
  \caption{The $a_1$ vibrational normal modes in the C${}_{3v}$ symmetry;
    $\nu_1$ symmetrical stretch (X${}_2$), $\nu_3$ asymmetrical stretch
    (X${}_4$), and $\nu_4$ umbrella (X${}_3$).}
  \label{fig:modes}
\end{figure}

\begin{figure}[p]
  \epsfig{file=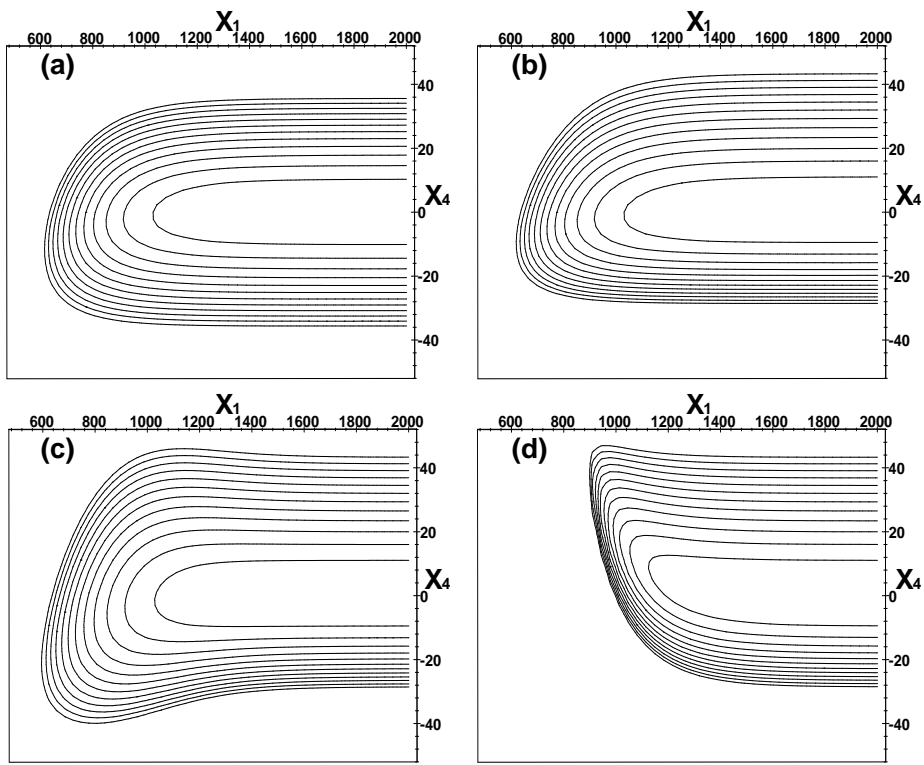, height=10cm}
    \caption{Contour plots in the energy range from 0 to 0.12 Hartree of
      a cross-section in the orientation with one hydrogen atom pointing
      towards the surface for the translational mode $X_1$ and the
      $\nu_3$ asymmetrical mode $X_4$ (an $a_1$ mode in C${}_{3v}$),
      both in atomic displacement, which show well the different
      behaviours of the four PESs; (a) harmonic
      PES, (b) an anharmonic intramolecular PES, (c)
      intramolecular PES with weakening C--H bond, and (d)
      intramolecular PES with elongation of the C--H bond. All
      other coordinates set to zero.} 
    \label{fig:potdif}
\end{figure}

If we write $V_{\rm surf}$ in terms of normal mode coordinates, then we
obtain for one hydrogen or deuterium pointing towards the surface 
\begin{equation}
  \label{vsurfone}
  V_{\rm surf}= A
  e^{-\alpha_1X_1}e^{-\alpha_2X_2}e^{-\alpha_3X_3}e^{-\alpha_4X_4},
\end{equation}
where $A$ as above, and $\alpha$'s as given in Tables \ref{tab:alpha}
and \ref{tab:alphaD}.
$X_2$, $X_3$ and $X_4$ correspond all to $a_1$ modes of the C${}_{3v}$
symmetry (see Fig.~\ref{fig:modes}). There is no coupling between the
modes $X_5$ to $X_{10}$ in the $V_{\rm surf}$ part of the PES, which are
all $e$ modes of the C${}_{3v}$ symmetry. Figure~\ref{fig:potdif}(a) shows
a contour plot of the cross-section of the total harmonic PES with one
hydrogen pointing towards the surface in the translational mode $X_1$
and the $\nu_3$ asymmetrical stretch mode $X_4$.

\begin{table}[t]
  \caption{$\alpha$ and $\beta$ values (in atomic units) of $V_{\rm
      surf}$ for CH${}_4$ with one, two or three hydrogens pointing
    towards the surface [see Eqs.\ (\ref{vsurfone}), (\ref{vsurftwo})
    and (\ref{vsurfthree})].}
  \label{tab:alpha}
  \begin{tabular}{l r r r}
    \hline
    & one & two & three \\
    \hline
    $\alpha_1$ & $6.281\cdot 10^{-3}$  & $6.281\cdot 10^{-3}$  &
    $6.281\cdot 10^{-3}$ \\ 
    $\alpha_2$ & $1.256\cdot 10^{-2}$  & $7.252\cdot 10^{-3}$  &
    $4.187\cdot 10^{-3}$ \\
    $\alpha_3$ & $4.226\cdot 10^{-3}$  & $-7.931\cdot 10^{-3}$ &
    $-1.198\cdot 10^{-2}$ \\
    $\alpha_4$ & $-2.040\cdot 10^{-2}$ & $-7.445\cdot 10^{-3}$ &
    $-3.128\cdot 10^{-3}$ \\
    $\alpha_5$ &                       & $-1.026\cdot 10^{-2}$  &
    \\
    $\beta_{1}$ & &                          & $5.921\cdot 10^{-3}$ \\
    $\beta_{2}$ & &                          & $1.026\cdot 10^{-2}$ \\
    $\beta_{3}$ & & $6.079\cdot 10^{-3}$ & $2.866\cdot 10^{-3}$ \\
    $\beta_{4}$ & &                      & $4.963\cdot 10^{-3}$ \\
    $\beta_{5}$ & & $6.476\cdot 10^{-3}$ & $3.053\cdot 10^{-3}$ \\
    $\beta_{6}$ & &                      & $5.288\cdot 10^{-3}$ \\
    \hline
  \end{tabular}
\end{table}

\begin{table}[t]
  \caption{$\alpha$ and $\beta$ values (in atomic units) of $V_{\rm
    surf}$ for CD${}_4$ with one, two or three deuteriums pointing
    towards the surface 
    [see Eqs.\  (\ref{vsurfone}), (\ref{vsurftwo}) and (\ref{vsurfthree})].} 
  \label{tab:alphaD}
  \begin{tabular}{l r r r}
    \hline
    & one & two & three \\
    \hline
    $\alpha_1$ & $5.617\cdot 10^{-3}$  & $5.617\cdot 10^{-3}$  &
    $5.617\cdot 10^{-3}$ \\ 
    $\alpha_2$ & $8.882\cdot 10^{-3}$  & $5.128\cdot 10^{-3}$  &
    $2.960\cdot 10^{-3}$ \\
    $\alpha_3$ & $4.703\cdot 10^{-3}$  & $-4.614\cdot 10^{-3}$ &
    $-7.720\cdot 10^{-3}$ \\
    $\alpha_4$ & $-1.353\cdot 10^{-2}$ & $-5.103\cdot 10^{-3}$ &
    $-2.295\cdot 10^{-3}$ \\
    $\alpha_5$ &                       & $-7.252\cdot 10^{-3}$  &
    \\
    $\beta_{1}$ & &                          & $4.187\cdot 10^{-3}$ \\
    $\beta_{2}$ & &                          & $7.252\cdot 10^{-3}$ \\
    $\beta_{3}$ & & $4.659\cdot 10^{-3}$ & $2.196\cdot 10^{-3}$ \\
    $\beta_{4}$ & &                      & $3.804\cdot 10^{-3}$ \\
    $\beta_{5}$ & & $4.212\cdot 10^{-3}$ & $2.295\cdot 10^{-3}$ \\
    $\beta_{6}$ & &                      & $3.439\cdot 10^{-3}$ \\
    \hline
  \end{tabular}
\end{table}

For two hydrogens or deuteriums we obtain
\begin{eqnarray}
  \label{vsurftwo}
  V_{\rm surf}=A&&e^{-\alpha_1X_1}e^{-\alpha_2X_2}
  e^{-\alpha_3X_3}e^{-\alpha_4X_4}e^{-\alpha_5X_5}\\
  \times{1\over2}\Big[
  &&e^{\beta_{3}X_7}
    e^{-\beta_{3}X_8}e^{-\beta_{5}X_9}e^{\beta_{5}X_{10}}
    \nonumber\\
  +&&e^{-\beta_{3}X_7}
    e^{\beta_{3}X_8}e^{\beta_{5}X_9}e^{-\beta_{5}X_{10}}
    \Big],\nonumber
\end{eqnarray}
with $A$ again as above, $\alpha$'s and $\beta$'s as given in Tables
\ref{tab:alpha} and \ref{tab:alphaD}.
$X_2$, $X_3$, $X_4$ and $X_5$ correspond all to $a_1$ modes of C${}_{2v}$. 
$X_7$, $X_8$, $X_9$ and $X_{10}$ correspond to $b_1$ and $b_2$ modes of
C${}_{2v}$. $X_6$ corresponds to the $a_2$ mode of C${}_{2v}$ and
has no coupling with the other modes in $V_{\rm surf}$.

For three hydrogens or deuteriums we obtain
\begin{eqnarray}
  \label{vsurfthree}
  V_{\rm surf}=A&&e^{-\alpha_1X_1}e^{-\alpha_2X_2}
  e^{-\alpha_3X_3}e^{-\alpha_4X_4}\\
  \times{1\over 3}\Big[
  &&e^{\beta_{1}X_5}e^{\beta_{2}X_6}e^{-\beta_{3}X_7}
    e^{-\beta_{4}X_8}e^{\beta_{5}X_9}e^{\beta_{6}X_{10}}
    \nonumber\\
  +&&e^{\beta_{1}X_5}e^{-\beta_{2}X_6}e^{-\beta_{3}X_7}
    e^{\beta_{4}X_8}e^{\beta_{5}X_9}e^{-\beta_{6}X_{10}}
    \nonumber\\
  +&&e^{-2\beta_{1}X_5}e^{2\beta_{3}X_7}e^{-2\beta_{5}X_9}
    \Big],\nonumber
\end{eqnarray}
with $A$ again as above, $\alpha$'s and $\beta$'s as given in Table
\ref{tab:alpha} and \ref{tab:alphaD}. $X_2$, $X_3$ and $X_4$ corresponds
to $a_1$ modes in the C${}_{3v}$ symmetry (see Fig.~\ref{fig:modes}).
Because these last six coordinates correspond to degenerate $e$ modes of
the C${}_{3v}$ symmetry, the $\beta$ parameters are not unique.

%
% Morse Potential
%
\subsection{An anharmonic intramolecular potential}
\label{sec:morse}

Even though we do not try to describe the dissociation of methane in
this paper, we do want to determine which internal vibration might be
important for this dissociation. The PES should
at least allow the molecule to partially distort as when
dissociating. The harmonic PES does not do this. A number of
changes have therefor been made. The first is that we have describe the
C--H/D bond by a Morse PES.
\begin{equation}
  \label{Vmorse_n}
  V_{\rm Morse}=D_e \sum_{i=1}^4\Big[1-e^{-\gamma \Delta r_i}\Big]^2 ,
\end{equation}
where $D_e=0.1828$ Hartree (the dissociation energy of methane in the
gas-phase) and $\Delta r_i$ the change in bond length from the
equilibrium distance. $\gamma$ was
calculated by equating the second derivatives along one bond of the
harmonic and the Morse PES. 
If we transform Eq.\ (\ref{Vmorse_n}) back into normal mode
coordinates, we obtain
\begin{equation}
  \label{Vmorse_d}
  V_{\rm Morse}=D_e
  \sum_{i=1}^4\Big[1-e^{\gamma_{i2}X_2}e^{\gamma_{i3}X_3}e^{\gamma_{i4}X_4}
  e^{\gamma_{i7}X_7}e^{\gamma_{i8}X_8}e^{\gamma_{i9}X_9}
  e^{\gamma_{i,10}X_{10}}\Big]^2,
\end{equation}
with $D_e$ as above. $\gamma$'s are given in Tables \ref{tab:gammaone},
\ref{tab:gammatwo}, \ref{tab:gammaoneD}, and \ref{tab:gammatwoD}. Note
that, although we have only changed the PES of the bond lengths, the
$\nu_4$ umbrella modes are also affected. This is because these modes
are not only bending, but also contain some changes of bond length.

\begin{table}[t]
  \caption{$\gamma$ values (in atomic units) of $V_{\rm Morse}$ for
    CH${}_4$ with one
    and three hydrogens pointing towards the surface [see
    Eq.\  (\ref{Vmorse_d})].} 
  \label{tab:gammaone}
  \begin{tabular}{l l r}
    \hline
    one & three & value\\
    \hline
    $\gamma_{12},\gamma_{22}, \gamma_{32}, \gamma_{42}$ &
    $\gamma_{12}, \gamma_{22}, \gamma_{32}, \gamma_{42}$ &
    $1.079\cdot 10^{-2}$ \\ 
    $\gamma_{13}, -3\gamma_{23}, -3\gamma_{33}, -3\gamma_{43}$ & 
    $-\gamma_{13}, 3\gamma_{23}, 3\gamma_{33}, 3\gamma_{43}$ &
    $1.359\cdot 10^{-3}$ \\
    $\gamma_{14}, -3\gamma_{24}, -3\gamma_{34}, -3\gamma_{44}$ &
    $-\gamma_{14}, 3\gamma_{24}, 3\gamma_{34}, 3\gamma_{44}$  &
    $-1.966 \cdot 10^{-2}$ \\
    $\gamma_{17}, \gamma_{18}, \gamma_{19}, \gamma_{1,10}, \gamma_{28}, 
    \gamma_{2,10}$ & 
    $\gamma_{17}, \gamma_{18}, \gamma_{19}, \gamma_{1,10}, \gamma_{28},  
    \gamma_{2,10}$  & $0.0$ \\
    $\gamma_{27}, -2\gamma_{37}, -2\gamma_{47}$ &
    $-\gamma_{27}, 2\gamma_{37}, 2\gamma_{47}$ & $1.282 \cdot 10^{-3}$
    \\
    $\gamma_{38}, -\gamma_{48}$ & $\gamma_{38}, -\gamma_{48}$ &
    $-1.110 \cdot 10^{-3}$ \\ 
    $\gamma_{29}, -2\gamma_{39}, -2\gamma_{49}$ &
    $-\gamma_{29}, 2\gamma_{39}, 2\gamma_{49}$ & $-1.853 \cdot
    10^{-2}$ \\
    $\gamma_{3,10}, -\gamma_{4,10}$ & $-\gamma_{3,10}, \gamma_{4,10}$ &
    $1.605 \cdot 10^{-2}$ \\  
    \hline
  \end{tabular}
\end{table}

\begin{table}[b]
  \caption{$\gamma$ values (in atomic units) of $V_{\rm Morse}$ for
    CH${}_4$ with two hydrogens pointing towards the surface [see Eq.\ 
    (\ref{Vmorse_d})].}
  \label{tab:gammatwo}
  \begin{tabular}{l  r}
    \hline
    two & value\\
    \hline $\gamma_{12}, \gamma_{22}, \gamma_{32}, \gamma_{42}$ &
    $1.079\cdot
    10^{-2}$ \\
    $\gamma_{13}, \gamma_{23}, -\gamma_{33}, -\gamma_{43}, \gamma_{17},\ 
    -\gamma_{27},$\\ $\ \gamma_{37}, -\gamma_{47}, \gamma_{18},
    -\gamma_{28},
    -\gamma_{38}, \gamma_{48}$ & $-7.849 \cdot 10^{-4}$  \\
    $\gamma_{14}, \gamma_{24}, -\gamma_{34}, -\gamma_{44}, \gamma_{19},
    -\gamma_{29},$\\$\ \gamma_{39}, -\gamma_{49}, \gamma_{1,10},
    -\gamma_{2,10}, -\gamma_{3,10}, \gamma_{4,10}$ & $1.135 \cdot
    10^{-2}$ \\
    \hline
  \end{tabular}
\end{table}

\begin{table}[t]
  \caption{$\gamma$ values (in atomic units) of $V_{\rm Morse}$ for
    CD${}_4$ with one and three deuteriums pointing towards the surface
    [see Eq.\  (\ref{Vmorse_d})].} 
  \label{tab:gammaoneD}
  \begin{tabular}{l l r}
    \hline
    one & three & value\\
    \hline
    $\gamma_{12},\gamma_{22}, \gamma_{32}, \gamma_{42}$ &
    $\gamma_{12}, \gamma_{22}, \gamma_{32}, \gamma_{42}$ &
    $7.629\cdot 10^{-3}$ \\ 
    $\gamma_{13}, -3\gamma_{23}, -3\gamma_{33}, -3\gamma_{43}$ & 
    $-\gamma_{13}, 3\gamma_{23}, 3\gamma_{33}, 3\gamma_{43}$ &
    $1.397\cdot 10^{-3}$ \\
    $\gamma_{14}, -3\gamma_{24}, -3\gamma_{34}, -3\gamma_{44}$ &
    $-\gamma_{14}, 3\gamma_{24}, 3\gamma_{34}, 3\gamma_{44}$  &
    $-1.454 \cdot 10^{-2}$ \\
    $\gamma_{17}, \gamma_{18}, \gamma_{19}, \gamma_{1,10}, \gamma_{28}, 
    \gamma_{2,10}$ & 
    $\gamma_{17}, \gamma_{18}, \gamma_{19}, \gamma_{1,10}, \gamma_{28},  
    \gamma_{2,10}$  & $0.0$ \\
    $\gamma_{27}, -2\gamma_{37}, -2\gamma_{47}$ &
    $-\gamma_{27}, 2\gamma_{37}, 2\gamma_{47}$ & $1.318 \cdot 10^{-3}$
    \\
    $\gamma_{38}, -\gamma_{48}$ & $\gamma_{38}, -\gamma_{48}$ &
    $-1.114 \cdot 10^{-3}$ \\ 
    $\gamma_{29}, -2\gamma_{39}, -2\gamma_{49}$ &
    $-\gamma_{29}, 2\gamma_{39}, 2\gamma_{49}$ & $-1.371 \cdot
    10^{-2}$ \\
    $\gamma_{3,10}, -\gamma_{4,10}$ & $-\gamma_{3,10}, \gamma_{4,10}$ &
    $1.187 \cdot 10^{-2}$ \\  
    \hline
  \end{tabular}
\end{table}

\begin{table}[b]
  \caption{$\gamma$ values (in atomic units) of $V_{\rm Morse}$ for
    CD${}_4$ with two deuteriums pointing towards the surface [see Eq.\
    (\ref{Vmorse_d})].}   
  \label{tab:gammatwoD}
  \begin{tabular}{l  r}
    \hline
    two & value\\
    \hline
    $\gamma_{12}, \gamma_{22}, \gamma_{32}, \gamma_{42}$ & $7.629\cdot
    10^{-3}$ \\ 
    $\gamma_{13}, \gamma_{23}, -\gamma_{33}, -\gamma_{43},
    \gamma_{17}, 
     -\gamma_{27},$\\ $\ \gamma_{37}, -\gamma_{47}, \gamma_{18}, -\gamma_{28},
    -\gamma_{38}, \gamma_{48}$ & $-8.070 \cdot 10^{-4}$  \\
    $\gamma_{14}, \gamma_{24}, -\gamma_{34}, -\gamma_{44}, \gamma_{19},
    -\gamma_{29},$\\ $\ \gamma_{39}, -\gamma_{49}, \gamma_{1,10},
    -\gamma_{2,10}, -\gamma_{3,10}, \gamma_{4,10}$ & $8.396 \cdot
    10^{-3}$ \\ 
    \hline
  \end{tabular}
\end{table}

The new intramolecular PES now becomes
\begin{equation}
  \label{Vintra_morse}
  V_{\rm intra}=V_{\rm harm}+V_{\rm Morse}-V_{\rm corr}, 
\end{equation}
where $V_{\rm harm}$ is as given in Eq.\ (\ref{Vharm}) and $V_{\rm
  corr}$ is the quadratic part of $V_{\rm Morse}$, which is already in
$V_{\rm harm}$.  Figure~\ref{fig:potdif}(b) shows a contour plot of the
cross-section of this total anharmonic PES with one hydrogen pointing
towards the surface in the translational mode $X_1$ and the $\nu_3$
asymmetrical stretch mode $X_4$.  (We would like to point out that there
are, of course, various anharmonic PESs for methane in the literature.
There are two reasons why we haven't use them. First, these PESs are not
in appropriate form to use then with the MCTDH method \cite{man92,jan93}.
Second, these PESs are generally quite complicated. We prefer to keep it
as simple as possible, because at this moment we're only interested in
qualitative effects.)

%
% Weak Potential
%
\subsection{Intramolecular potential with weakening bonds}
\label{sec:weak}

When the methane molecule approach the surface the overlap of substrate
orbitals and anti-bonding orbitals of the molecule weakens the C--H bonds.
We want to include this effect for the C--H bonds of the hydrogens
pointing towards the surface. We have redefined the $V_{\rm Morse}$ 
given in Eq.\ (\ref{Vmorse_d}) and also replace it in
Eq.\ (\ref{Vintra_morse}). A sigmoidal function is used to switch from the
gas phase C--H bond to a bond close to the surface. We have used the
following, somewhat arbitrary, approximations. 
(i) The point of inflection should be at a reasonable distance from the
surface. It is set to the turnaround point for a rigid methane molecule
with translation energy 93.2 kJ/mol plus twice the fall-off distance of
the interaction with the surface. (ii) The depth of the PES of the C--H
bond is 480 kJ/mol in the gas phase, but only 93.2 kJ/mol near the
surface. The value 93.2 kJ/mol corresponds to the height of the
activation barrier used in our dissociation \cite{jan95}. (iii) The
exponential factor is the same as for the interaction with the surface. 

If we transform to normal-mode coordinates for the particular
orientations, we then obtain
\begin{equation}
  \label{Vweak_d}
  V_{\rm weak}=D_e
  \sum_{i=1}^4W_i\Big[1-e^{\gamma_{i2}X_2}e^{\gamma_{i3}X_3}
  e^{\gamma_{i4}X_4}
  e^{\gamma_{i7}X_7}e^{\gamma_{i8}X_8}e^{\gamma_{i9}X_9}
  e^{\gamma_{i,10}X_{10}}\Big]^2,
\end{equation}
where $W_i=1$ for non-interacting bonds and
\begin{equation}
  \label{Wi}
  W_i= { {1 + \Omega e^{-\alpha_1X_1 + \omega}} \over {1 + 
  e^{-\alpha_1X_1 + \omega}} }  
\end{equation}
for the interacting bonds pointing towards the surface.  $\alpha_1$ is
as given in Table \ref{tab:alpha}, $\gamma$'s are given in Tables
\ref{tab:gammaone}, \ref{tab:gammatwo}, \ref{tab:gammaoneD}, and
\ref{tab:gammatwoD}, $\Omega=1.942 \cdot10^{-1}$ and $\omega=7.197$.
Figure~\ref{fig:potdif}(c) shows a contour plot of the cross-section of
this total anharmonic PES with weakening C--H bond for the orientation
with one hydrogen pointing towards the surface in the translational mode
$X_1$ and the $\nu_3$ asymmetrical stretch mode $X_4$.

%
% The Shift PES
%
\subsection{Intramolecular potential with elongation of the bond}
\label{sec:shift}

A weakened bond generally has not only a reduced bond strength, but also
an increased bond length.  We include the effect of the elongation of
the C--H/D bond length of the hydrogens or deuteriums that point towards
the surface due to interactions with the surface.  We have redefined the
$V_{\rm Morse}$ given in Eq.\ (\ref{Vmorse_d}) and also replace it in
Eq.\ (\ref{Vintra_morse}) for this type of PES.  We have used the
following approximations: (i) The transition state, as determined by
Refs.\cite{bur93b,bur93c}, has a C--H bond that is 0.54 {\AA} longer
than normal. This elongation should occur at the turn around point for a
rigid methane molecule with a translation energy of 93.2 kJ/mol.  (ii)
The exponential factor is again the same as for the interaction with the
surface.

\begin{figure}[p]
  \epsfig{file=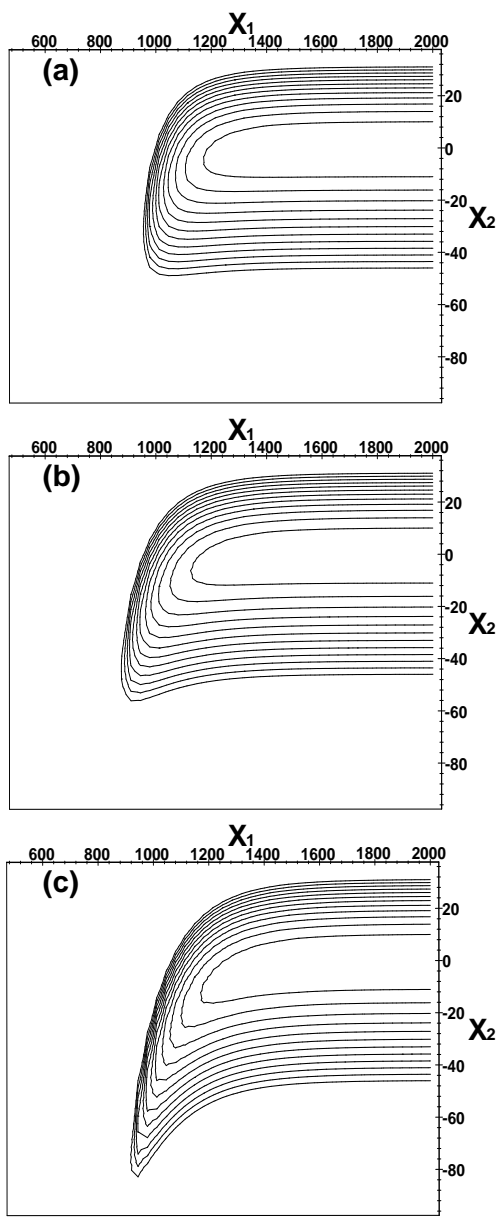, height=16cm}
    \caption{Contour plots in the energy range from 0 to 0.12 Hartree
      of a cross-section of the PES with  
      elongation of the C--H bond in the translational $X_1$
      and the $\nu_1$ symmetrical stretch mode $X_2$, both in
      atomic displacement; 
      (a) one, (b) two, and (c) three hydrogens pointing towards the
      surface. All other coordinates set to zero.} 
    \label{fig:symshift}
\end{figure}

If we transform to normal-mode coordinates for the particular
orientations, then we obtain
\begin{equation}
  \label{Vshift_d}
  V_{\rm shift}=D_e
  \sum_{i=1}^4\Big[1-e^{\gamma_{i2}X_2}e^{\gamma_{i3}X_3}e^{\gamma_{i4}X_4}
  e^{\gamma_{i7}X_7}e^{\gamma_{i8}X_8}e^{\gamma_{i9}X_9}
  e^{\gamma_{i,10}X_{10}}\exp[S_ie^{-\alpha_1X_1}]\Big]^2 ,
\end{equation}
where $\alpha_1$ is as given in Tables \ref{tab:alpha} and
\ref{tab:alphaD}, $\gamma$'s are given in Tables \ref{tab:gammaone},
\ref{tab:gammatwo}, \ref{tab:gammaoneD}, and \ref{tab:gammatwoD}. For
orientation with one hydrogen or deuterium towards the surface we
obtain; $S_1=2.942 \cdot 10^{2}$ and $S_2=S_3=S_4=0$, with two hydrogens
or deuteriums; $S_1=S_2=0$ and $S_3=S_4=1.698 \cdot 10^{2}$, and with
three hydrogens or deuteriums; $S_1=0$ and $S_2=S_3=S_4=2.942 \cdot
10^{2}$.  Figure~\ref{fig:potdif}(d) shows a contour plot of the
cross-section of this total anharmonic PES with elongation of the C--H
bond for the orientation with one hydrogen pointing towards the surface
in the translational mode $X_1$ and the $\nu_3$ asymmetrical stretch
mode $X_4$, and Fig.~\ref{fig:symshift} shows a contour plot of a
cross-section in the translational $X_1$ and the $\nu_1$ symmetrical
stretch mode $X_2$ in the three different orientations.

Finally we like to note that the monotonic behaviour of all PES types in
the translational mode does not contradict with the fact that the
dissociative adsorption of methane is activated. The activation barrier
for dissociative adsorption in the translational mode is situated in a
region with very high excitations of the stretch modes, which we do not
reach in these simulations.

\section{Initial States}
\label{sec:states}

All initial states in the simulations start with the vibrational
ground state. The initial translational part $\psi^{({\rm tr})}$ is
represented by a Gaussian wave-packet, 
\begin{equation}
  \label{Gausswave}
  \psi^{({\rm tr})}(X_1)=(2\pi\sigma^2)^{-1/4}
  \exp\left[-{(X_1-X_0)^2\over 4\sigma^2}+
  iP_1X_1\right],
\end{equation}
where $\sigma$ is the width of the wave-packet (we used $\sigma=320.248$
atomic units), $X_0$ is the initial position (we used $X_0=11\sigma$,
which is far enough from the surface to observe no repulsion) and $P_1$
is the initial momentum. We performed simulations in the energy range of
32 to 96 kJ/mol. We present in Chapters \ref{chap:ch4scat} and
\ref{chap:iso} only the results of 96 kJ/mol (equivalent to $P_1=0.2704$
atomic units), because they showed the most obvious excitation
probabilities for $V_{\rm Morse}$. We used seven natural single-particle
states, 512 grid points and a grid-length of $15\sigma$ for the
translational coordinate. With this grid-width we can perform simulation
with a translational energy up to 144 kJ/mol.

Gauss-Hermite discrete-variable representations (DVR) \cite{light85}
were used to 
represent the wavepackets of the vibrational modes.
We used for all simulations 5 DVR points for the $\nu_2$ bending modes and 8
DVR points for the $\nu_4$ umbrella, $\nu_3$ asymmetrical
stretch, and $\nu_1$ symmetrical stretch mode for an numerical exact
integration, except for the simulations with $V_{\rm shift}$,
where we used 16 DVR points for the $\nu_1$ symmetrical stretch mode,
because of the change in the equilibrium position.

We did the simulation with one hydrogen pointing towards the
surface in eight dimensions, because the $\nu_2$ bending modes $X_5$ and
$X_6$ do not couple with the other modes. We needed four natural
single-particle states for modes $X_2$, $X_3$ and $X_4$, and just one
for the others. So the number of configurations was $7^1 \cdot 4^3 \cdot
1^4 = 448$.  
The simulation with two hydrogens pointing towards the surface was
performed in nine dimensions. One of the $\nu_2$ bending mode ($X_6$)
does not couple with the other modes, but for the other mode $X_5$ we
needed four natural single-particle states. The number of configurations was
$7^1 \cdot 4^4 \cdot 1^4 = 1792$, because we needed the same number of
natural single-particle states as mentioned above for the other modes.
We needed ten dimensions to perform the simulation with three hydrogens
pointing towards the surface. We used here one natural single-particle
state for the modes $X_5$ to $X_{10}$ and four natural single-particle
states for $X_2$ to $X_4$, which gave us $7^1 \cdot 4^3 \cdot 1^6 = 448$
configurations. 

%\bibliography{Methane}
%\bibliographystyle{prstyfull}

%% file: ch4scattering.tex
\chapter{Vibrational scattering of methane}
\label{chap:ch4scat}

\begin{quote}
  {\it We present results of wavepacket simulations of scattering of an
    oriented methane molecule from a flat surface including all nine
    internal vibrations. At a translational energy up to 96 kJ/mol we
    find that the scattering is almost completely elastic. Vibrational
    excitations when the molecule hits the surface and the corresponding
    deformation depend on generic features of the potential energy
    surface. In particular, our simulation indicate that for methane to
    dissociate the interaction of the molecule with the surface should
    lead to an elongated equilibrium C--H bond length close to the
    surface.}\footnote{This chapter has been published as a part of {\sl
      Ten-dimensional wave packet simulations of methane scattering}
    \cite{mil98}, \copyright \ {\sl 1998 American Institute of Physics}.}
\end{quote}

%\vspace*{2cm}

\section{Introduction}

At the time I started my research, our group had just performed a
multi-configurational time-dependent Hartree (MCTDH) study of CH${}_4$
dissociation on Ni(111)\cite{jan95} with a potential energy surface
(PES) based on earlier density functional theory (DFT) calculations
within the group \cite{bur93a,bur93b,bur93c,bur94,bur95a}. The
simulations included the distance of the methane molecule to the
surface, a C--H distance, and the orientation of methane as coordinates.
Previous wavepacket simulations focused on the molecule-surface and C--H
distance too, and combination it with lattice motion on several
metals\cite{lun91,lun92,lun95}. When I finished the simulations
described in this chapter, a wave packet simulation of methane
dissociation on a Ni(111) atop site including all these coordinates
\cite{car98}. None of the wavepacket simulations published so far have
looked at the role of the internal vibrations of methane. It was
observed experimentally at the time that vibrationally hot ${\rm CH}_4$
dissociates more readily than cold ${\rm CH}_4$, with the energy in the
internal vibrations being about as effective as the translational energy
in inducing dissociation \cite{ret85,ret86,lee87,hol95,lun89}. However,
a more detailed assessment of the importance of the internal vibrations
could not be made, because of the large number of internal vibrations.
A DFT calculation at that time also showed that the transition state for
CH${}_4$ dissociation on Ni(111) involves considerable internal
excitation of the molecule \cite{kra96}.

In this chapter we report on wavepacket simulations that we have done to
determine which and to what extent internal vibrations are important for
the dissociation of methane on transition-metals. We are not able yet to
simulate the dissociation including all internal vibrations. Instead we
have simulated the scattering of methane, for which all internal
vibrations can be included. By looking at vibrational excitations and
the deformation of the molecule when it hits the surface we can derive
information that is relevant for the dissociation.  We have used model
PESs that have been developed with Ni(111) in mind, but our results
should hold for other surfaces as well. The various model PESs we have
derived are described in Section~\ref{sec:pes} of this thesis. A harmonic
intramolecular PES [$V_{\rm intra}$, see Eq.(\ref{Vharm})] is adapted to
include anharmonicities in the C--H distance [$V_{\rm Morse}$, see
Eq.(\ref{Vmorse_d})], the decrease of the C--H bond energy due to
interactions with the surface [$V_{\rm weak}$, see Eq.(\ref{Vweak_d})],
and the increase of the C--H bond length also due to interactions with
the surface [$V_{\rm shift}$, see Eq.(\ref{Vshift_d})].  We have used
the multi-confi\-guratio\-nal time-depen\-dent Hartree (MCTDH) method
for our wave packet simulation \cite{man92,jan93}. This method can deal
with a large number of degrees of freedom and with large grids.  (See
Ref.~\cite{bec00} for a recent review, or Section~\ref{sec:mctdh} of this
thesis for a survey.)  Initial translational energy has been chosen in
the range of 32 to 96 kJ/mol.  We present only the results of the
simualtions at 96 kJ/mol here (equivalent to $P_1=-0.2704$ mass-weighted
atomic units), because they showed the most obvious excitation
probabilties for $V_{\rm Morse}$.  The initial state has been written as
a product state of ten functions; one for the normally incident
translational coordinate, and one for each internal vibration. All
vibrations were taken to be in the ground state.  Grid methods, number
of points, and the configurational basis have been described in
Section~\ref{sec:states}. 
The results of the simulations are presented and
discussed next. We focus on excitation probabilities and deformation of
the molecule when it hits the surface. The implications for the
dissociation are discussed separately. We end with a summary and some
general conclusions.

\section{Scatter Probabilities}

We found that the scattering is predominantly elastic. The elastic
scattering probability is larger than 0.99 for all orientations and PESs
at a translational energy of 96 kJ/mol, except for the PES with $V_{\rm
shift}$ and three hydrogens pointing towards the surface for which it is 0.956.
This agrees with the observation that the translation-vibration coupling
is generally small\cite{lev87}.

If we would have wanted to determine the role of the internal vibrations
from the scattering probabilities, we would have to do quite accurate
simulations. We have opted instead to look at the molecule when it hits
the surface, which enables us to obtain good results with much less
costly simulations.

\section{Excitation Probabilities}

The surface PES has C${}_{3v}$ (with one or three hydrogens
towards the surface) or C${}_{2v}$ symmetry (with two hydrogens towards
the surface). If we expand this PES in a Taylor-series of internal
vibrations, we see that the linear terms contain only those vibrations
that transform as $a_1$ in C${}_{3v}$, respectively, C${}_{2v}$. These
are therefore easier to excite than others; We did not find any
appreciable excitation of $e$ modes of C${}_{3v}$ and the $a_2$, $b_1$,
and $b_2$ modes of C${}_{2v}$.

We will not present results of the simulations with the harmonic PES,
because they give almost the same excitation probabilities as the PES
with $V_{\rm Morse}$. 
The maximum excitation probabilities at the surface for the PES with
$V_{\rm Morse}$ are given in Table \ref{tab:exprop}. We have observed
the highest excitations for this PES in the $\nu_4$ umbrella
and $\nu_2$ bending modes in the orientation with two hydrogens, and
for the $\nu_4$ umbrella mode in the orientation with three
hydrogens pointing towards the surface. 
The excitation probabilities for the $\nu_1$ and $\nu_3$
stretch modes at this orientations are a factor of magnitude lower.  

\begin{table}[t]
  \caption{Excitation probabilities, at an initial translational energy
    of 96 kJ/mol and all initial vibrational states in the ground state,
    for the three different PESs in
    the $a_1$ modes of the C${}_{3v}$  and C${}_{2v}$ symmetry, with
    one, two or three hydrogens pointing towards the surface. 
    These modes are a $\nu_1(a_1)$ symmetrical stretch, a $\nu_2(e)$
    bending , a $\nu_3(t_2)$ asymmetrical stretch, and a $\nu_4(t_2)$
    umbrella in the T${}_d$ symmetry. The PESs are:
    An anharmonic intramolecular PES [Morse, see Eq.\ (\ref{Vmorse_d})],
    an intramolecular PES with weakening C--H bonds [weak, see
    Eq.\ (\ref{Vweak_d})], and an intramolecular PES with elongation of
    the C--H bonds [shift, see Eq.\ (\ref{Vshift_d})].}
  \label{tab:exprop}
  \begin{tabular}[t]{l l c c c c}
    \hline
      orientation & PES & $\nu_1(a_1)$ & $\nu_2(e)$  &
      $\nu_3(t_2)$ & $\nu_4(t_2)$ \\
      & &  stretch & bending & stretch & umbrella \\
    \hline
      one   & Morse & 0.023 &       & 0.054 & 0.030  \\
            & weak  & 0.135 &       & 0.308 & 0.075  \\
            & shift & 0.340 &       & 0.727 & 0.067 \\
    \hline
      two   & Morse & 0.010 & 0.108 & 0.009 & 0.104  \\
            & weak  & 0.067 & 0.135 & 0.073 & 0.107  \\
            & shift & 0.707 & 0.065 & 0.768 & 0.331 \\
    \hline
      three & Morse & 0.003 &       & 0.006 & 0.102  \\
            & weak  & 0.038 &       & 0.019 & 0.208  \\
            & shift & 0.819 &       & 0.674 & 0.214 \\
    \hline
  \end{tabular}
\end{table}

We have observed higher excitation probabilities for the $\nu_1$ and
$\nu_3$ stretch modes in the one hydrogen orientation. The excitation
probability of the $\nu_4$ umbrella mode is here lower then for the
other orientations, but still in the same order as the stretch modes for
this orientation.  This can be explained by the values of the $\alpha$'s
of $V_{\rm surf}$ [see Table \ref{tab:alpha}] and the force constants of
$V_{\rm harm}$ [see Table \ref{tab:gen}], because $V_{\rm harm}$ is
approximately $V_{\rm Morse}$ for the ground state. The force constants
of $V_{\rm intra}$ for the $\nu_1$ symmetrical stretch and $\nu_3$
asymmetrical stretch are of the same order, but the $\alpha$ parameter
in $V_{\rm surf}$ in the orientation with one hydrogen pointing towards
the surface for the $\nu_3$ asymmetrical stretch is around twice as
large as for the $\nu_1$ symmetrical stretch, which explains why $\nu_3$
is more excited as $\nu_1$. The surface repulsion on the $\nu_4$
umbrella mode is even three times lower then on the $\nu_1$ symmetrical
stretch mode, but the force constant is also much lower. It results in a
little more excitation of the $\nu_4$ umbrella then in the $\nu_1$
symmetrical stretch.

For the orientation with two hydrogens pointing towards the surface the
repulsion on the vibrational modes is for all modes in the same order
($\nu_2$ bending a little higher), so the difference in excitation
probabilities correlate here primarily with the force constants. The force
constants of the $\nu_2$ bending and $\nu_4$ umbrella modes are
of the same order, as are those of $\nu_1$ and $\nu_3$. The stretch
force constants are higher, however, so that the excitation
probabilities are lower. 
The repulsion on the $\nu_4$ umbrella mode is the largest in the
orientation with three hydrogens pointing towards the surface. The force
constants is lower for this modes then for the $\nu_1$ and
$\nu_3$ stretch modes, so the primary excitation is seen in the
$\nu_4$ umbrella mode. 

For the orientation with three hydrogens pointing towards the
surface the repulsion on the $\nu_4$ umbrella mode is the highest in
combination with a low force constant, so this mode has a much higher
excitation probability than the $\nu_1$ and $\nu_3$ stretch modes.
Another interesting detail is that the
$\alpha_3$ parameter is higher for the orientation with two than 
with three hydrogens pointing toward the surface, but the
excitation probabilities for the $\nu_4$ umbrella in this
orientations are equal. This is caused by a coupling between
the excitation of the $\nu_2$ bending and the $\nu_4$ umbrella
mode in the orientation with two hydrogens pointing towards the surface.

We observed with $V_{\rm weak}$ [see Table \ref{tab:exprop}] that all
excitation probabilities become much higher than with $V_{\rm Morse}$,
except for the $\nu_4$ umbrella mode with two hydrogens towards the
surface, which stays almost the same.  
It is caused by the fact that, although the $\nu_1$ symmetrical stretch
and $\nu_3$ asymmetrical stretch contribute almost completely to
$V_{\rm Morse}$, the $\nu_4$ umbrella does so just for a small part. The
$\nu_4$ umbrella contributes primarily, and the $\nu_2$ bending
completely, to the harmonic terms of intramolecular PES
$V_{\rm intra}$. $V_{\rm weak}$ gives only a lowering in the $V_{\rm
  Morse}$ terms of $V_{\rm intra}$, so we should expect primarily an
higher excitation probability in the $\nu_1$ and $\nu_3$ stretch
modes. This will also cause a higher excitation probability of
the other modes, because the turn-around point will be some what later,
which give effectively more repulsion on the other modes.

We also observed that for both stretches the excitation
probabilities shows the following trend; three hydrogens $<$ two hydrogens
$<$ one hydrogen pointing towards the surface. For one hydrogen pointing
towards the surface the excitation probability of the $\nu_3$
asymmetrical stretch is around twice that of the $\nu_1$
symmetrical stretch, they are almost equal for two hydrogens, and for
three hydrogens the $\nu_1$ is twice the $\nu_3$
stretch. Some of these trends can also be found for the PES with $V_{\rm
  Morse}$, but are not always that obvious. They follow the same trends
as the $\alpha$ parameters of $V_{\rm surf}$ in Table \ref{tab:alpha}.

$V_{\rm shift}$ [see Table \ref{tab:exprop}] gives also higher
excitation probabilities for almost all modes than $V_{\rm Morse}$, 
but for the $\nu_2$ bending mode in the two hydrogens pointing
towards the surface orientation it became lower. This means that the
repulsion of the surface is here caused for a large part by the $V_{\rm
  shift}$ terms [see Eq.\  (\ref{Vshift_d})], where the
$\nu_2$ bending modes don't contribute. It is also in agreement with
the fact that we have observed the excitation maximum earlier, so the
$V_{\rm surf}$  repulsion on the $\nu_2$ bending modes will be
lower. Fig.~\ref{fig:potdif} shows that the repulsion for the PES with
$V_{\rm shift}$ is stronger than for the other three PESs.

The excitations in the $\nu_1$ and $\nu_3$ stretch modes are
extremely high with $V_{\rm shift}$, because they contribute strongly to
the C--H bond elongation.  
The trend in the excitation probabilities of the $\nu_1$ symmetrical
stretch mode is caused by the number of bonds pointing towards the surface,
because the $\gamma_{i2}$ parameters are for all bonds and orientations
the same, so only the strength of the coupling with the shift-factor
dominates. This effect is illustrated in Fig.~\ref{fig:symshift}.
We didn't observe this trend in the $\nu_3$ asymmetrical
stretch mode, because there is also difference between the $\gamma_{i4}$
parameters as can be seen in Table \ref{tab:gammaone} and
\ref{tab:gammatwo}.

\section{Structure Deformation}
\label{sec:strdef}

\begin{figure}[p]
  \epsfig{file=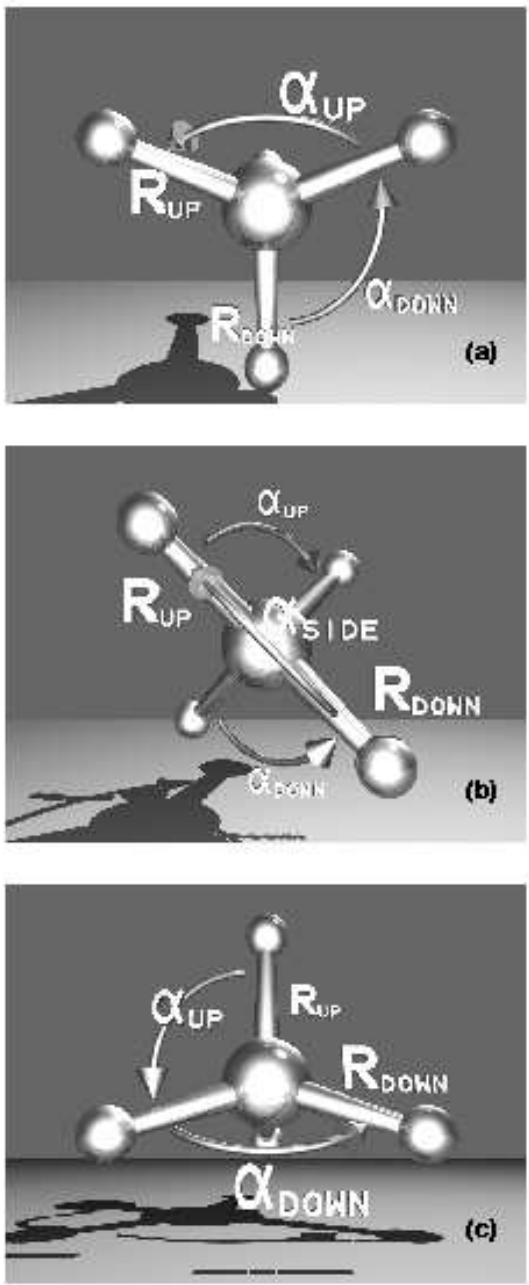, height=17cm}
    \caption{Schematic representation of the $R_{\rm down}$ and
      $R_{\rm up}$ bonds, and the $\alpha_{\rm down}$, $\alpha_{\rm
        side}$ and $\alpha_{\rm up}$ angles for the three orientations:
      (a) one, (b) two, and (c) three hydrogens pointing towards the
      surface.}
    \label{fig:orient}
\end{figure}

If we put the methane molecule far from the surface, then all PESs are
identical. For this situation we calculated a bond distance of 1.165
{\AA} and bond angle of 109.5 degrees in the ground state. 
The results of the maximum structure deformations are shown in the Table
\ref{tab:strdef}. Figure \ref{fig:orient} shows the names of the bonds and
angles for the three orientations.

\begin{table}[t]
  \caption{Structure deformation, at an initial translational energy
    of 96 kJ/mol and all initial vibrational states in the ground state,
    with one, two, and three hydrogen 
    pointing towards the surface for three different PESs:
    An anharmonic intramolecular PES [Morse, see Eq.\ (\ref{Vmorse_d})],
    an intramolecular PES with weakening C--H bonds [weak, see
    Eq.\ (\ref{Vweak_d})], and an intramolecular PES with elongation of
    the C--H bonds [shift, see Eq.\ (\ref{Vshift_d})].
    See Fig.\ref{fig:orient} for the meaning $R_{\rm down}$, $R_{\rm
    up}$, $\alpha_{\rm down}$, $\alpha_{\rm up}$, and $\alpha_{\rm
    side}$. All bond distances are given in 
    {\AA} and all bond angles are given in degrees. The gas phase values
    correspond to the bond distance and angles far from the surface, and
    are the same for all three PES types.}
  \label{tab:strdef}
  \begin{tabular}[t]{l l c c c c c}
    \hline
      Orientation & PES & $R_{\rm down}$ & $R_{\rm up}$ & $\alpha_{\rm
      down}$ & $\alpha_{\rm up}$ &  $\alpha_{\rm side}$ \\
    \hline
    gas phase & all & 1.165 & & 109.5 & & \\
    \hline
      one   & Morse & 1.123 & 1.166 & 108.5 & 110.4 & \\
            & weak  & 1.051 & 1.167 & 108.2 & 110.7 & \\
            & shift & 1.397 & 1.158 & 109.9 & 109.0 & \\
    \hline
      two   & Morse & 1.157 & 1.165 & 116.7 & 108.9 & 107.8 \\
            & weak  & 1.137 & 1.163 & 117.6 & 109.0 & 107.5 \\
            & shift & 1.386 & 1.159 & 115.7 & 106.9 & 108.5 \\
    \hline
      three & Morse & 1.164 & 1.160 & 111.2 & 107.7 & \\
            & weak  & 1.167 & 1.155 & 112.4 & 106.3 & \\
            & shift & 1.389 & 1.144 & 111.2 & 107.7 & \\
    \hline
  \end{tabular}
\end{table}

We observed that the PESs with $V_{\rm Morse}$ and $V_{\rm weak}$ give
again the same trends, but that for the PES with $V_{\rm weak}$ these
trends are much stronger. This is in agreement with the observations
discussed for the excitation probabilities above. The deformations for
the PES with $V_{\rm shift}$ are dominated
by the change of the bond distances for the bonds which are pointing
towards the surface. These bonds become longer for all orientations in
the same order. So there is no orientational effect and this will be
probably caused completely by the  $V_{\rm shift}$ terms [see
Eq.\  (\ref{Vshift_d})]. 

For the orientation with one hydrogen pointing towards the surface [see
Table \ref{tab:strdef}], we observed that the bond pointing towards the
surface becomes shorter, as expected, for the PES with $V_{\rm Morse}$
and even more with $V_{\rm weak}$. This is caused by the repulsion of
the $V_{\rm surf}$ terms, which works in the direction of the bond
axes. We observed also some small bond angle deformation. This is
probably a secondary effect of the change in the bond distance and
correlates to excitation of the $\nu_4$ umbrella mode.
Remarkably, for $V_{\rm shift}$ the bond length increases. Clearly the
effect of the change in equilibrium bond length is more effective than
the repulsion with the surface. The shift effect may be somewhat too
large, but the observed change (0.232 {\AA}) is much lower
than the shift of the equilibrium (0.54 {\AA}) on which we fitted
$V_{\rm shift}$.

We also observed for all orientations with $V_{\rm shift}$ an earlier
turn-around point, which is in agreement with the earlier observed
excitation maximum. This is caused by the extra repulsion contribution
in the $V_{\rm shift}$ terms and by the longer bond length; a longer
bond gives a higher repulsion for the same position of the center of
mass. Fig. \ref{fig:potdif} shows this effect in the $\nu_3$
asymmetrical stretch mode.

The orientation with two hydrogens pointing towards the surface [see
Table \ref{tab:strdef}] gives also shorter bond distances for the
bonds pointing towards the surface for the PESs with $V_{\rm Morse}$ and
$V_{\rm weak}$, although very little in comparison to the one hydrogen
orientation. We expected this, because just half of the repulsion is now
in the direction of the bonds. 
The other half is perpendicular on the C--H bonds and this makes the
bond angle larger. The bond angles between the bonds pointing towards
the surface show a quite large deformation. This was already expected
from the excitation of the $\nu_2$ bending and $\nu_4$ umbrella
modes. 

We also observed that a smaller bond length of the bonds pointing
towards the surface correlates with a larger bond angle of the angles
between the bonds pointing towards the surface and a higher excitation
probability of the $\nu_2$ bending mode. This can be explained as
follows. When the bonds become shorter the center of mass can come
closer to the surface. Because of this there will be more repulsion on
the $\nu_2$ bending mode, which will cause a larger bond angle
deformation.

Most of the deformations of the
other bond angles can be explained as an indirect effect of the
deformation of this bond angle, but the deformation of the angle between
the bonds pointing away from the surface which was seen at the PES with
$V_{\rm shift}$ must be caused by the excitation of the $\nu_4$ umbrella
mode. 

We observed almost no change in bond length for the PESs with $V_{\rm
Morse}$ and $V_{\rm weak}$ at the three hydrogens towards the surface
orientation [see Table \ref{tab:strdef}]. Almost all energy is absorbed
by the $\nu_4$ umbrella mode, so this gives quite large bond angle
deformations. The bond angle deformation is larger for the PES with
$V_{\rm weak}$ than with $V_{\rm Morse}$ PES; in agreement with higher
excitation probability for the PES with $V_{\rm weak}$. The PES with
$V_{\rm shift}$  has around the same excitation probabilities for the
$\nu_4$ umbrella  mode as the PES with $V_{\rm weak}$, but the bond
angle deformation is the same as for the PES with $V_{\rm Morse}$. This
may be caused by the higher excitations of the $\nu_1$ and $\nu_3$
stretch modes and the longer bonds pointing towards the surface at the
PES with $V_{\rm shift}$, which can make it harder to deform the bond
angles.

\section{Dissociation models}
\label{sec:diss}

Several dissociation mechanisms for direct methane dissociation on
transition metals have been suggested. All proposals can be related to
two main ideas. One of them is the breaking of a single C--H bond in the
initial collision.\cite{lee87,lun91} This model is most suggested in
literature and also all wavepacket simulations have focussed on the effects
of this model.\cite{jan95,lun91,lun92,lun95,car98} The other mechanism
is often called ``splats'' and suggests that the critical requirement
for methane dissociation is angular deformation of methane which allows
a Ni--C bond to form.\cite{lee87} Even though we did not try to describe
the dissociation itself, we would like to discuss the implication of our
simulation for the dissociation.

The ``splats'' model seems easiest to discuss. Angular deformation is
related to the $\nu_2$ bending and the $\nu_4$ umbrella mode. The
excitation probabilities in Table~\ref{tab:exprop} seem to indicate that
these modes are easy to excite, and that angular deformation should be
large. These excitation probabilities are misleading, however.
Table~\ref{tab:strdef} shows that, although the excitation probabilities
depend on the PES, the changes in the bond angles
do not. Moreover, these changes are quite small. They are largest for
the orientation with two hydrogen atoms pointing towards to surface, for
which they are about $8\,$degrees at most. This seems much too small to
enable the formation of a Ni--C bond.  A previous estimate of the bond
angle deformation gave an energy of $68.1\,$kJ/mol to get three
hydrogens and the carbon in one plane. This correspond to a bond angle
of $120\,$degrees.\cite{lee87} We find with three hydrogens pointing
towards the surface that the bond angle only changes to about
$112\,$degrees at the higher energy of $96\,$kJ/mol. The reason for this
difference is that for the older estimate it was assumed that all
translational energy can be used to deform the bond angle, whereas
Tables~\ref{tab:exprop} and \ref{tab:strdef} clearly show that this is
not correct.

The excitation probabilities for the stretch modes depend strongly on
the orientation of the molecule and on the PES. For
the PESs without a change of the equilibrium bond
length the excitation probabilities are only appreciably with one
hydrogen pointing towards the surface. This is, however, not a
favourable orientation for the dissociation. Moreover,
Table~\ref{tab:strdef} shows that in those cases the repulsion with the
surface shortens the C--H bond. The same holds for the orientation with
two hydrogens pointing towards the surface, although to a lesser extent,
whereas the bond length changes hardly at all with three hydrogens
pointing towards the surface. This indicates that the intramolecular
PES needs the bond elongation to overcome the
repulsion of the surface that shortens the C--H bond, and to get
dissociation. This agrees completely with electronic structure
calculations that yield a late barrier for dissociation with a very
elongated C--H bond.\cite{bur93a,kra96} The large excitation
probabilities for the PES with the elongated
equilibrium bond length should not be over-interpreted, however. They are
to be expected even if the molecule stays in its (shifted) vibrational
ground state. More telling is that for this PES we do find at
least some inelastic scattering.

%
%%%%%%%%%%%%%%%
% CONCLUSIONS %
%%%%%%%%%%%%%%%
\section{Conclusions}

We have done wavepacket simulations on the scattering of methane from a
flat Ni(111) model surface with a fixed orientation with one, two, or
three hydrogens pointing towards the surface. We used the MCTDH method
and four different model PESs for each orientation.
We used a translational energy of up to $96\,$kJ/mol and all internal
vibrations in the ground state. The scattering was in all cases
predominantly elastic.

When the molecule hits the surface, we always observe vibrational
excitations of the $\nu_4$ umbrella and $\nu_2$ bending modes,
especially in the orientations with two or three hydrogens pointing
towards the surface. This is due to a favorable coupling that originates
from the repulsive interaction with the surface, and the low excitation
energies. Deformations of the molecule are predominantly in the bond
angles. The changes in the bond angles are, however, too small to allow
for the formation of a Ni--C bond, as suggested in the ``splats'' model
of methane dissociation.

Appreciable excitations of the $\nu_1$ and $\nu_3$ stretch modes when
methane hits the surface are only observed when one hydrogen atom points
towards the surface, or when the intramolecular PES has an
elongated equilibrium C--H bond length close to the surface. The
repulsion of the surface shortens the C--H bond. This can only be
overcome when the intramolecular PES incorporates the effect of a
longer equilibrium C--H bond length caused by overlap of occupied
surface orbitals with the antibonding orbitals of methane. This agrees
with quantum chemical calculations, which show a late barrier for
dissociation.

The simulations with these model PESs show that the internal
vibrations play an important role in the dissociation mechanism.
Excitation probabilities when the molecule hits the surface show how the
translational energy is converted into vibrational energy and it is
distributed over the internal modes. These probabilities vary strongly
with the PES. As only few internal vibrations contribute to the
dissociation, it is important to obtain more information on the real
PES for this system.

%\bibliography{/home/tgakrm/writings/BibTeX/Methane}
%\bibliographystyle{/home/tgakrm/writings/revtex3.1/prstyfull}
%\bibliography{Methane}
%\bibliographystyle{prstyfull}

%% file: isotope.tex
\chapter{The isotope effect}
\label{chap:iso}

\begin{quote}
  {\it The isotope effect in the scattering of methane is studied by
    wave packet simulations of oriented CH${}_4$ and CD${}_4$ molecules
    from a flat surface including all nine internal vibrations. At a
    translational energy up to 96 kJ/mol we find that the scattering is
    still predominantly elastic, but less so for CD${}_4$.  Energy
    distribution analysis of the kinetic energy per mode and the
    potential energy surface terms, when the molecule hits the surface,
    are used in combination with vibrational excitations and the
    corresponding deformation. They indicate that the orientation with
    three bonds pointing towards the surface is mostly responsible for
    the isotope effect in the methane dissociation.}\footnote{This
    chapter has been published as a part of {\sl Energy distribution
      analysis of the wave packet simulations of CH${}_4$ and CD${}_4$
      scattering}\cite{mil00a}, \copyright \ {\sl 2000 Elsevier
      Science}.}
\end{quote}

\section{Introduction}

A nice way to study reaction dynamics is the use of isotopes. The most
recent wave packet simulation on the dissociation probability of CH${}_4$
and CD${}_4$ showed a semiquantitative agreement with the molecular beam
experiments of Ref.\cite{hol95}, except for the isotope effect 
and the extracted vibrational efficacy \cite{car98}. The molecular beam
study with laser excitation of the $\nu_3$ asymmetrical stretch mode
shows that the incorrect vibrational efficacy is caused by the
assumptions in the fit procedure that both stretch modes behaves
indentical \cite{juur99}. One of the possible explanation of the
incorrect isotope effect can be the role played by the non-included
intramolecular vibrations. 

In this chapter we report wave packet simulations of CD${}_4$ scattering
including all internal vibrations for fixed orientations, performed on
the same model PESs as in the previous chapter for CH${}_4$, which have
been described in Section~\ref{sec:pes}. A harmonic intramolecular PES
[$V_{\rm intra}$, see Eq.(\ref{Vharm})] is adapted to include
anharmonicities in the C--H distance [$V_{\rm Morse}$, see
Eq.(\ref{Vmorse_d})], the decrease of the C--H bond energy due to
interactions with the surface [$V_{\rm weak}$, see Eq.(\ref{Vweak_d})],
and the increase of the C--H bond length also due to interactions with
the surface [$V_{\rm shift}$, see Eq.(\ref{Vshift_d})].  We have used
the multi-confi\-guratio\-nal time-depen\-dent Hartree (MCTDH) method
for our wave packet simulation \cite{man92,jan93}. This method can deal
with a large number of degrees of freedom and with large grids.  (See
Ref.~\cite{bec00} for a recent review, or Section~\ref{sec:mctdh} of this
thesis for a survey.)  Initial translational energy has been chosen in
the range of 32 to 96 kJ/mol.  We present only the results of the
simualtions at 96 kJ/mol here (equivalent to $P_1=-0.2704$ mass-weighted
atomic units), because they showed the most obvious excitation
probabilties for $V_{\rm Morse}$.  The initial state has been written as
a product state of ten functions; one for the normally incident
translational coordinate, and one for each internal vibration. All
vibrations were taken to be in the ground state.  Grid methods, number
of points, and the configurational basis have been described in
Section~\ref{sec:states}.

We will discuss the vibrational excitation and the deformation of the
CD${}_4$ molecule when it hits the surface and compare it with CH${}_4$.
Later on we will look at the energy distribution of the kinetic energy
per mode and the potential energy in some terms of our model PESs for
both isotopes.  The transfer of translational kinetic energy towards
vibrational kinetic energy gives an indication about the dissociation
probability, since vibrational kinetic energy helps in overcoming the
dissociation barrier.  It gives a better idea too about which modes are
essential to include in a more accurate wave packet simulation of methane
dissociation. After that we will discuss the implications of this for
the dissociation and give a summary with some general conclusions.

\section[Excitations and structure deformation]{Excitation probabilities
  and structure deformation of CD${}_4$} 

The scattering probabilities for CD${}_4$ are predominantly elastic, as
we also found in our previous simulations of CH${}_4$ scattering. (See
Chapter~\ref{chap:ch4scat} and Ref.~\cite{mil98}) The elastic scattering
probability is larger than 0.99 for all orientation of the PESs with
$V_{\rm Morse}$ and $V_{\rm weak}$ at a translational energy of 96
kJ/mol. For the PES with $V_{\rm shift}$ we observe an elastic
scattering probability of 0.981 for the orientation with one, 0.955 with
two and 0.892 with three deuteriums pointing towards the surface. This
is lower than we have found for CH${}_4$, which is 0.956 for the
orientation with three hydrogens pointing towards the surface and larger
than 0.99 for the others. The higher inelastic scattering probabilities
of CD${}_4$ was expected, because the force constants $k_i$ of CD${}_4$
are decreased up to 50\% with respect to those of CH${}_4$ and the
translational surface repulsion fall-off differs only little.

Since the inelastic scattering probabilities are generally small, we
have opted to look at the molecule when it hits the surface. Excitation
probabilities and structure deformation enable us to get good results
with much less costly simulations. The surface PES has C${}_{3v}$ (with
one or three bonds towards the surface) or C${}_{2v}$ symmetry (with two
bonds towards the surface). If we expand this PES in a Taylor-series of
internal vibrations, we see that the linear terms contain only those
vibrations that transform as $a_1$ in C${}_{3v}$, respectively,
C${}_{2v}$. These are therefore easier to excite than others; We have not
find any appreciable excitation of $e$ modes of C${}_{3v}$ and the
$a_2$, $b_1$, and $b_2$ modes of C${}_{2v}$.

We did not present results of the simulations with the harmonic PES,
because they give almost the same excitation probabilities as the PES
with $V_{\rm Morse}$. 
When we compare the excitation probabilities of CD${}_4$ in
Table \ref{tab:expropD} with CH${}_4$ in Table \ref{tab:exprop} (see
also Ref.\cite{mil98}), we see
that most of the modes at the different orientations give also a higher
excitation probability similar to the scattering probability. We  
also see that some modes give a lower or equal excitation probability than
for CH${}_4$ scattering. This is caused by the fact that the force
constants $k_i$, the surface repulsion parameters $\alpha_i$ and
$\beta_i$, and the $\gamma_{i,j}$ Morse bond-potential values changed
relatively to each other. As a consequence of that some modes are more 
preferred for absorbing energy, which leaves less energy for others. So
the isotope effect gives besides a general higher excitation probability
also a different excitation probability distribution over the modes.
We will now discuss this in more detail for the three different
orientations and model PESs in combination with the structure
deformation.  

\begin{table}[t]
  \caption{Excitation probabilities at the surface for CD${}_4$, at an
    initial translational energy of 96 kJ/mol and all initial
    vibrational states in the ground state, for the three different PESs
    in the $a_1$ modes of the C${}_{3v}$ and C${}_{2v}$ symmetry, with
    one, two or three deuteriums pointing towards the surface.  These
    modes are a $\nu_1(a_1)$ symmetrical stretch, a $\nu_2(e)$ bending ,
    a $\nu_3(t_2)$ asymmetrical stretch, and a $\nu_4(t_2)$ umbrella. In
    parenthesis is the irreducible representation in T${}_d$ symmetry.
    The PESs are: An anharmonic intramolecular PES [Morse, see Eq.\ 
    (\ref{Vmorse_d})], an intramolecular PES with weakening C--D bonds
    [weak, see Eq.\ (\ref{Vweak_d})], and an intramolecular PES with
    elongation of the C--D bonds [shift, see Eq.\ (\ref{Vshift_d})].}
  \label{tab:expropD}
  \begin{tabular}[t]{l l c c c c}
    \hline
       & PES & $\nu_1(a_1)$ & $\nu_2(e)$ & $\nu_3(t_2)$ & $\nu_4(t_2)$ \\
      orientation &     & stretch & bending & stretch & umbrella \\
    \hline
      one   & Morse & 0.052 &       & 0.056 & 0.094  \\
            & weak  & 0.220 &       & 0.362 & 0.203  \\
            & shift & 0.460 &       & 0.910 & 0.174 \\
    \hline
      two   & Morse & 0.026 & 0.164 & 0.011 & 0.094  \\
            & weak  & 0.099 & 0.192 & 0.080 & 0.080  \\
            & shift & 0.792 & 0.092 & 0.830 & 0.495 \\
    \hline
      three & Morse & 0.014 &       & 0.003 & 0.242  \\
            & weak  & 0.046 &       & 0.021 & 0.308  \\
            & shift & 0.868 &       & 0.756 & 0.387  \\
    \hline
  \end{tabular}
\end{table}

For the PESs with $V_{\rm Morse}$ and $V_{\rm weak}$ we find that the
excitation probabilities of the $\nu_3$ stretch for CD${}_4$ stays
almost the same for all orientations compared to CH${}_4$, except for
the orientation with one deuterium 
pointing towards the surface with the $V_{\rm weak}$ PES for which
it becomes 0.054 higher (see Table \ref{tab:expropD}). In orientation
with C${}_{3v}$ symmetry most of 
the extra excitation is absorbed by the $\nu_4$ umbrella
mode. Remarkable is that for the orientation with two deuteriums
pointing towards the surface excitation probabilities of this mode
become lower than for the CH${}_4$ scattering. Even the excitation
probability for the $V_{\rm weak}$ PES of this mode is lower than for the
$V_{\rm Morse}$ PES, while normally this excitation is preferred for
the PES with $V_{\rm Morse}$. One of the reasons is that for this
orientation most of the extra excitation probability is taken by the
$\nu_2$ bending mode. We also see that there is a higher excitation
probability in $\nu_1$ symmetrical stretch mode for these PESs in all
orientations. If we observe an extra excitation probability for a mode
then this is relatively more for the PES with $V_{\rm Morse}$ than with
$V_{\rm weak}$.  

\begin{table}[t]
  \caption{Structure deformation at the surface for CD${}_4$, at an
    initial translational energy of 96 kJ/mol and all initial
    vibrational states in the ground state, with one, two, and three
    deuterium pointing towards the surface for three different PESs: An
    anharmonic intramolecular PES [Morse, see Eq.\ (\ref{Vmorse_d})], an
    intramolecular PES with weakening C--D bonds [weak, see Eq.\ 
    (\ref{Vweak_d})], and an intramolecular PES with elongation of the
    C--D bonds [shift, see Eq.\ (\ref{Vshift_d})].  See
    Fig.\ref{fig:orient} for the meaning $R_{\rm down}$, $R_{\rm up}$,
    $\alpha_{\rm down}$, $\alpha_{\rm up}$, and $\alpha_{\rm side}$. All
    bond distances are given in {\AA} and all bond angles are given in
    degrees. The gas phase values correspond to the bond distance and
    angles far from the surface, and are the same for all three PES
    types.}
  \label{tab:strdefD}
  \begin{tabular}[t]{l l c c c c c}
    \hline
      Orientation & PES & $R_{\rm down}$ & $R_{\rm up}$ & $\alpha_{\rm
      down}$ & $\alpha_{\rm up}$ &  $\alpha_{\rm side}$ \\
    \hline
    gas phase & all & 1.161 & & 109.5 & & \\
    \hline
      one   & Morse & 1.117 & 1.159 & 108.0 & 110.9 & \\
            & weak  & 1.046 & 1.161 & 107.7 & 111.2 & \\
            & shift & 1.395 & 1.157 & 109.8 & 109.2 & \\
    \hline
      two   & Morse & 1.152 & 1.156 & 116.4 & 109.7 & 107.7 \\
            & weak  & 1.133 & 1.158 & 116.9 & 109.9 & 107.5 \\
            & shift & 1.380 & 1.158 & 115.7 & 107.0 & 108.5 \\
    \hline
      three & Morse & 1.160 & 1.153 & 112.4 & 106.4 & \\
            & weak  & 1.160 & 1.152 & 112.8 & 105.9 & \\
            & shift & 1.357 & 1.134 & 111.2 & 107.7 & \\
    \hline
  \end{tabular}
\end{table}

As a consequence of the extra excitation in the $\nu_4$
umbrella mode in the orientations with C${}_{3v}$ symmetry we observe
also more bond angle deformation for these PESs (see Table
\ref{tab:strdefD}). It is also remarkable that the difference between the
bond angle deformation between the PES with $V_{\rm Morse}$ and $V_{\rm
  weak}$ in the orientation with three deuterium pointing towards the
surface is much smaller than we saw for CH${}_4$ scattering. So the
weakening of the bond strength is less effective for CD${}_4$ than
CH${}_4$. 

We also see this effect in the orientation with two deuteriums pointing
towards the surface. The bond angle deformation of the angle between the
C--D bond pointing towards the surface has also become lower than we had
observed for CH${}_4$. The loss of excitation of the $\nu_4$ umbrella
mode is not recovered by the extra excitation of the $\nu_2$ bending
mode. On the other hand, as a consequence of this, the angle between the
C--D bonds pointing away from the surface becomes more deformed, because
excitation of the $\nu_4$ umbrella mode decreases this bond angle and
the $\nu_2$ bending mode increases it.

The bond distances for all orientations and the PESs with $V_{\rm
  Morse}$ and $V_{\rm weak}$ are generally somewhat lower for CD${}_4$
than CH${}_4$, but this in the same order as we get for the gas phase
bond distance ($0.004$ {\AA}). So the isotope effect primary effects the
bond angle deformation for these PES types.

For the PES with $V_{\rm shift}$ we do not observe this effect on the
bond angle deformation. The bond angle deformation for the orientation
with two and three deuteriums pointing towards the surface is the same
as for CH${}_4$ and it is just $0.1^{\circ}$ less for the bond angle at
the surface side in the orientation with one deuterium pointing towards
the surface. The excitation probabilities for the $\nu_2$ bending and
$\nu_4$ umbrella modes become higher for all orientations for CD${}_4$,
which is necessary for getting the same bond angle deformations as
CH${}_4$. 

The changes in the bond distances for the orientations with one and two
bonds pointing towards the surface is for CD${}_4$ almost the same as
for CH${}_4$. For the orientation with three bonds pointing towards the
surface, we found that the maximum bond lengthening of the bonds on the
surface side was $0.032$ {\AA} less for CD${}_4$ than CH${}_4$. We also
found that the bond shortening of the bond pointing away from the
surface is $0.010$ {\AA} more for CD${}_4$. These are only minimal
differences, which also only suggest that the bond deformation for CD${}_4$
has been influenced slightly more by the $\nu_3$ asymmetrical stretch
mode than the $\nu_1$ symmetrical stretch mode. The observed excitation
probabilities for these modes do not contradict this, but are not
reliable enough for hard conclusions because of their high magnitude.
It is also not clear, beside of this problem, what they really
represent. Is the excitation caused by a different equilibrium position
of the PES at the surface in a mode or is caused by extra energy in this
mode? To answer these questions we decided to do an energy distribution
analysis during the scattering for both isotopes. 

\section{Energy distribution in CH${}_4$ and CD${}_4$}

For an analysis of the isotope effect in the scattering of methane in
more detail and to be able to discuss some possible implications of this
for the dissociation probabilities of both isotopes, we did an analysis
of the energy distribution during the scattering of CD${}_4$ and
CH${}_4$ for all presented orientations and potentials in
Section~\ref{sec:pes}. (See also Refs.~\cite{mil98,mil00a}).

We do this by calculating the expectation values of the important term
of the Hamiltonian $H$ [see Eq.\ (\ref{Ham})] for the
wave-function $\Psi (t)$ at a certain time $t$.
We can obtain good information about the energy distribution per mode by
looking at the kinetic energy expectation values
$\langle\Psi (t)\vert h_j \vert\Psi (t)\rangle$ per mode $j$ (see Tables
\ref{tab:kinch4} and \ref{tab:kincd4}), because the kinetic energy
operators $h_j$ have no cross terms like the PESs have. When we discuss the
kinetic energy of a mode we normally refer to the $a_1$ mode of the 
C${}_{3v}$ or C${}_{2v}$ symmetry, because in these modes we have
observed the highest excitation probabilities and the change in kinetic
energy in the other modes is generally small.  

\begin{table}[t]
  \caption{Expectation values of the kinetic energy per mode in
    mHartree at the surface for CH${}_4$, at an initial translational energy 
    of 96 kJ/mol and all initial vibrational states in the ground state,
    for the three different PESs in
    the $a_1$ modes of the C${}_{3v}$  and C${}_{2v}$ symmetry, with
    one, two or three deuteriums pointing towards the surface. 
    These modes are a $\nu_1(a_1)$ symmetrical stretch, a $\nu_2(e)$
    bending , a $\nu_3(t_2)$ asymmetrical stretch, and a $\nu_4(t_2)$
    umbrella. In parenthesis is the irreducible representation in 
    T${}_d$ symmetry. The PESs are: 
    An anharmonic intramolecular PES [Morse, see Eq.\ (\ref{Vmorse_d})],
    an intramolecular PES with weakening C--H bonds [weak, see
    Eq.\ (\ref{Vweak_d})], and an intramolecular PES with elongation of
    the C--H bonds [shift, see Eq.\ (\ref{Vshift_d})].}
  \label{tab:kinch4}
  \begin{tabular}[t]{l l c c c c c}
    \hline
      orientation & PES &  translation & $\nu_1(a_1)$ &
      $\nu_2(e)$ & 
      $\nu_3(t_2)$ & $\nu_4(t_2)$ \\
      & &  &  stretch & bending & stretch & umbrella \\
    \hline
    initial & all   & 36.57 & 3.30 & 1.75 & 3.39 & 1.50 \\
    \hline
      one   & Morse &  9.65 & 3.44 &      & 3.78 & 1.51 \\
            & weak  &  9.37 & 3.15 &      & 2.86 & 1.51 \\
            & shift & 16.76 & 3.56 &      & 4.53 & 1.51 \\
    \hline
      two   & Morse &  9.42 & 3.36 & 1.80 & 3.52 & 1.54 \\
            & weak  & 10.98 & 2.69 & 1.80 & 2.75 & 1.53 \\
            & shift & 14.59 & 3.50 & 1.79 & 4.67 & 1.57 \\
    \hline
      three & Morse & 18.81 & 3.46 &      & 3.47 & 1.58 \\
            & weak  & 12.71 & 2.34 &      & 3.16 & 1.57 \\
            & shift & 20.53 & 5.32 &      & 4.39 & 1.58 \\
    \hline
  \end{tabular}
\end{table}

\begin{table}[t]
  \caption{Expectation values of the kinetic energy per mode in
    mHartree at the surface for CD${}_4$, at an initial translational energy 
    of 96 kJ/mol and all initial vibrational states in the ground state,
    for the three different PESs in
    the $a_1$ modes of the C${}_{3v}$  and C${}_{2v}$ symmetry, with
    one, two or three deuteriums pointing towards the surface. 
    These modes are a $\nu_1(a_1)$ symmetrical stretch, a $\nu_2(e)$
    bending , a $\nu_3(t_2)$ asymmetrical stretch, and a $\nu_4(t_2)$
    umbrella. In parenthesis is the irreducible representation in T${}_d$
    symmetry. The PESs are: 
    An anharmonic intramolecular PES [Morse, see Eq.\ (\ref{Vmorse_d})],
    an intramolecular PES with weakening C--D bonds [weak, see
    Eq.\ (\ref{Vweak_d})], and an intramolecular PES with elongation of
    the C--D bonds [shift, see Eq.\ (\ref{Vshift_d})].}
  \label{tab:kincd4}
  \begin{tabular}[t]{l l c c c c c}
    \hline
      orientation & PES &  translation & $\nu_1(a_1)$ &
      $\nu_2(e)$ & 
      $\nu_3(t_2)$ & $\nu_4(t_2)$ \\
      & &  &  stretch & bending & stretch & umbrella \\
    \hline
    initial & all   & 36.57 & 2.33 & 1.24 & 2.52 & 1.12 \\
    \hline
      one   & Morse &  8.29 & 2.46 &      & 2.82 & 1.14 \\
            & weak  &  7.93 & 2.23 &      & 2.16 & 1.13 \\
            & shift & 16.17 & 2.61 &      & 4.09 & 1.18 \\
    \hline
      two   & Morse &  8.22 & 2.41 & 1.28 & 2.56 & 1.13 \\
            & weak  &  9.25 & 1.93 & 1.28 & 2.10 & 1.14 \\
            & shift & 14.00 & 2.78 & 1.27 & 4.05 & 1.27 \\
    \hline
      three & Morse &  8.31 & 2.40 &      & 2.61 & 1.17 \\
            & weak  & 10.20 & 1.64 &      & 2.28 & 1.19 \\
            & shift & 20.06 & 4.37 &      & 3.80 & 1.28 \\
    \hline
  \end{tabular}
\end{table}

\begin{table}[t]
  \caption{Expectation values of the potential energy terms in
    mHartree at the surface for CH${}_4$, at an initial translational energy 
    of 96 kJ/mol and all initial vibrational states in the ground state,
    for the three different PESs.
    $V_{\rm surf}$ is the total surface hydrogen repulsion; $V_{\rm
    harm}(\nu_2)$ and $V_{\rm harm}(\nu_4)$ are the harmonic
    terms of the intramolecular PES of the $a_1$ modes in the C${}_{3v}$
    and C${}_{2v}$ symmetry corresponding to a $\nu_2(e)$ bending and
    $\nu_4(t_2)$ umbrella modes respectively in the T${}_d$ symmetry.
    $V_{\rm bond}(R_{\rm up})$ and $V_{\rm bond}(R_{\rm down})$ are the
    potential energy in a single C--H bond pointing respectively towards
    and away from the surface. 
    The different PESs are: An anharmonic intramolecular PES [Morse, see Eq.\
    (\ref{Vmorse_d})], an intramolecular PES with weakening C--H bonds
    [weak, see Eq.\ (\ref{Vweak_d})], and an intramolecular PES with
    elongation of the C--H bonds [shift, see Eq.\ (\ref{Vshift_d})].}
  \label{tab:potch4}
  \begin{tabular}[t]{l l c c c c c}
    \hline
      orient. & PES &  $V_{\rm surf}$ & $V_{\rm harm}(\nu_2)$ & 
       $V_{\rm harm}(\nu_4)$ & $V_{\rm bond}(R_{\rm up})$ &
      $V_{\rm bond}(R_{\rm down})$ \\  
    \hline
    initial & all   &  0.00 & 1.75 & 1.50 & 3.39 & 3.39 \\
    \hline
      one   & Morse & 25.69 &      & 1.63 & 3.39 & 4.06 \\
            & weak  & 26.83 &      & 1.92 & 3.41 & 4.46 \\
            & shift & 18.20 &      & 1.87 & 3.25 & 3.85 \\
    \hline
      two   & Morse & 25.47 & 2.50 & 2.07 & 3.38 & 3.37 \\
            & weak  & 28.58 & 2.71 & 2.10 & 3.75 & 1.76 \\
            & shift & 18.55 & 2.18 & 4.01 & 2.75 & 3.45 \\
    \hline
      three & Morse & 17.05 &      & 2.04 & 3.32 & 3.38 \\
            & weak  & 31.13 &      & 2.80 & 3.32 & 1.87 \\
            & shift &  9.22 &      & 2.94 & 3.00 & 3.74 \\
    \hline
  \end{tabular}
\end{table}

\begin{table}[t]
  \caption{Expectation values of the potential energy terms in
    mHartree at the surface for CD${}_4$, at an initial translational energy 
    of 96 kJ/mol and all initial vibrational states in the ground state,
    for the three different PESs.
    $V_{\rm surf}$ is the total surface deuterium repulsion; $V_{\rm
    harm}(\nu_2)$ and $V_{\rm harm}(\nu_4)$ are the harmonic
    terms of the intramolecular PES of the $a_1$ modes in the C${}_{3v}$
    and C${}_{2v}$ symmetry corresponding to a $\nu_2(e)$ bending and
    $\nu_4(t_2)$ umbrella modes respectively in the T${}_d$ symmetry.
    $V_{\rm bond}(R_{\rm up})$ and $V_{\rm bond}(R_{\rm down})$ are the
    potential energy in a single C--D bond pointing respectively towards
    and away from the surface. 
    The different PESs are: An anharmonic intramolecular PES [Morse, see Eq.\
    (\ref{Vmorse_d})], an intramolecular PES with weakening C--D bonds
    [weak, see Eq.\ (\ref{Vweak_d})], and an intramolecular PES with
    elongation of the C--D bonds [shift, see Eq.\ (\ref{Vshift_d})].}
  \label{tab:potcd4}
  \begin{tabular}[t]{l l c c c c c}
    \hline
      orient. & PES &  $V_{\rm surf}$ & $V_{\rm harm}(\nu_2)$ & 
       $V_{\rm harm}(\nu_4)$  & $V_{\rm bond}(R_{\rm up})$ &
      $V_{\rm bond}(R_{\rm down})$ \\  
    \hline
    initial & all   &  0.00 & 1.24 & 1.12 & 2.48 & 2.48 \\
    \hline
      one   & Morse & 26.88 &      & 1.47 & 2.47 & 3.29 \\
            & weak  & 27.40 &      & 2.04 & 2.48 & 4.05 \\
            & shift & 17.94 &      & 1.89 & 2.43 & 3.44 \\
    \hline
      two   & Morse & 26.75 & 2.09 & 1.48 & 2.46 & 2.50 \\
            & weak  & 29.20 & 2.26 & 1.42 & 2.71 & 1.30 \\
            & shift & 18.45 & 1.68 & 4.52 & 2.29 & 2.74 \\
    \hline
      three & Morse & 26.91 &      & 2.22 & 2.46 & 2.46 \\
            & weak  & 31.04 &      & 2.62 & 2.47 & 1.31 \\
            & shift &  8.71 &      & 3.49 & 2.28 & 3.21 \\
    \hline
  \end{tabular}
\end{table}
 
By looking at the expectation values of some terms of the PES
$\langle\Psi (t)\vert V_{\rm term} \vert\Psi (t)\rangle$ (see
Tables \ref{tab:potch4} and \ref{tab:potcd4}), we obtain information
about how the kinetics of the scattering is driven by the PES. 
The $V_{\rm surf}$ PES [see Eqs.\ (\ref{vsurfone}), (\ref{vsurftwo}) and
(\ref{vsurfthree})] is the surface hydrogen/deuterium 
repulsion for a given orientation. $V_{\rm harm}(\nu_2)$ and
$V_{\rm harm}(\nu_4)$ [see Eq.\ (\ref{Vharm})] are the pure harmonic
terms of the intramolecular 
PES of the $a_1$ modes in the C${}_{3v}$ and C${}_{2v}$ symmetry
corresponding to a $\nu_2$ bending and $\nu_4$ umbrella modes,
respectively. The pure harmonic correction terms of $V_{\rm corr}$ [see
Eq.\ (\ref{Vintra_morse})] are included in them. $V_{\rm bond}(R_{\rm up})$ and
$V_{\rm bond}(R_{\rm down})$ are the potential energy in a single C--H
or C--D bond pointing respectively towards and away from the surface,
and we give the expectation value of one bond term of $V_{\rm Morse}$
[see Eq.\ (\ref{Vmorse_d})], $V_{\rm weak}$ [see Eq.\ (\ref{Vweak_d})]
or $V_{\rm shift}$ [see Eq.\ (\ref{Vshift_d})]. 
All given expectation values are the maximum deviation of the initial
values, which effectively means the values at the moment the molecule hits
the surface.

The largest changes in expectation values are, of course, in
the kinetic energy of the translational mode. The translational kinetic
energy does not become zero as we should expect in classical dynamics.
The loss of translational kinetic energy is primary absorbed by the
$V_{\rm surf}$ terms of the PESs. The expectation values of the $V_{\rm
  surf}$ terms show  the ability of the hydrogens or deuteriums to come
close to the metal surface, since in real space their exponential
fall-offs are the same for both isotopes. For a rigid molecule the sum
of the translational kinetic energy and $V_{\rm surf}$ should be
constant, so all deviations of this sum have to be found back in the
intramolecular kinetic energy and other PES terms. 

The minimum expectation values of the translational kinetic energy for a
given orientation and PES is always larger for CH${}_4$ than CD${}_4$. 
We observe that for the PES with $V_{\rm Morse}$ and $V_{\rm weak}$ the
expectation value of $V_{\rm surf}$ is higher for CD${}_4$ than
CH${}_4$, so it is easier for the deuterium to come close to the surface
then for the hydrogens. This is the other way around for the PES with $V_{\rm
  shift}$. So we find with $V_{\rm shift}$ CH${}_4$ has both a higher
$V_{\rm surf}$ and kinetic energy expectation value than CD${}_4$.
 
We see for CD${}_4$ with the $V_{\rm Morse}$ PES in all orientations
about the
same translational energy. For CH${}_4$ this is only true for the
orientation with one and two hydrogens pointing towards the surface,
because we found in the orientation with three hydrogens pointing
towards the surface the translational energy at the surface is twice
that in the other orientations. This is the largest difference we have
found between the isotopes. We observe for all modes a little more
kinetic vibrational energy than in the groundstate for both isotopes,
but in 
the $\nu_1$ stretch term the extra kinetic energy is twice as high for
CH${}_4$ as CD${}_4$. The absolute value of the $V_{\rm harm}(\nu_4)$
PES term of CD${}_4$ has become even higher than CH${}_4$ in the
orientation with three bonds pointing towards the surface. This explains
the 1.2 degrees higher bond angle deformation of CD${}_4$ in this
orientation.

For the orientations with one and two bonds pointing towards the surface
we observe slightly more kinetic energy in the
vibrational modes and the harmonic PES terms become higher. In the
orientation with one bond pointing towards the surface we observe a
higher value for the $V_{\rm bond}(R_{\rm down})$ PES term. This is of
course caused by the repulsion of $V_{\rm surf}$, which also lead to a
shorter bond distance and because of that to higher contribution of the
repulsion term of $V_{\rm Morse}$.

The PES with $V_{\rm weak}$ shows for both isotopes an upward trend in
kinetic energy in the translational mode and the surface repulsion
$V_{\rm surf}$ terms with the number bonds pointing towards the surface.
Although the sum of these values for the orientation with one bond
pointing towards the surface are still lower than the value of the initial
translational energy, the kinetic energy in the stretch modes becomes
lower. This is also observed in the $e$ modes in the C${}_{3v}$ symmetry,
and $b_1$ and $b_2$ modes in the C${}_{2v}$ symmetry of the $\nu_3$
asymmetrical stretch modes. It is caused by the flattening of the bond
potential close to the surface. The $V_{\rm bond}(R_{\rm down})$ and
$V_{\rm  harm}(\nu_4)$ terms have to compensate this lost of energy by
increasing more than with the PES with $V_{\rm Morse}$. The increase for
CD${}_4$ in these terms is higher than for CH${}_4$. The value of the
$V_{\rm harm}(\nu_4)$ became even the highest for CD${}_4$. 

For the orientations with two and three bonds pointing towards the
surface we observe that the sum of the translational
kinetic energy and $V_{\rm surf}$ terms become higher than the initial
kinetic energy . This is compensated by a loss of vibrational kinetic
energy in all stretch modes and a lost of the $V_{\rm bond}(R_{\rm
  down})$ terms of the PES. The $V_{\rm harm}(\nu_2)$ and $V_{\rm
  harm}(\nu_4)$ become higher or stay around the same value compared
with the PES with $V_{\rm Morse}$. In the orientation with two bonds
pointing towards the surface we also observe a higher value for the
$V_{\rm bond}(R_{\rm up})$ term for both isotopes. 

Since we found for the PES with $V_{\rm shift}$ both the translational
kinetic energy and the $V_{\rm surf}$ terms were higher for CH${}_4$
than CD${}_4$, we have to find more increase in energy in the
intramolecular modes and PES terms for CD${}_4$ than CH${}_4$. We indeed
do so and that can be one of the reasons we found higher inelastic
scatter probabilities for CD${}_4$ with this type of model PES.

For the orientations with one and two bonds pointing towards the surface
we observe a large increase of the kinetic energy in the $\nu_3$
asymmetrical stretch mode. The translational kinetic energy becomes much
higher than with the other model PESs for these orientations, especially
in the orientation with one bond pointing towards the surface. If we
compare this with the excitation probabilities, we find that the kinetic
energy analysis gives indeed a different view on the dynamics. For the
orientation with two bond pointing towards the surface we have found for
both isotopes very high excitation probabilities in the $\nu_1$ and
$\nu_3$ stretch modes. We know now from the kinetic energy distribution
that for the $\nu_1$ symmetrical stretch mode the high excitation
probability is caused by the change of the equilibrium position of the
$\nu_1$ mode in the PES and that for the $\nu_3$ stretch mode probably
the PES also has become narrower.

For the orientation with three bonds pointing towards the surface we also
obtain an large increase of the kinetic energy of the $\nu_3$
asymmetrical stretch mode, but we also find an even larger increase in
the kinetic energy of the $\nu_1$ symmetrical stretch mode. The total
kinetic energy was extremely large, because the kinetic energy of the
translational mode becomes also much larger than for 
the other orientations. Because of this the $V_{\rm surf}$ terms had to
be around twice as low as for the other orientations. This means that it is
unfavourable to get the hydrogens and deuterium close to the surface in
this orientation with this PES type. 

All $V_{\rm bond}(R_{\rm up})$ terms of the PES with $V_{\rm shift}$
become lower compared to the initial value, especially in the
orientation with two bond pointing towards the surface. In the
orientation with one bond pointing towards the surface, the $V_{\rm
  bond}(R_{\rm down})$ term became higher. Again this is caused by the
repulsion of $V_{\rm surf}$ in the direction of the bond. For CH${}_4$
however it stays lower than the values for the other model PESs. The
increase of this PES term value is higher for CD${}_4$ and we can find
it above the value of the PES with $V_{\rm Morse}$, but still much lower
than for the PES with $V_{\rm weak}$. 

In the orientation with three bond pointing towards the surface we also
observe a higher $V_{\rm bond}(R_{\rm down})$ value, with also the
highest increase for CD${}_4$. In relation with the somewhat shorter
bond distance for the $R_{\rm down}$ for CD${}_4$ compared with
CH${}_4$, which was also a bit lower compared with the other
orientations, we know now that the hydrogens and especially the
deuterium have problems in following the minimum energy path of the PES
with $V_{\rm shift}$ during the scattering dynamics. This leads to higher
kinetic energy in the vibrational modes, which results in more inelastic
scattering. 

The $V_{\rm harm}(\nu_2)$ term of the PESs with $V_{\rm shift}$
increases in respect to the initial value, but not as much as with the 
other PES types. The increase of the $V_{\rm harm}(\nu_4)$ is for the
orientation with one bond pointing towards the surface in between the
values of the other PES types. In the other orientations it is higher,
especially in the orientation with two hydrogens pointing towards the
surface. The values for the CD${}_4$ isotopes are also higher than for
CH${}_4$. We observe also a larger increase of the kinetic energy in the
$\nu_4$ umbrella mode for CD${}_4$ than for CH${}_4$.

\section{Dissociation hypotheses}

We like now to discuss some possible implications of the scattering
simulation for the isotope effect on the dissociation of methane.  In
Chapter~\ref{chap:ch4scat} we have already drawn some conclusions about
the possible reaction mechanism and which potential type would be
necessary for dissociation. (See also Ref.~\cite{mil98}). We found the
direct breaking of a single C--H bond in the initial collision more
reasonable than the splats model with single bond breaking after an
intermediary Ni--C bond formation as suggested by Ref.~\cite{lee87},
because the bond angle deformations seems to small to allow a Ni--C to
form. From the simulations with CD${}_4$ we can draw the same
conclusions. The PES with $V_{\rm shift}$ gives the same angle
deformations for both isotopes, which is not sufficient for the splats
model. The other potentials give higher bond angle deformations for
orientation with three deuteriums pointing towards the surface. If the
Ni--C bond formation would go along this reaction path, then the
dissociation of CD${}_4$ should be even more preferable than CH${}_4$,
which is not the case.  So we only have to discuss the implication of
the scattering simulation for the dissociation probabilities of CH${}_4$
and CD${}_4$ for a direct breaking of a single bond reaction mechanism.
This reaction mechanism can be influenced by what we will call a direct
or an indirect effect.

A direct effects is the expected changes in the dissociation probability
between CH${}_4$ and CD${}_4$ for a given orientation. Since we expect
that we need for dissociation a PES with an elongation of the bonds
pointing towards the surface, we only have to look at the isotope effect
in the simulation for the PES with $V_{\rm shift}$ for different
orientations to discuss some direct effect. It is clear from our
simulations that the bond lengthening of CD${}_4$ is smaller than
CH${}_4$ for the orientation with three bonds pointing towards the
surface. If this orientation has a high contribution to the dissociation
of methane, then this can be the reason of the higher dissociation
probability of CH${}_4$. In this case our simulations also explain why
Ref.\cite{car98} did not observe a high enough isotope effect in
the dissociation probability of their simulation with CH${}_4$ and CD${}_4$
modelled by a diatomic, because we do not observe a change in bond
lengthening between the isotopes for the orientation with one bond
pointing towards the surface.

The orientation with three bonds pointing towards the surface is also
the orientation with the highest increase of the total vibrational
kinetic energy for the PES with $V_{\rm shift}$, because the energy
distribution analysis shows besides an high increase of the kinetic
energy in the $\nu_3$ asymmetrical stretch mode also an high increase in
the $\nu_1$ symmetrical stretch mode. Since vibrational kinetic energy
can be used effectively to overcome the dissociation barrier, the
orientation with three bonds indicates to be a more preferable orientation
for dissociation. Moreover the relative difference in kinetic energy
between both isotopes is for the $\nu_1$ stretch mode larger than
for the $\nu_3$ stretch mode. If the kinetic energy in the $\nu_1$
stretch mode contributes significantly to overcoming the dissociation
barrier, then it is another explanation for the low isotope effect in
Ref.\cite{car98}. 

An indirect effect is the expected changes in the dissociation
probability between CH${}_4$ and CD${}_4$ through changes in the
orientations distribution caused by the isotope effect in the
vibrational modes. This can be the case if the favourable
orientation for dissociation is not near the orientation with three
bonds pointing towards the surface, but more in a region where one or
two bonds pointing towards the surface. These orientations do not show a
large difference in deformation for the PES with $V_{\rm shift}$. We can
not draw immediate conclusion about the indirect effect from our
simulations, since we did not include rotational motion, but our
simulation show that an indirect isotope effect can exist. For the PES
with $V_{\rm Morse}$ in the orientation with three bonds pointing
towards the surface, we observe that CD${}_4$ is able to come closer to
the surface than CH${}_4$. So this rotational orientation should be more
preferable for CD${}_4$ than for CH${}_4$. On the other hand, if the PES
is in this orientation more like $V_{\rm shift}$ the dissociation
probability in other orientation can be decreased for CD${}_4$ through
higher probability in inelastic scattering channels. 

So for both effects the behaviour of the orientation with three bonds
pointing towards the surface seems to be essential for a reasonable
description of the dissociation mechanism of methane. 
A wave packet simulation of methane scattering including one or more
rotational degrees of freedom and the vibrational stretch modes will be
a good starting model to study the direct and indirect effects, since
most of the kinetic energy changes are observed in the stretch modes and
so the bending and umbrella modes are only relevant with accurate PESs.
Eventually dissociation paths can be introduced in the PES along one or
more bonds. 

Beside of our descriptions of the possible isotopes effect for
the dissociation extracted of the scatter simulations we have to keep
in mind that also a tunneling mechanism can be highly responsible for the
higher observed isotope effect in the experiment and that a different
dissociation barrier in the simulations can enhance this effect of tunneling. 

\section{Conclusions}

We have performed wave packet simulations on the scattering of CD${}_4$
from a flat surface with a fixed orientation with one, two, or three
hydrogens pointing towards the surface. We used the MCTDH method and
three different model PESs for each orientation. We used a translational
energy of up to 96 kJ/mol and all internal vibrations in the
groundstate. The scattering was in all cases predominantly
elastic. However, the observed inelastic scattering of CD${}_4$ compared
with simulation on CH${}_4$ is higher for the PES with an
elongated equilibrium bond length close to the surface.

When the molecule hits the surface, we observed in general a higher
vibrational excitation for CD${}_4$ than CH${}_4$. The PES with an
elongated equilibrium bond length close to the surface gives for both
isotopes almost the same deformations, although we observe a somewhat
smaller bond lengthening for CD${}_4$ in the orientation with three bonds
pointing towards the surface.   
The other model PESs show differences in the bond angle deformations and
in the distribution of the excitation probabilities of CD${}_4$ and
CH${}_4$, especially for the PES with only an anharmonic intramolecular
potential.

Energy distribution analysis proofs observations and hypotheses obtained
from excitation probabilities and structure deformation, and contributes
new information on the scattering dynamics.
A high increase of vibrational kinetic energy results in higher
inelastic scattering. The highest increase of vibrational kinetic energy
is found in the $\nu_3$ asymmetrical stretch modes for all orientations
and in the $\nu_1$ symmetrical stretch mode for the orientation with
three bonds pointing towards the surface, when the PES has an elongated
equilibrium bond length close to the surface.

Our simulations give an indications that the isotope effect in the
methane dissociation is caused mostly by the difference in the
scattering behaviour of the molecule in the orientation with three bonds
pointing towards the surface. At least multiple vibrational stretch
modes should be included for a reasonable description of isotope effect
in the methane dissociation reaction.

%\bibliography{Methane}
%\bibliographystyle{prstyfull}

%% file: excitations.tex
\chapter{The role of vibrational excitations}
\label{chap:exc}
%\vspace*{2cm}

\begin{quote}
  {\it The role of vibrational excitation of a single mode in the
    scattering of methane is studied by wave packet simulations of
    oriented CH${}_4$ and CD${}_4$ molecules from a flat surface. All
    nine internal vibrations are included. In the translational energy
    range from 32 up to 128 kJ/mol we find that initial vibrational
    excitations enhance the transfer of translational energy towards
    vibrational energy and increase the accessibility of the entrance
    channel for dissociation.  Our simulations predict that initial
    vibrational excitations of the asymmetrical stretch ($\nu_3$) and
    especially the symmetrical stretch ($\nu_1$) modes will give the
    highest enhancement of the dissociation probability of methane.
    }\footnote{This chapter has been published as a part of {\sl Bond
      breaking in vibrationally excited methane on transition-metal
      catalysts}\cite{mil00b}, \copyright \ {\sl 2000 The American
      Physical Society}.}
\end{quote}

\section{Introduction}

The dissociative adsorption of methane on transition metals is an
important reaction in catalysis; it is the rate limiting step in steam
reforming to produce syngas, and it is prototypical for catalytic C--H
activation. Although the reaction mechanism has been studied
intensively, it is not been fully understood yet. A number of molecular
beam experiments in which the dissociation probability was measured as a
function of translational energy have observed that vibrationally hot
CH${}_4$ dissociates more readily than cold CH${}_4$, with the energy in
the internal vibrations being about as effective as the translational
energy in inducing
dissociation \cite{ret85,ret86,lee87,lun89,hol95,lar99,walker99}. Two
independent bulb gas experiment with laser excitation of the $\nu_3$
asymmetrical stretch and $2\nu_4$ umbrella modes on the Rh(111)
surface \cite{yates79}, and laser excitation of the $\nu_3$ and $2\nu_3$
modes on thin films of rhodium \cite{brass79} did not reveal any
noticeable enhancement in the reactivity of CH${}_4$.  A recent
molecular beam experiment with laser excitation of the $\nu_3$ mode did
succeed in measuring a strong enhancement of the dissociation on a
Ni(100) surface. However, this enhancement was still much too low to
account for the vibrational activation observed in previous studies and
indicated that other vibrationally excited modes contribute
significantly to the reactivity of thermal samples \cite{juur99}.

Wave packet simulations of the methane dissociation reaction on
transition metals have treated the methane molecule always as a diatomic
molecule up to now \cite{har91,lun91,lun92,lun95,jan95,car98}. Apart
from one C--H bond (a pseudo-$\nu_3$ stretch mode) and the molecule
surface distance, either (multiple) rotations or some lattice motion
were included. None of them have looked at the role of the other
internal vibrations, so there is no model that describes which
vibrationally excited mode might be responsible for the experimental
observed vibrational activation. In previous chapters we have reported
on wave packet simulations to determine which and to what extent
internal vibrations are important for the dissociation of CH${}_4$ in
the vibrational ground state (Chapter \ref{chap:ch4scat}, see also
Ref.~\cite{mil98}), and the isotope effect of CD${}_4$ (Chapter
\ref{chap:iso}, see also Ref.~\cite{mil00a}). We were not able yet to
simulate the dissociation including all internal vibrations. Instead we
simulated the scattering of methane, for which all internal vibrations
can be included, and used the results to deduce consequences for the
dissociation. These simulations indicate that for methane to dissociate
the interaction of the molecule with the surface should lead to an
elongated equilibrium C--H bond length close to the surface. In this
chapter we report on new wave packet simulations of the role of
vibrational excitations for the scattering of CH${}_4$ and CD${}_4$
molecules with all nine internal vibrations. The dynamical features of
these simulations give new insight into the initial steps of the
dissociation process. The conventional explanation is that vibrations
help dissociation by adding energy needed to overcome the dissociation
barrier.  We find that two other new explanations play also a role. One
of them is the enhanced transfer of translational energy into the
dissociation channel by initial vibrational excitations. The other more
important explanation is the increased accessibility of the entrance
channel for dissociation.

We have used the multi-confi\-guratio\-nal time-depen\-dent Hartree
(MCTDH) method for our wave packet simulation \cite{man92,jan93}. This
method can deal with a large number of degrees of freedom and with large
grids.  (See Ref.~\cite{bec00} for a recent review, or
Section~\ref{sec:mctdh} of this thesis for a survey.)  Initial
translational energy has been chosen in the range of 32 to 128 kJ/mol.
The initial state has been written as a product state of ten functions;
one for the normally incident translational coordinate, and one for each
internal vibration. All vibrations were taken to be in the ground state
except one that was put in the first excited state. The orientation of
the CH${}_4$/CD${}_4$ was fixed, and the vibrationally excited state had
$a_1$ symmetry in the symmetry group of the molecule plus surface
(C${}_{3v}$ when one or three H/D atoms point towards the surface, and
C${}_{2v}$ when two point towards the surface.) The potential-energy
surface is characterised by an elongation of the C--H bonds when the
molecule approaches the surfaces [$V_{\rm shift}$, see
Eq.(\ref{Vshift_d})], no surface corrugation, and a molecule-surface
part appropriate for Ni(111). It has been shown to give reasonable
results, and is described in Section~\ref{sec:pes} (and in
Refs.~\cite{mil98,mil00a}).  The computational details about the
configurational basis and number of grid points are discussed in
Section~\ref{sec:states}.  Figure \ref{fig:orient} illustrates the
orientations and Fig.~\ref{fig:modes} the important vibrational modes.

\section{Energy distribution analysis}
\subsection{Translational kinetic energy}

\begin{figure}[t]
  \epsfig{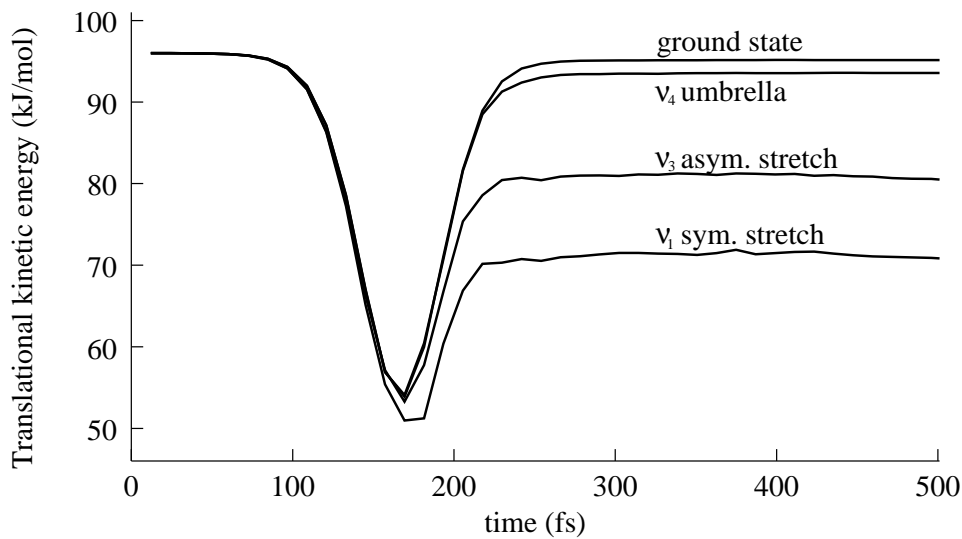}
  \caption{Translational kinetic energy versus time for a CH${}_4$
    molecule with three bonds pointing towards the surface. The initial
    translational kinetic energy is 96 kJ/mol.}
  \label{fig:trans96}
\end{figure}

We can obtain a good idea about the overall activation of a mode by
looking at the kinetic energy expectation values $\langle\Psi (t)\vert
T_j \vert\Psi (t)\rangle$ for each mode $j$. During the scattering
process the change in the translational kinetic energy is the largest.
It is plotted in Fig.~\ref{fig:trans96} as a function of time for
CH${}_4$ in the orientation with three bonds pointing towards the
surface with an initial kinetic energy of 96 kJ/mol and different
initial vibrational states. When the molecule approaches the surface the
kinetic energy falls down to a minimum value.  This minimum value varies
only slightly with the initial vibrational states of the molecule. 

The total loss of translational kinetic energy varies substantially,
however. The initial translational kinetic energy is not conserved. This
means that the vibrational excitation enhances inelastic scattering.
Especially an excitation of the $\nu_1$ symmetrical stretch mode and to
a lesser extent the $\nu_3$ asymmetrical stretch mode results in an
increased transfer of kinetic energy towards the intramolecular
vibrational energy. The inelastic scatter component (the initial minus
the final translational energy) for both isotopes in the orientation
with three bonds pointing towards the surface, shows the following trend
for the initial vibrational excitations of the modes; $\nu_1$ $>$
$\nu_3$ $>$ $\nu_4$ $>$ ground state. CH${}_4$ scatters more inelastically
than CD${}_4$ over the whole calculated range of translational kinetic
energies, if the molecule has an initial excitation of the $\nu_3$
stretch mode.  CH${}_4$ scatters also more inelastically than CD${}_4$
in the $\nu_1$ symmetrical stretch mode at higher energies , but at
lower energies it scatters slightly less inelastically.  For the
molecules with the non-excited state or an excitation in the $\nu_4$
umbrella mode CD${}_4$ has a higher inelastic scattering component than
CH${}_4$. At an initial translational kinetic energy of 128 kJ/mol the
excitation of the $\nu_4$ umbrella mode results in a strong enhancement
of the inelastic scattering component. For CD${}_4$ the inelastic
scattering component for the initial excited $\nu_4$ umbrella mode can
become even larger than that for the initial excited $\nu_3$ stretch mode.

For the orientation with two bonds pointing towards the surface we
observe the same trends for the relation between the inelastic scatter
components and the excited initial vibrational modes, but the inelastic
scatter component are less than half of the values for the orientation
with three bonds pointing towards the surface. Also the excitation of
the $\nu_3$ asymmetrical stretch modes results now in a higher inelastic
scattering component for CD${}_4$ than for CH${}_4$. Excitation of the
$\nu_2$ bending mode gives a slightly higher inelastic scatter component
than the vibrational ground state.

For the orientation with one bond pointing towards the surface we
observe an even lower inelastic scattering component. At an initial
kinetic energy of 128 kJ/mol we find that both the $\nu_1$ and $\nu_3$
stretch modes have an inelastic component of around 6.5 kJ/mol for
CD${}_4$ and 4.0 kJ/mol for CH${}_4$. 

At an initial translational energy of 32 kJ/mol we observe for both
isotopes in all orientations a very small increase of translational
kinetic energy for the vibrational excited molecule, which means that
there is a net transfer from intramolecular vibrational energy through
the surface repulsion into the translational coordinate.

\subsection{Vibrational kinetic energy}

\begin{figure}[t]
  \epsfig{file=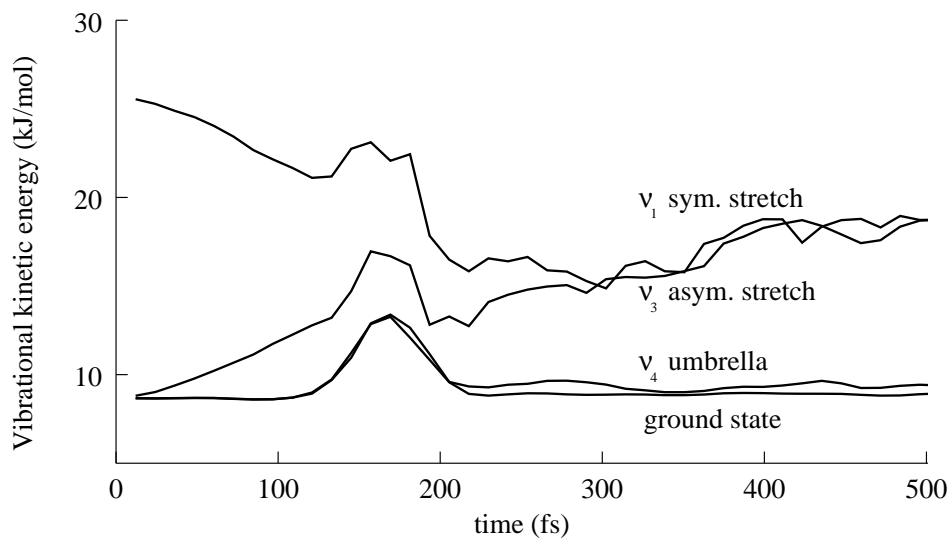, height=7.0cm}
  \caption{Vibrational kinetic energy in the $\nu_1$ {\em symmetrical stretch}
    mode versus time for a CH${}_4$ molecule with three bonds pointing
    towards the surface. The initial translational kinetic energy is 96
    kJ/mol.}
  \label{fig:vibnu1_96}
\end{figure}

\begin{figure}[t]
  \epsfig{file=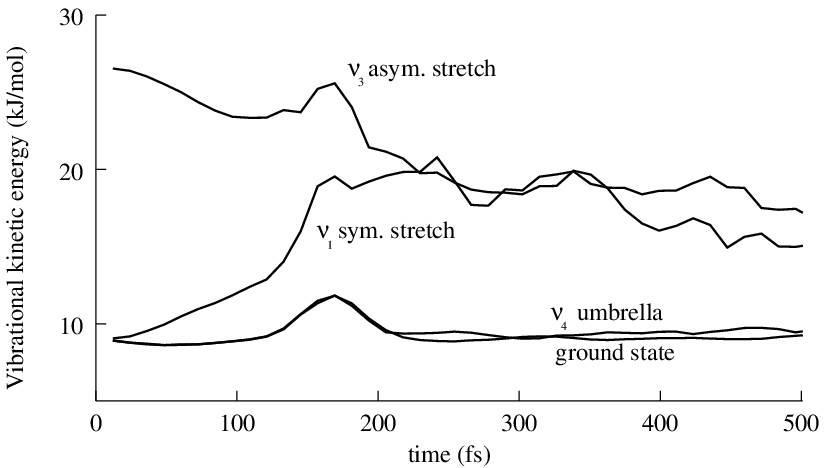, height=7.0cm}
  \caption{Vibrational kinetic energy in the $\nu_3$ {\em asymmetrical stretch}
    mode versus time for a CH${}_4$ molecule with three bonds pointing
    towards the surface. The initial translational kinetic energy is 96
    kJ/mol.}
  \label{fig:vibnu3_96}
\end{figure}

\begin{figure}[t]
  \epsfig{file=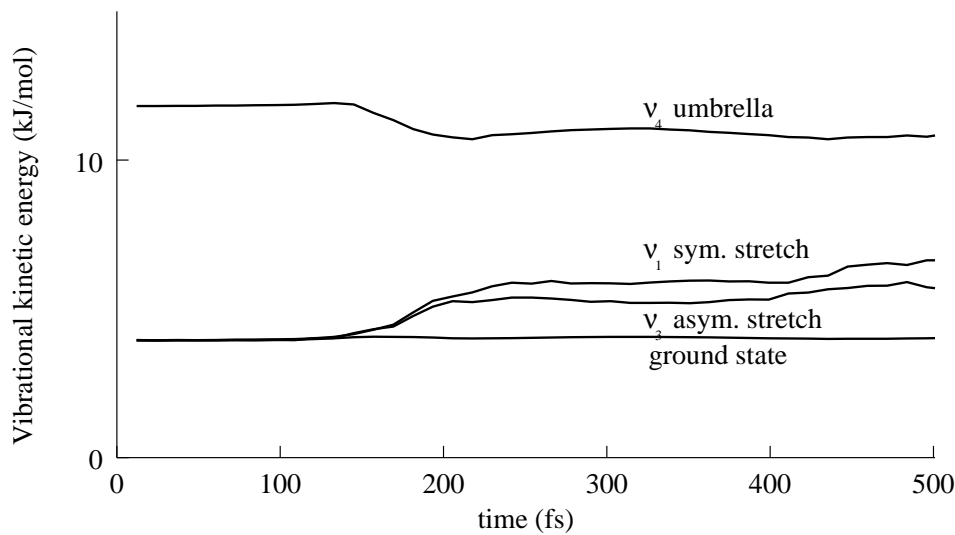, height=7.0cm}
  \caption{Vibrational kinetic energy in the $\nu_4$ {\em umbrella}
    mode versus time for a CH${}_4$ molecule with three bonds pointing
    towards the surface. The initial translational kinetic energy is 96
    kJ/mol.}
  \label{fig:vibnu4_96}
\end{figure}

There seem to be two groups of vibrations with different qualitative
behavior with respect to (de)excitation when the molecule hits the
surface. The first group, which we call the ``stretch'' group, consists
of the $\nu_3$ asymmetric stretch mode in any orientation and the
$\nu_1$ symmetric stretch mode in the orientation with three
hydrogen/deuterium atoms pointing to the surface. Figures
\ref{fig:vibnu1_96} and \ref{fig:vibnu3_96} show the vibrational kinetic
energy in the $\nu_1$ and $\nu_3$ stretch modes respectively versus time
for CH${}_4$ in the orientation with three bonds to the surface.  The
second, called the ``bending'' group, consists of all bending vibrations
and the $\nu_1$ in other orientations. Figure \ref{fig:vibnu4_96} shows
the vibrational kinetic energy in the $\nu_4$ umbrella mode versus time
for the same orientation of CH${}_4$.

When the molecule is initially in the vibrational ground state the
kinetic energy in the vibrations increases, reaches a maximum at the
turn-around point, and then drops back almost to the initial level
except for a small contribution due to the inelastic scattering
component. The vibrations within a group have very similar amounts of
kinetic energy, but the ``stretch'' group has clearly a larger inelastic
component than the ``bending'' group, and also the kinetic energy at the
turn-around point is larger.  When the molecule has initially an
excitation of a vibration of the ``stretch'' group then the kinetic
energy of that vibration increases, reaches a maximum at the turn-around
point, and drops to a level lower than it was initially. For an
excitation of a vibration of the ``bending'' group there is no maximum,
but its kinetic energy simply drops to a lower level. We see that in all
cases there is not only a transfer of energy from the translation to
vibrations, but also an energy flow from the initially excited vibration
to other vibrations. However, the total energy of the vibrational
kinetic energy and the intramolecular potential energy increases,
because it has to absorb the inelastic scattering component.

\subsection{Potential energy of the surface repulsion term}

\begin{figure}[t]
  \epsfig{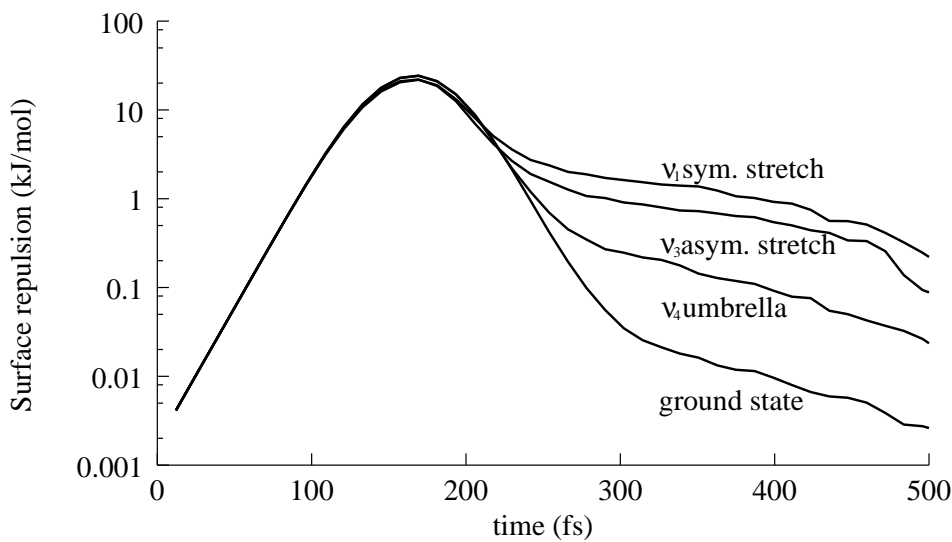}
  \caption{Surface repulsion versus time during the scattering dynamics
    of CH${}_4$ at an initial translational energy of 96 kJ/mol in the
    orientation with three bonds pointing towards the surface. }
   \label{fig:vsurf}
\end{figure}

Figure \ref{fig:vsurf} shows the (repulsive) interaction with the
surface during the scattering process of CH${}_4$ at an initial kinetic
energy of 96 kJ/mol and different initial vibrational excitations for
the orientation with three hydrogens pointing towards the surface. Since
this is a repulsive term with a exponential fall-off changes in the
repulsion indicate the motion of the part of the wave packet closest to
the surface.  At the beginning of the simulation the curves are almost
linear in a logarithmic plot, because the repulsion hardly changes the
velocity of the molecule.  After some time the molecule enters into a
region with a higher surface repulsion and the slopes of the curves
drop. This results in a maximum at the turn-around point, where most of
the initial translational kinetic energy is transferred into potential
energy of the surface repulsion.  For a classical simulation it would
have meant no translational kinetic energy, but it corresponds with the
minimum kinetic energy for our wave packet simulations.  Past the
maximum, a part of the wave packet will accelerate away from the
surface, and the slope becomes negative.  The expectation value of the
translational kinetic energy (see Fig.\ref{fig:trans96}) increases at
the same time.  The slope of the curves in Fig.~\ref{fig:vsurf} becomes
less negative towards the end of the simulation, although the
expectation value of the translational kinetic energy in this time
region is almost constant.  The reason for this is that a part of the
wave packet with less translational kinetic energy is still in a region
close to the surface. We see also that the height of the plateaus for
the different initial vibrational excitations is again in the order;
$\nu_1$ $>$ $\nu_3$ $>$ $\nu_4$ $>$ ground state. This again indicates
that a larger part of the wave packet is inelastically scattered when
$\nu_1$ is excited than when $\nu_3$ is excited, etc.

At lower initial translational kinetic energies the plateaus have a
lower position and the main gap exists between the plateaus of the
$\nu_1$ and $\nu_3$ stretch modes and the lower positioned plateaus of
the $\nu_4$ umbrella mode and the ground state. At an initial translational
kinetic energy of 128 kJ/mol the positions of the plateaus are higher
and the differences between the initial vibrational excitations are also
smaller.  The plateau of the $\nu_3$ stretch mode is even around the
same position as the $\nu_4$ umbrella mode for CD${}_4$ in the
orientation with three bonds pointing towards the surface at this
initial energy.  The orientation with two bonds pointing towards the
surface shows the same trends. The plateaus of the initial excited
$\nu_2$ bending mode are located slightly above the ground state for
both isotopes.  For the orientation with one bond pointing towards the
surface the relative positions of the plateaus of the different initial
excitations are the same as at low energies in the orientation with
three bonds pointing towards the surface.
 
\section{Dissociation hypotheses}

Even though we did not try to describe the dissociation itself, the
scattering simulation do yield indications for the role of vibrational
excitations on the dissociation of methane, and we compare these with
experimental observations.  The dissociation of methane occurs over a
late barrier, because it is enhanced by vibrational energy \cite{lev87}.
Conventionally, the role of vibrational excitation on the enhancement of
dissociation probability was discussed as an effect of the availability
of the extra (vibrational kinetic) energy for overcoming the
dissociation barrier. Our simulations show that such a process might
play a role, but they show also that two other processes occur through
vibrational excitation.

Firstly, an initial vibrational excitation increases translational
kinetic energy transfer towards the intramolecular vibrational energy.
The simulations show that this inelastic scatter component can be seen
in an large enhancement of the vibrational kinetic energy in the stretch
modes at the turn-around point.  This increase is larger for higher
initial translational kinetic energy and is most effective in the
orientation with three bonds pointing towards the surface. If the
dissociation of methane occurs primarily in this orientation, then we
would expect, based on the total available vibrational energy after
hitting the surface, that excitation of the $\nu_1$ symmetrical stretch
mode is the most effective for enhancing the dissociation probability.
The $\nu_3$ asymmetrical stretch mode appears to be less so. An
explanation of the enhanced inelastic scatter compound by vibrational
excitation is that through excitation the bonds are weakened, which will
ease excitation in the initial non-excited modes. Excitations other than
the $\nu_2$, $\nu_3$, or $\nu_4$ with $a_1$ symmetry for a particular
orientation can possibly result in higher energy transfers, but we think
that the difference with $\nu_1$ (which has always $a_1$ symmetry) would
be still large.

Second, the accessibility of the dissociation channel also enhances the
dissociation probability. We have concluded in Chapter~\ref{chap:ch4scat}
(and Ref.~\cite{mil98}) that our potential mimics reasonably the
entrance channel for dissociation.  In this chapter we find that a part
of the wave packet has a longer residence time at the surface. It is
this part of the wave packet that accesses the dissociation channel, and
it is also this part that is able to come close to the transition state
for dissociation.  From Figs.~\ref{fig:trans96} and \ref{fig:vsurf} we
conclude that the $\nu_1$ stretch mode will enhance the dissociation
probability the most.  The enhanced accessibility by vibrational
excitation is explained by the spread of the wave packet along a C--H
bond, which gives a higher probability for the system to be atop the
dissociation barrier.

The molecular beam experiment with excitations of the $\nu_3$
asymmetrical stretch mode of CH${}_4$ of Ref.~\cite{juur99} shows that a
single excitation of the $\nu_3$ asymmetrical stretch mode enhances
dissociation, but the measured reactivity of the $\nu_3$ stretch mode is
too low to account for the total vibrational activation observed in the
molecular beam study of Ref.~\cite{hol95}. It means that excitation of a
mode other than the $\nu_3$ stretch mode will be more effective for
dissociation. Our simulations show that indeed excitation of the $\nu_3$
stretch mode will enhance dissociation, but predict that excitation of
the $\nu_1$ symmetrical stretch mode will be more effective if the
dissociation occurs primary in the orientation with multiple bonds
pointing towards the surface. The contribution of the $\nu_1$
symmetrical stretch mode cannot be measured directly, because it has no
infra-red activity. However, the contribution of the $\nu_1$ mode can be
estimated using a molecular beam study as follows. The contribution of
the $\nu_3$ stretch mode has already been determined \cite{juur99}.
Similarly the contribution of the $\nu_4$ umbrella mode can be
determined. The contribution of the $\nu_2$ bending can be estimated
from our simulations to be somewhat lower than the $\nu_4$ umbrella
contribution. The total contribution of all vibrations is known from
Ref.~\cite{hol95}, and a simple subtraction will give us then the
contribution of the $\nu_1$ stretch mode.  At high translational
energies the accessibility of the dissociation channel for molecules
with an excited $\nu_4$ umbrella mode is close to that of the molecules
with excited stretch modes, and for CD${}_4$ the inelastic scattering is
also enhanced.  So the excitation of the $\nu_4$ umbrella mode can still
contribute significantly to the vibrational activation, because it also
has higher Boltzmann population in the molecular beam than the stretch
modes.

\section{Conclusions}

We have performed wavepacket simulation of the scattering of fixed
oriented, vibrationally excited CH${}_4$ and CD${}_4$ from a flat
surface. We used initial translational energies in the range of 32 to
128 kJ/mol. A single vibrational excitation was put in one of the
vibrational modes with $a_1$ symmetry in the symmetry group of the molecule
plus surface, while the other vibrational modes were kept in the
groundstate. The potential-energy surface is characterised by an
elongation of the C--H bonds when the molecule approaches the surfaces,
and a molecule-surface part appropriate for Ni(111).

We find that initial vibrational excitations enhance the transfer of
translational energy towards vibrational energy and increase the
accessibility of the entrance channel for dissociation, which means an
increase of inelastic scattering. The largest effects are observed in
the orientation with three bonds pointing to the surface.  Our
simulations predict that initial vibrational excitations of the
asymmetrical stretch ($\nu_3$) and especially the symmetrical stretch
($\nu_1$) modes will give the highest enhancement of the dissociation
probability of methane.

%\bibliography{Methane}
%\bibliographystyle{prstyfull}

%% file: trajectory.tex
\chapter{Classical trajectory simulations}
\label{chap:traj}

\begin{quote}
  {\it We present classical trajectory calculations of the rotational
    vibrational scattering of a non-rigid methane molecule from a
    Ni(111) surface. Energy dissipation and scattering angles have been
    studied as a function of the translational kinetic energy, the
    incidence angle, the (rotational) nozzle temperature, and the
    surface temperature.  Scattering angles are somewhat towards the
    surface for the incidence angles of 30${}^{\circ}$, 45${}^{\circ}$,
    and 60${}^{\circ}$ at a translational energy of 96 kJ/mol. Energy
    loss is primarily from the normal component of the translational
    energy. It is transfered for somewhat more than half to the surface
    and the rest is transfered mostly to rotational motion.  The spread
    in the change of translational energy has a basis in the spread of
    the transfer to rotational energy, and can be enhanced by raising of
    the surface temperature through the transfer process to the surface
    motion.\footnote{This chapter will be published as {\sl Energy
        dissipation and scattering angle distribution analysis of the
        classical trajectory calculations of methane scattering from a
        Ni(111) surface} \cite{mil01}, preprint available at
      \texttt{http://arXiv.org/abs/physics/0103053}.} }
\end{quote}

%\copyright \ {\sl 2001 American Institute of Physics}

\section{Introduction}
\label{dyn:intro}

It is very interesting to simulate the dynamics of the dissociation,
because of the direct dissociation mechanism, and the role of the
internal vibrations. In previous Chapters \ref{chap:ch4scat} and
\ref{chap:iso} we have reported on wave packet simulations to determine
which and to what extent internal vibrations are important for the
dissociation in the vibrational ground state of CH${}_4$ (see also
Ref.~\cite{mil98}), and CD${}_4$ (see also Ref.~\cite{mil00a}). We were
not able yet to simulate the dissociation including all internal
vibrations. Instead we simulated the scattering of methane in fixed
orientations, for which all internal vibrations can be included, and
used the results to deduce consequences for the dissociation. These
simulations indicate that to dissociate methane the interaction of the
molecule with the surface should lead to an elongated equilibrium C--H
bond length close to the surface, and that the scattering was almost
elastic. In Chapter \ref{chap:exc} we reported on wave packet
simulations of the role of vibrational excitations for the scattering of
CH${}_4$ and CD${}_4$ (see also Ref.~\cite{mil00b}). We predicted that
initial vibrational excitations of the asymmetrical stretch ($\nu_3$)
but especially the symmetrical stretch ($\nu_1$) modes will give the
highest enhancement of the dissociation probability of methane.
Although we have performed these wave packet simulations in ten
dimensions, we still had to neglect two translational and three
rotational coordinates of the methane molecule and we did not account
for surface motion and corrugation. It is nowadays still hard to include
all these features into a wave packet simulation, therefore we decided
to study these with classical trajectory simulations.

In this chapter we will present full classical trajectory simulations of
methane from a Ni(111) surface. We have especially interest in the
effect of the molecular rotations and surface motion, which we study as
a function of the nozzle and surface temperature. The methane molecule
is flexible and able to vibrate. We do not include vibrational kinetic
energy at the beginning of the simulation, because a study of
vibrational excitation due to the nozzle temperature needs a special
semi-classical treatment. Besides its relevance for the dissociation
reaction of methane on transition metals, our scattering simulation can
also be of interest as a reference model for the interpretation of
methane scattering itself, which have been studied with molecular beams
on Ag(111) \cite{asada81,asada82}, Pt(111)
\cite{yagyu99,yagyu99b,yagyu00,hiraoka00}, and Cu(111) surfaces
\cite{Andersson00}. It was reported in Refs.~\cite{yagyu99,yagyu99b}
that the scattering angles are in some cases in disagreement with the
outcome of the classical Hard Cube Model (HCM) described in
Ref.~\cite{logan66}.  We will show in this article that the assumptions
of this HCM model are too crude for describing the processes obtained
from our simulation. The time-of-flight experiments show that there is
almost no vibrational excitation during the scattering
\cite{yagyu00,hiraoka00}, which is in agreement with our current
classical simulations and our wave packet simulations discussed in the
previous Chapters \ref{chap:ch4scat} and \ref{chap:iso}.

The rest of this chapter is organized as follows. We start with a
description of our model potential, and an explanation of the 
simulation conditions. The results and discussion are presented next. We
start with the scattering angles, and relate them to the energy
dissipation processes. Next we will compare our simulation with other
experiments and theoretical models. We end with a summary and some
general conclusions.

\section{Computational details}
\label{dyn:comp}

We have used classical molecular dynamics for simulating the scattering
of methane from a Ni(111) surface. The methane molecule was modelled as
a flexible molecule. The forces on the carbon, hydrogen, and Ni atoms
are given by the gradient of the model potential energy surface
described below.  The first-order ordinary differential equations for
the Newtonian equations of motion of the Cartesian coordinates were
solved with use of a variable-order, variable-step Adams
method \cite{nag17}. We have simulated at translational energies of 24,
48, 72, and 96 kJ/mol at normal incidence, and at 96 kJ/mol for incidence
angles of 30${}^{\circ}$, 45${}^{\circ}$, and 60${}^{\circ}$ with the
surface normal. The surface temperature and (rotational) nozzle temperature for
a certain simulation were taken independently between 200 and 800 K.

\subsection{Potential energy surface}
\label{dyn:pes}

The model potential energy surface used for the classical dynamics is
derived from one of our model potentials with elongated C--H bond
lengths towards the surface, previously used for wave packet simulation
of the vibrational scattering of fixed oriented methane on a flat
surface [see Eq.~\ref{Vshift_d} in Section \ref{sec:pes}, and also
Refs.~\cite{mil98} and \cite{mil00a}]. In this original potential there
is one part responsible for the repulsive interaction between the
surface and the hydrogens, and another part for the intramolecular
interaction between carbon and hydrogens.

We have rewritten the repulsive part [see Eq.~\ref{gen_Vsurf}] in pair
potential terms between top layer surface Ni atoms and hydrogens in such
a way that the surface integral over all these Ni atoms give the same
overall exponential fall-off as the original repulsive PES term for a
methane molecule far away from the surface in an orientation with three
bonds pointing towards the surface.  The repulsive pair interaction term
$V_{\rm rep}$ between hydrogen $i$ and Ni atom $j$ at the surface is
then given by
\begin{equation}
  \label{Vrep}
  V_{\rm rep}=\frac{A\ e^{-\alpha Z_{ij}}} {Z_{ij}},
\end{equation}
where $Z_{ij}$ is the distance between hydrogen atom $i$ and Ni atom $j$.

The intramolecular potential part is split up in bond, bond angle, and
cross potential energy terms. The single C--H bond energy is given
by a Morse function with bond lengthening towards the surface
\begin{equation}
  \label{Vbond}
  V_{\rm bond}=  D_e \ \Big[1-e^{ -\gamma (R_i-R_{eq})}\Big]^2 ,
\end{equation}
where $D_e$ is the dissociation energy of methane in the gas phase, and
$R_i$ is the length of the C--H bond $i$. Dissociation is not possible
at the surface with this potential term, but the entrance channel for
dissociation is mimicked by an elongation of the equilibrium bond length
$R_{eq}$ when the distance between the hydrogen atom and the Ni atoms in
the top layer of the surface become shorter. This is achieved by
\begin{equation}
  \label{Req}
  R_{eq}= R_0 + S~\sum\limits_{j} \frac{e^{-\alpha Z_{ij}}} {Z_{ij}} ,
\end{equation}
where $R_0$ is the equilibrium C--H bond length in the gas phase. The
bond elongation factor $S$ was chosen in such a way that the elongation
is 0.054 nm at the classical turning point of 93.2 kJ/mol incidence
translational energy for a rigid methane molecule, when the molecule
approach a surface Ni atom atop with one bond pointing towards the
surface.  The single angle energy is given by the harmonic expression
\begin{equation}
  \label{Vangle}
  V_{\rm angle} = k_{\theta}\ (\theta_{ij} - \theta_{0})^2 ,
\end{equation}
where $\theta_{ij}$ is the angle between C--H bond $i$ and $j$, and
$\theta_{0}$ the equilibrium bond angle.
Furthermore, there are some cross-term potentials between bonds and
angles. The interaction between two bonds are given by
\begin{equation}
  \label{Vbb}
  V_{\rm bb} = k_{RR}\ (R_i-R_0)(R_j-R_0) .
\end{equation}
The interaction between a bond angle and the bond angle on the other side is
given by
\begin{equation}
  \label{Vaa}
  V_{\rm aa} = k_{\theta\theta}\ 
             (\theta_{ij}-\theta_{0})(\theta_{kl}- \theta_{0}).
\end{equation}
The interaction between a bond angle and one of its bonds is given by
\begin{equation}
  \label{Vab}
  V_{\rm ab} = k_{\theta R}\ (\theta_{ij}-\theta_{0})(R_i- R_{0}).
\end{equation}
The parameters of the intramolecular potential energy terms
were calculated by fitting the second derivatives of these terms on the
experimental vibrational frequencies of CH${}_4$ and
CD${}_4$ in the gas phase \cite{gray79,lee95}.

The Ni-Ni interaction between nearest-neighbours is given by the
harmonic form
\begin{eqnarray}
  \label{Vnini}
  V_{\rm Ni-Ni} & = & \frac{1}{2} \lambda_{ij} [(\mathbf{u}_i -
                  \mathbf{u}_j) \cdot \mathbf{\hat r}_{ij} ]
                  \nonumber\\ 
                  & & + \frac{1}{2} \mu_{ij}  \Big\{(\mathbf{u}_i -
                  \mathbf{u}_j)^2 - [(\mathbf{u}_i -
                  \mathbf{u}_j) \cdot \mathbf{\hat r}_{ij} ]^2 \Big\}.
\end{eqnarray}
The $\mathbf{u}$'s are the displacements from the equilibrium positions,
and $\mathbf{\hat r}$ is a unit vector connecting the equilibrium
positions. The Ni atoms were placed at bulk positions with a
nearest-neighbour distance of 0.2489 nm. The parameters $\lambda_{ij}$
and $\mu_{ij}$ were fitted on the elastic constants \cite{Landolt} and
cell parameters \cite{Ashcroft} of the bulk. The values of all parameters
are given in Table \ref{tab:traj_par}.

\begin{table}
  \caption{Parameters of the potential energy surface.}
  \label{tab:traj_par}
  \begin{tabular}{l l r l}
     \hline
     \hline
    Ni--H & $A$ & 971.3 & kJ nm mol${}^{-1}$ \\
     & $\alpha$ & 20.27   & nm${}^{-1}$ \\
     & $S$      & 0.563     & nm${}^2$  \\
     \\
    CH${}_4$    & $\gamma$ & 17.41 & nm${}^{-1}$\\
     & $D_e$    & 480.0 & kJ mol${}^{-1}$\\
     & $R_0$    & 0.115 & nm \\
     & $k_{\theta}$ & 178.6 & kJ mol${}^{-1}$ rad${}^{-2}$ \\
     & $\theta_0$   & 1.911 & rad \\
     & $k_{RR}$     & 4380 & kJ mol${}^{-1}$ nm${}^{-2}$ \\
     & $k_{\theta\theta}$ & 11.45 & kJ mol${}^{-1}$ rad${}^{-2}$\\
     & $k_{\theta R}$ & -472.7 & kJ mol${}^{-1}$ rad${}^{-1}$ nm${}^{-1}$\\
     \\
    Ni--Ni & $\lambda_{nn}$ & 28328 & kJ mol${}^{-1}$ nm${}^{-2}$ \\
     & $\mu_{nn}$ & -820 & kJ mol${}^{-1}$ nm${}^{-2}$ \\
     \hline
     \hline
  \end{tabular}
\end{table}

\subsection{Simulation model}

The surface is modelled by a slab consisting of four layers of eight
times eight Ni atoms. Periodic boundary conditions have been used in the
lateral direction for the Ni-Ni interactions. The methane molecule has
interactions with the sixty-four Ni atoms in the top layer of the slab.
The surface temperature is set according to the following procedure. The
Ni atoms are placed in equilibrium positions and are given random
velocities out of a Maxwell-Boltzmann distribution with twice the
surface temperature. The velocities are corrected such that the total
momentum of all surface atoms is zero in all directions, which fixes the
surface in space. Next the surface is allowed to relax for 350 fs.
We do the following ten times iteratively. If at the end of previous
relaxation the total kinetic energy is above or below the given surface
temperature, then all velocities are scaled down or up with a factor of
$\sqrt 1.1$ respectively. Afterwards a new relaxation simulation is
performed. The end of each relaxation run is used as the begin condition
of the surface for the actual scattering simulation.

The initial perpendicular carbon position was chosen 180 nm above the
equilibrium $z$-position of the top layer atoms and was given randomly
parallel ($x$, $y$) positions within the central surface unit cell of
the simulation slab for the normal incidence simulations. The methane
was placed in a random orientation with the bonds and angles of the
methane in the minimum of the gas phase potential. The initial
rotational angular momentum was generated randomly from a
Maxwell-Boltzmann distribution for the given nozzle temperature for all
three rotation axis separately. No vibrational kinetic energy was given
initially. Initial translational velocity was given to all methane
atoms according to the translational energy. The simulations under an
angle were given parallel momentum in the [110] direction. The parallel
positions have been translated according to the parallel velocities in
such a way that the first collision occurs one unit cell before the
central unit cell of the simulation box. We tested other directions, but
did not see any differences for the scattering.

Each scattering simulation consisted of 2500 trajectories with a
simulation time of 1500 fs each. We calculated the (change of)
translational, total kinetic, rotational and vibrational kinetic,
intramolecular potential, and total energy of the methane molecule; and
the scattering angles at the end of each trajectory. We calculated for
them the averages and standard deviations, which gives the spread for
the set of trajectories, and correlations coefficients from which we can
abstract information about the energy transfer processes.

\section{Results and discussion}
\label{dyn:results}

We will now present and discuss the results of our simulations. We begin
with the scattering angle distribution. Next we will explain this in
terms of the energy dissipation processes. Finally we will compare our
simulation with previous theoretical and experimental scattering
studies, and discuss the possible effects on the dissociation of methane
on transition metal surfaces.

\subsection{Scattering angles}

\begin{figure}[p]
\epsfig{file=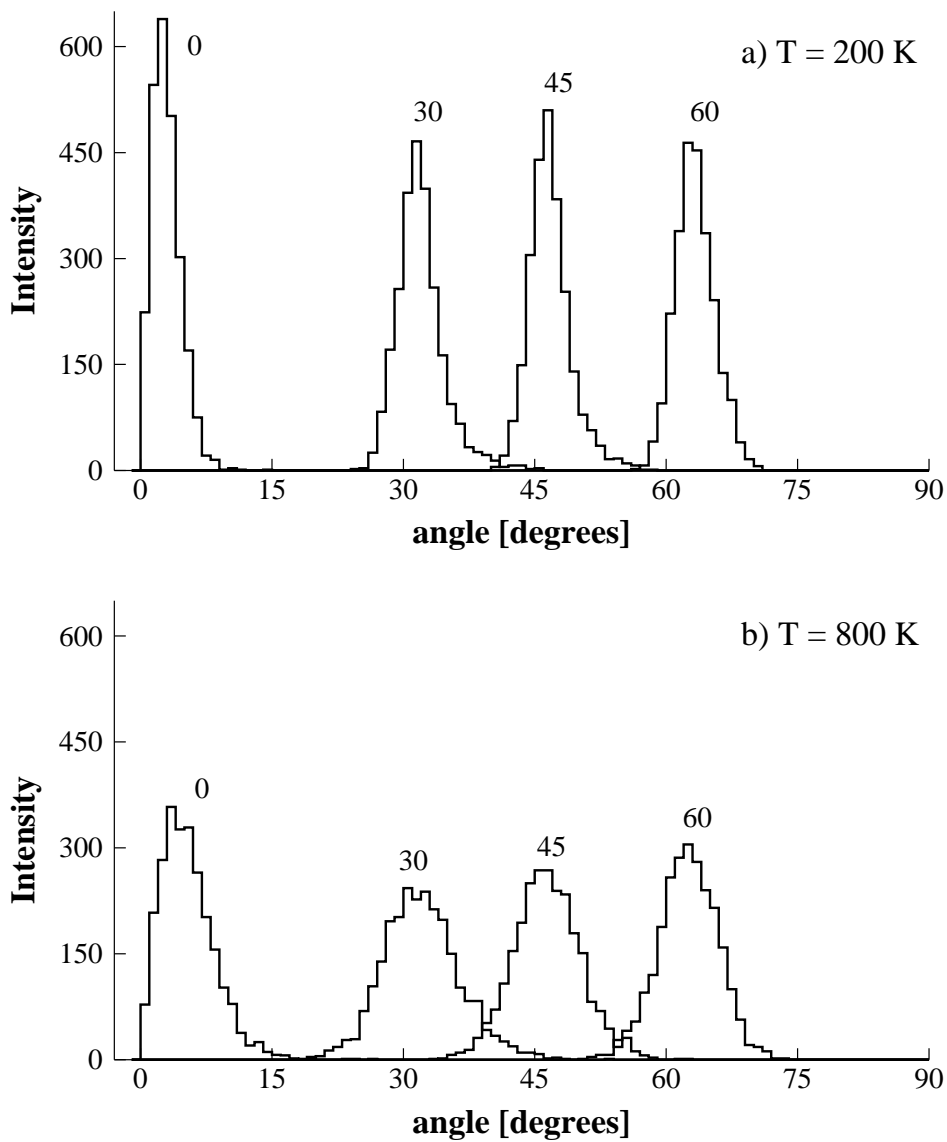, width=12.5cm}%
\caption{The distribution of the scattering angle for a total initial
  translational energy of 96 kJ/mol with incidence angles of
  0${}^{\circ}$, 30${}^{\circ}$, 45${}^{\circ}$, and 60${}^{\circ}$ with
  the surface normal. Both the nozzle and surface temperature are: a) 200K,
  and b) 800K.}
\label{fig:sctangdst}
\end{figure}

Figure \ref{fig:sctangdst} shows the scattering angle distribution for
different incidence angles with a initial total translational energy of
96 kJ/mol at nozzle and surface temperatures of both 200 and 800 K. The
scatter angle is calculated from the ratio between the normal and the
total parallel momentum of the whole methane molecule. We observe that
most of the trajectories scatter some degrees towards the surface from
the specular. This means that there is relatively more parallel momentum
than normal momentum at the end of the simulation compared with the
initial ratio. This ratio change is almost completely caused by a
decrease of normal momentum.

The higher nozzle and surface temperatures have almost no influence on
the peak position of the distribution, but give a broader distribution.
The standard deviation in the scattering angle distribution goes up from
$2.7^{\circ}$, $2.4^{\circ}$, and $2.2^{\circ}$ at 200K to
$4.4^{\circ}$, $3.8^{\circ}$, and $3.4^{\circ}$ at 800K for incidence
angles of $30^{\circ}$, $45^{\circ}$, and $60^{\circ}$ respectively.
This means that the angular width is very narrow, because the full width
at half maximum (FWHM) are usually larger than
$20^{\circ}$.\cite{wiskerke95} (The FWHM is approximately somewhat more
than twice the standard deviation.)  The broadening is caused almost
completely by raising the surface temperature, and has again
primarily an effect on the spread of the normal momentum of the
molecule. This indicates that the scattering of methane from Ni(111) is
dominated by a thermal roughening process.

We do not observe an average out-of-plane diffraction for the non normal
incidence simulations, but we do see some small out-of-plane
broadening. The standard deviations in the out-of-plane angle were
0.9${}^{\circ}$, 1.8${}^{\circ}$, 3.4${}^{\circ}$ at a surface
temperature of 200K, and 1.7${}^{\circ}$, 3.3${}^{\circ}$, and
6.0${}^{\circ}$ at 800K for incidence angles of 30${}^{\circ}$,
45${}^{\circ}$, and 60${}^{\circ}$ with the surface normal. Raising the
(rotational) nozzle temperature has hardly any effect on the
out-of-plane broadening. 

\subsection{Energy dissipation processes}

\subsubsection{Translational energy}

\begin{figure}[p]
\epsfig{file=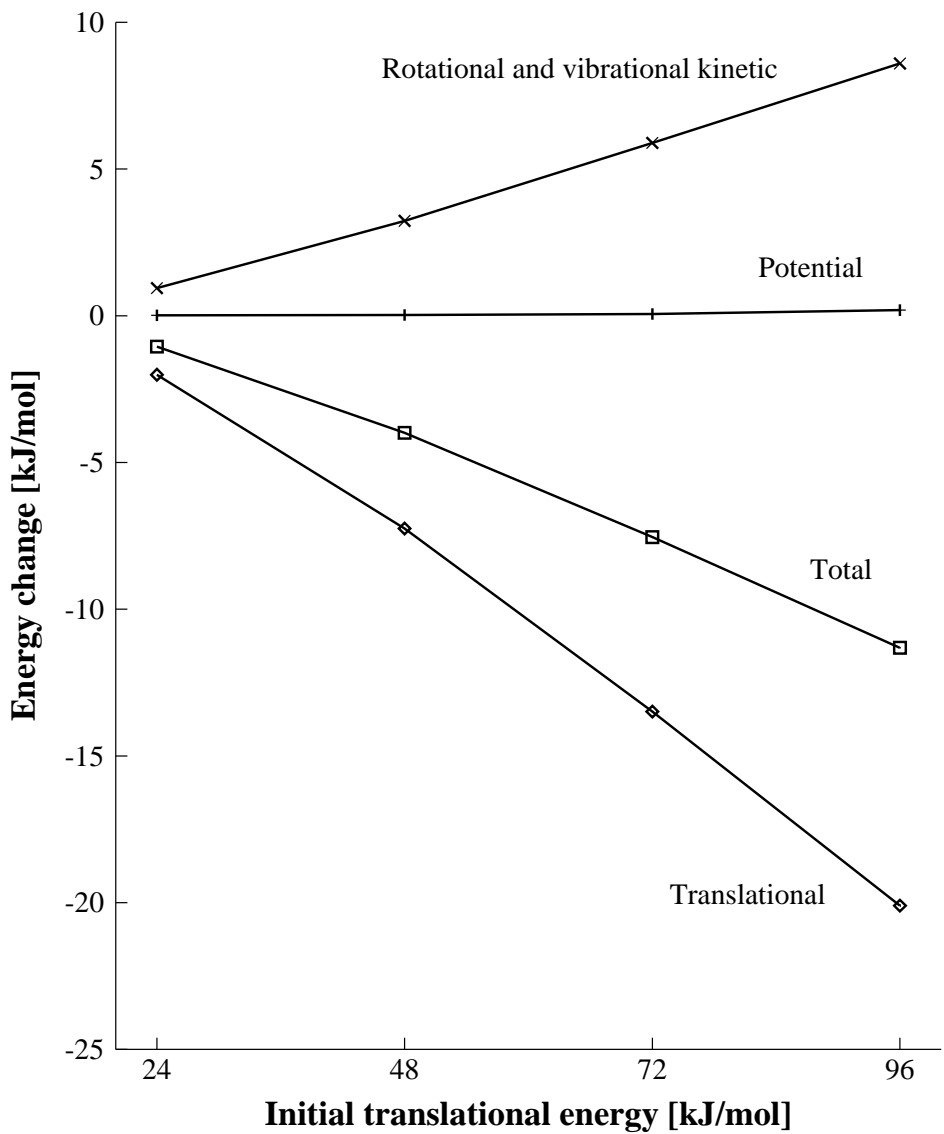, width=12.5cm}%
\caption{The average energy change  (kJ/mol) of the methane translational
  energy, the  methane total energy, the methane potential energy,
  and the methane rotational and vibrational kinetic energy as a function
  translational kinetic energy (kJ/mol) at normal incidence. The nozzle
  and surface temperature were 400K. }
\label{fig:dtotav}
\end{figure}

Figure \ref{fig:dtotav} shows the average energy change of some energy
components of the methane molecule between the end and the begin of the
trajectories as a function of the initial total translational energy.
The incoming angle for all is 0${}^{\circ}$ (normal incidence), and both
the nozzle and surface are initially 400K. If we plot the normal incidence
translational energy component of the simulation at 96 kJ/mol for the
different incidence angles, then we see a similar relation. This means
that there is normal translational energy scaling for the scattering
process in general, except for some small differences discussed later
on.

%\subsubsection{Translational energy loss}

Most of the initial energy of methane is available as translational
energy, so it cannot be surprising that we see here the highest energy
loss. The translational energy loss takes a higher percentage of the
initial translational energy at higher initial translational energies.
Since almost all of the momentum loss is in the normal direction, we
also see that the loss of translational energy can be found back in the
normal component of the translational energy for the non-normal incidence
simulations.

%\subsubsection{Transfer to the surface}

The average change of the total energy of the methane molecule is less
negative than the average change in translational energy, which 
means that there is a net transfer of the initial methane energy towards
the surface during the scattering. This is somewhat more than half of
the loss of translational energy. The percentage of transfered energy to
the surface related to the normal incidence translational energy is also
enhanced at higher incidence energies.  There is somewhat more
translational energy loss, and energy transfer towards the surface for
the larger scattering angles, than occurs at the comparable normal
translational energy at normal incidence.  This is caused probably by
interactions with more surface atoms, when the molecule scatters under
an larger angle with the surface normal.

%\subsubsection{rotational and vibrational excitations}

In Fig.~\ref{fig:dtotav} we also plotted the average change of methane
potential energy and the change of rotational and vibrational kinetic
energy of methane.  We observe that there is extremely little energy
transfer towards the potential energy, and a lot of energy transfer
towards rotational and vibrational kinetic energy. Vibrational motion
gives an increase of both potential and kinetic energy. Rotational
motion gives only an increase in kinetic energy. So this means that
there is almost no vibrational inelastic scattering, and very much
rotational inelastic scattering.

\begin{figure}[t]
\epsfig{file=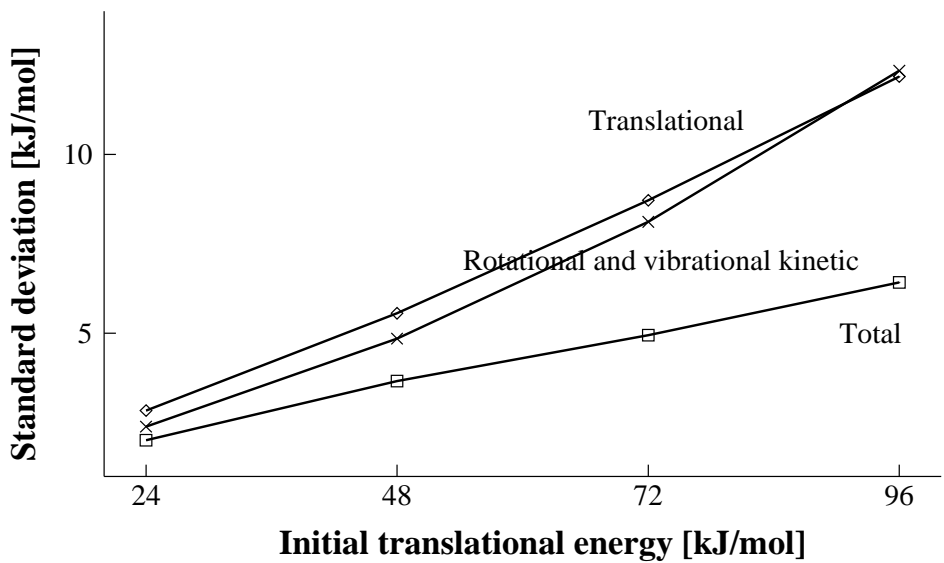, width=12.5cm}%
\caption{The standard deviation in the energy change (kJ/mol) of the
  methane translational energy, the methane total energy, and the methane
  rotational and vibrational kinetic energy as a function of the initial
  translational energy (kJ/mol) at normal incidence. The surface and
  nozzle temperature are both 200K. }
\label{fig:st_dev_trans}
\end{figure}

Figure \ref{fig:st_dev_trans} shows the standard deviations in the
energy change of some energy components of methane versus the initial
translational energy at normal incidence for a nozzle and surface
temperature of 200K. (The temperature effects will be discussed below.)
The standard deviations in the energy changes are quite large compared
to the average values. The standard deviations in the change of the
methane translational energy and in the change of methane rotational and
vibrational kinetic energy increase more than the standard deviation in the
change of methane total energy, when the initial translational energy is
increased.  We find again an identical relation if we plot the standard
deviations versus the initial normal energy component of the scattering
at different incidence angles. The standard deviations are much smaller
in the parallel than in the normal component of the translational
energy, so again only the normal component of the translational energy
is important.  Although the standard deviations in the translational
energy is smaller at larger incidence angles than at smaller incidence
angles, we see in Fig.~\ref{fig:sctangdst} that the spread in the angle
distribution is almost the same. This is caused by the fact that at
large angles deviations in the normal direction has more effect on the
deviation in the angle than at smaller angles with the normal.

%FIGURE

\subsubsection{Surface temperature}

An increase of surface temperature gives a small reduction of average
translational energy loss (around 5 $\%$ from 200K to 800K at 96 kJ/mol
normal incidence). This is the reason why we do not observe a large shift
of the peak position of the scattering angle distribution. However, an
increase of surface temperature does have a larger effect on the average
energy transfer to the surface, but this is in part compensated through
a decrease of energy transfer to rotational energy.

\begin{figure}[t]
\epsfig{file=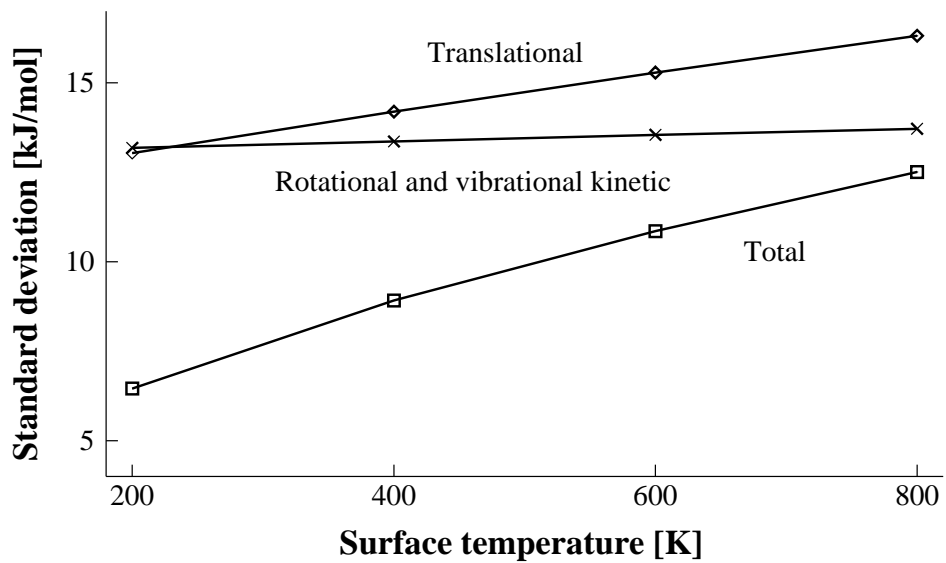, width=12.5cm}%
\caption{The standard deviation in the energy change (kJ/mol) of the
  methane translational energy, the methane total energy, and the methane
  rotational and vibrational kinetic energy as a function of the surface
  temperature (K). The nozzle temperature is 400K, and the translational
  energy is 96 kJ/mol at normal incidence. }
\label{fig:st_dev_surfT}
\end{figure}

Figure \ref{fig:st_dev_surfT} shows the standard deviations in the
energy change of the translational energy, the methane total energy, and
the methane rotational and vibrational kinetic energy as a function of
the surface temperature. We observe that the standard deviation in the
change of rotational and vibrational kinetic energy hardly changes at
increasing surface temperature. At a low surface temperature it is much
higher than the standard deviation in the change of the methane total
energy. So the baseline broadening of translational energy is caused by
the transfer of translational to rotational motion.  The standard
deviation in the change of the methane total energy increases much at
higher surface temperature. This results also in an increase of the
standard deviation in the change of translational energy, which means
that the surface temperature influences the energy transfer process
between translational and surface motion.  The spread in the change of
translational energy is related to the spread in the scattering angle
distributions. It is now clear that the observed broadening of the
scattering angle distribution with increasing surface temperature is
really caused by a thermal roughening process.

%FIGURE

\subsubsection{Nozzle temperature}

\begin{figure}[t]
\epsfig{file=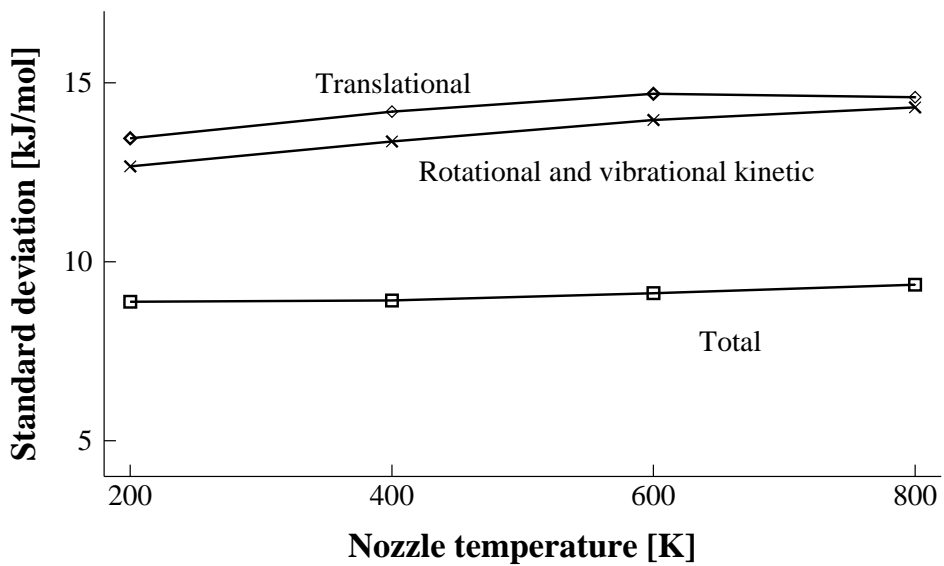, width=12.5cm}%
\caption{The standard deviation in the energy change (kJ/mol) of the
  methane translational energy, the methane total energy, and the methane
  rotational and vibrational kinetic energy as a function of the nozzle
  temperature (K). The surface temperature is 400K, and the translational
  energy is 96 kJ/mol at normal incidence. }
\label{fig:st_dev_nozzleT}
\end{figure}

Figure \ref{fig:st_dev_nozzleT} shows the dependency of the standard
deviations for the different energy changes on the nozzle temperature.
From this figure it is clear that the nozzle temperature has relative
little influence on the standard deviations in the different energy
changes. Therefore we observe almost no peak broadening in the
scattering angle distribution due to the nozzle temperature.
The nozzle temperature has also no influence on the average change of
rotational and vibrational kinetic energy, which means that this part of the
energy transfer process is driven primarily by normal incidence
translational energy.

%FIGURE

We have to keep in mind that we only studied the rotational heating by
the nozzle temperature, and that we did not take vibrational excitation
by nozzle heating into account. From our wave packet simulations in
Chapter \ref{chap:iso} we know that vibrational excitations can
contribute to a strong enhancement of vibrational inelastic scattering
(see also Ref.~\cite{mil00b}). So the actual effect of raising the
nozzle temperature can be different than sketched here.

\subsection{Comparison with other studies}

\subsubsection{Scattering angles and the Hard Cube Model}

The angular dependence of scattered intensity for a fixed total
scattering angle has only been measured at Pt(111)
\cite{yagyu99,yagyu99b}. The measurement has been compared with the
predictions of the Hard Cube Model (HCM) as described in
Ref.~\cite{logan66}. There seems to be more or less agreement for low
translational energies under an angle around 45${}^{\circ}$ with the
surface, but is anomalous at a translational energy of 55 kJ/mol. The
anomalous behaviour has been explained by altering the inelastic
collision dynamics through intermediate methyl fragments.

Although our simulations are for Ni(111) instead of Pt(111) and we
calculate real angular distributions, we will show now that the HCM is
insufficient for describing the processes involved with the scattering
of methane in our simulation. The HCM neglects the energy transfer to
rotational excitations, and overestimates the energy transfer to the
surface. This is not surprising, because the HCM is constructed as a
simple classical model for the scattering of gas atoms from a solid
surface. The basic assumptions are that (1) the interaction of the gas
atom with a surface atom is represented by an impulsive force of
repulsion, (2) the gas-surface intermolecular potential is uniform in
the plane of the surface, (3) the surface is represented by a set of
independent particles confined by square well potentials, (4) the
surface particles have a Maxwellian velocity distribution
\cite{logan66}. Assumption 1 excludes inelastic rotational scattering,
because the gas particle is an atom without moment of inertia.  So the
HCM misses a large part of inelastic scattering. However, it still
predicts scattering angles much more below the incidence angles than we
found from our simulation. For example: The HCM predicts an average
scattering angle with the surface normal of 64${}^{\circ}$ from Ni(111),
at an incidence angle of 45${}^{\circ}$ at a surface temperature four
times lower than the gas temperature. This is much more than for
Pt(111), because the mass ratio between the gas particle and the surface
atom is higher for Ni(111).  There are several explanations for this
error.  First, the assumption 3 is unreasonable for atomic surfaces with
low atom weight, because the surface atoms are strongly bound to each
other. This means that effectively the surface has a higher mass than
assumed \cite{grimmelmann80}. Second, there is no one-on-one interaction
between surface atom and methane molecule, but multiple hydrogen atoms
interacting with different Ni atoms.  Third, the methane molecule is not
rigid in contrast to assumption 1.  We have followed the energy
distribution during the simulation for some trajectories and find that
the methane molecule adsorbs initial rotational and translational energy
as vibrational energy in its bonds and bond angles when close the
surface, which is returned after the methane moves away from it.

It would be nice to test our model with molecular beam experiment of the
scattering angles on surfaces with relatively low atom weight, which
also try to look at rotational inelastic scattering.

\subsubsection{Wave packet simulations}

Let us now compare the full classical dynamics with our fixed oriented
wave packet simulations as discussed in the previous chapters (see also
Refs.~\cite{mil98,mil00a,mil00b}), because this was initial the reason
to perform the classical dynamics simulations.  Again we observe very
little vibrational inelastic scattering. This is in agreement with the
observations in the time-of-flight experiments on
Pt(111) \cite{yagyu00,hiraoka00}.

Since we used our wave packet simulations to deduce consequences for the
dissociation of methane, we have to wonder whether the observed
inelastic scattering in our classical simulations changes the 
picture of the dissociation in our previous publications.  Therefore we
have to look at what happens at the surface. We did so by following some
trajectories in time. 

We find approximately the same energy rearrangements for the classical
simulations as discussed for the wave packet simulations for the
vibrational groundstate in Chapters \ref{chap:ch4scat} and
\ref{chap:iso} (see also Refs.~\cite{mil00a} and \cite{mil00b}).  Again
most of the normal translational energy is transfered to the potential
energy terms of the surface repulsion [see Eq.~\ref{Vrep}].  This
repulsive potential energy was only given back to translational energy
in the wave packet simulations, because the orientations and surface
were fixed. For the classical trajectory simulations presented in this
article, the repulsive potential energy is transfered to translational,
rotational, and surface energy through the inherent force of the
repulsive energy terms. We observe almost no energy transfers to
translational energy parallel to the surface, so exclusion of these
translational coordinates in the wave packet simulations do not effect
our deduction on the dissociation. The energy transfers to the
rotational and surface energy during the collision make it harder for
the molecule to approach the surface. This will have a quantitative
effect on the effective bond lengthening near the surface, but not a
qualitative.

The remaining problem deals with the effect of rotational motion on the
dissociation probability and steering. Our first intension was to look
for the favourable orientation at the surface, but from following some
trajectories it is clear that steering does not seem to occur. There
is always some rotational motion, and the molecule leaves the surface
often with another hydrogen pointing towards to surface than when it
approaches the surface. This indicates that multiple bonds have a chance
to dissociate during one collision. However, it will be very speculative
to draw more conclusion on the dissociation of methane based on the
scattering in these classical trajectory simulations. Classical trajectory
simulation with an extension of our potentials with an exit channel for
dissociation can possibly learn us more.

\section{Conclusions}
\label{dyn:concl}

We have performed classical dynamics simulations of the rotational
vibrational scattering of non-rigid methane from a corrugated Ni(111)
surface. Energy dissipation and scattering angles have been studied as a
function of the translational kinetic energy, the incidence angle, the
(rotational) nozzle temperature, and the surface temperature.

We find the peak of the scattering angle distribution somewhat below the
incidence angle of 30${}^{\circ}$, 45${}^{\circ}$, and 60${}^{\circ}$ at
a translational energy of 96 kJ/mol. This is caused by an average energy
loss in the normal component of the translational energy. An increase of
initial normal translational energy gives an enhancement of inelastic
scattering. The energy loss is transfered for somewhat more than half to
the surface and the rest mostly to rotational motion.  The vibrational
scattering is almost completely elastic.

The broadening of the scattering angle distribution is mainly caused by
the energy transfer process of translational energy to rotational
energy. Heating of the nozzle temperature gives no peak broadening.
Heating of the surface temperature gives an extra peak broadening
through thermal roughening of the surface.

The Hard Cube Model seems to be insufficient for describing the
scattering angles of methane from Ni(111), if we compare its assumptions
with the processes found in our simulations.

%\bibliography{Methane}
%\bibliographystyle{prstyfull}

%% file: conclusion.tex
\chapter{Concluding remarks}
\label{chap:concl}

\begin{quote}
  {\it I will give an overview and brief discussion of the research
  described in this thesis, and make some suggestion for further research.}
\end{quote}
%\vspace*{3cm}

\section{Introduction}

This thesis is mainly a research article collection.  I have tried to
keep the contains of the separate articles as close as possible to the
originals, when it deals with the results and discussion sections and
put the overlap between some articles, dealing with methods and
potentials, in a separate chapter.  By doing so, I have left the
research within the history of time. Over the years our thoughts have
been changed by new experimental and theoretical studies. I will give
therefore now a point by point overview of the phenomena discussed
within different chapters, and give some suggestions for further
research on the dissociation of methane on transition metal surfaces
based on our findings and recent new developments.

\section{Scattering}

We have performed scattering simulation of methane scattering on
a Ni(111) surface with wave packet and classical trajectory dynamics. 

We have been using the multiconfigurational time-dependent Hartree
(MCTDH) method for our wave packet simulation, because it can deal with
a large number of degrees of freedom and with large grids. We used four
different PESs for fixed molecular orientations with one, two, and three
bonds pointing towards the surface, which are described in Chapter
\ref{chap:wp}. (A harmonic intramolecular PES is adapted to include
anharmonicities in the C--H/D distance, the decrease of C--H/D bond
energy due to interactions with the flat surface, and the increase of
C--H/D bond length also due to interactions with the flat surface.)
We used in Chapters \ref{chap:ch4scat} and \ref{chap:iso} a
translational energy of up to 96 kJ/mol normal incidence and all
internal vibrations in the ground state. We did not include parallel
translational and rotational coordinates in our model.  The scattering
was in all cases predominantly elastic.  However, the observed inelastic
scattering of CD${}_4$ compared with simulation on CH${}_4$ is higher
for the PES with an elongated equilibrium bond length close to the
surface.
Our wave packet simulations in Chapter \ref{chap:exc} showed that
vibrational excitation of a single mode enhanced inelastic scattering
with transfer of translational energy to vibrational energy, especially
for the stretch modes in the orientation with three bonds pointing
towards the surface.

We have performed classical dynamics simulations of the rotational
vibrational scattering of non-rigid methane from a corrugated Ni(111)
surface, which is described in Chapter \ref{chap:traj}. Energy
dissipation and scattering angles have been studied as a function of the
translational kinetic energy, the incidence angle, the (rotational)
nozzle temperature, and the surface temperature.  We find the peak of
the scattering angle distribution somewhat towards the surface for the
incidence angle of 30${}^{\circ}$, 45${}^{\circ}$, and 60${}^{\circ}$ at
a translational energy of 96 kJ/mol. This is caused by an average energy
loss in the normal component of the translational energy. An increase of
initial normal translational energy gives an enhancement of inelastic
scattering. The energy loss is transfered for somewhat more than half to
the surface and the rest mostly to rotational motion.  The vibrational
scattering is almost completely elastic, which is in agreement with the
wave packet simulations in the vibrational ground state.  The broadening
of the scattering angle distribution is mainly caused by the energy
transfer process of translational energy to rotational energy. Heating
of the nozzle temperature gives no peak broadening.  Heating of the
surface temperature gives an extra peak broadening through thermal
roughening of the surface.

\section{Dissociation}

The dissociative adsorption of methane on transition metals is an
important reaction in catalysis; it is the rate limiting step in steam
reforming to produce syngas, and it is prototypical for catalytic C--H
activation. Therefore the dissociation is of high interest for many
surface scientists. I gave an overview of the experimental and
theoretical studies upon methane dissociation on transition-metal
surfaces in Chapter \ref{chap:exper}.  Molecular beam experiments in
which the dissociation probability was measured as a function of
translational energy have observed that the dissociation probability is
enhanced by the normal incidence component of the incidence
translational energy.
This suggests that the reaction occurs primarily through a direct
dissociation mechanism at least for high translational kinetic energies.
Some experiments have also observed that vibrationally hot CH${}_4$
dissociates more readily than cold CH${}_4$, with the energy in the
internal vibrations being about as effective as the translational energy
in inducing dissociation.  A molecular beam
experiment with laser excitation of the $\nu_3$ mode did succeed in
measuring a strong enhancement of the dissociation on a Ni(100) surface.
However, this enhancement was still much too low to account for the
vibrational activation observed in previous studies and indicated that
other vibrationally excited modes contribute significantly to the
reactivity of thermal samples.

It is very interesting to simulate the dynamics of the dissociation,
because of the direct dissociation mechanism, and the role of the
internal vibrations. Wave packet simulations of the methane dissociation
reaction on transition metals have treated the methane molecule always
as a diatomic up to now.
Apart from one C--H bond (a pseudo $\nu_3$ stretch mode) and the
molecule surface distance, either (multiple) rotations or some lattice
motion were included. None of these studies have looked at the role of
the other internal vibrations, so there is no model that describes which
vibrationally excited mode might be responsible for the experimental
observed vibrational activation.  We were not able yet to simulate the
dissociation including all internal vibrations with wave packet
dynamics. Instead we have deduced from our scattering simulations
consequences for the dissociation. We did so by looking at vibrational
excitation, the structure deformation, and the energy distribution of
the methane molecule, when it hits the surface.

\subsection{Reaction mechanism and paths}

When the molecule hits the surface, we always observe vibrational
excitations of the $\nu_4$ umbrella and $\nu_2$ bending modes for the
CH${}_4$ molecule in Chapter \ref{chap:ch4scat}, especially in the
orientations with two or three hydrogens pointing towards the surface.
This is due to a favorable coupling that originates from the repulsive
interaction with the surface, and the low excitation energies.
Deformations of the molecule are predominantly in the bond angles. The
changes in the bond angles are, however, too small to allow for the
formation of a Ni--C bond, as suggested in the ``splats'' model of
methane dissociation \cite{lee87}.

Appreciable excitations of the $\nu_1$ and $\nu_3$ stretch modes when
methane hits the surface are only observed when one hydrogen atom points
towards the surface, or when the intramolecular PES has an
elongated equilibrium C--H bond length close to the surface. The
repulsion of the surface shortens the C--H bond. This can only be
overcome when the intramolecular PES incorporates the effect of a
longer equilibrium C--H bond length caused by overlap of occupied
surface orbitals with the antibonding orbitals of methane. This agrees
with quantum chemical calculations, which show a late barrier for
dissociation.

%%I like to begin with the re-invention of the ``splats''
%%model\cite{lee87}, which we have rejected in Chapter \ref{chap:ch4scat}
%%and \ref{chap:iso}. Our main argument, which is discussed in Section
%%\ref{sec:diss}, states that the angular deformation is too small to
%%enable the formation of a Ni--C bond. So the rejection is primary based
%%on the angular deformation picture of the ``splats'' model as
%%represented in Ref. \cite{lee87}. I shall have a closer look now on how the
%%``splats'' model entered the discussion of reaction mechanism in the
%%first place. Very important were the collision induced experiments,
%%which show that at low surface temperature physisorped methane
%%dissociates under argon bombardment. 

\subsection{The isotope effect}

A nice way to study reaction dynamics is the use of isotopes. The most
recent wave packet simulation on the dissociation probability of
CH${}_4$ and CD${}_4$ showed a semiquantitative agreement with the
molecular beam experiments of Ref.\cite{hol95}, except for the isotope
effect and the extracted vibrational efficacy \cite{car98}. Therefore we
started a study on the scattering of CD${}_4$, which is described in
Chapter \ref{chap:iso}.

When the molecule hits the surface, we observed in general a higher
vibrational excitation for CD${}_4$ than CH${}_4$. The PES with an
elongated equilibrium bond length close to the surface gives for both
isotopes almost the same deformations, although we observe a somewhat
smaller bond lengthening for CD${}_4$ in the orientation with three
bonds pointing towards the surface.  The other model PESs show
differences in the bond angle deformations and in the distribution of
the excitation probabilities of CD${}_4$ and CH${}_4$, especially for
the PES with only an anharmonic intramolecular potential.

Energy distribution analysis proofs observations and hypotheses obtained
from excitation probabilities and structure deformation, and contributes
new information on the scattering dynamics.
A high increase of vibrational kinetic energy results in higher
inelastic scattering. The highest increase of vibrational kinetic energy
is found in the $\nu_3$ asymmetrical stretch modes for all orientations
and in the $\nu_1$ symmetrical stretch mode for the orientation with
three bonds pointing towards the surface, when the PES has an elongated
equilibrium bond length close to the surface.

\subsection{The role of vibrational excitation}

In Chapter \ref{chap:exc} we discuss the effect of vibrational
excitation on the dissociation. We used initial translational energies
in the range of 32 to 128 kJ/mol. A single vibrational excitation was
put in one of the vibrational modes with $a_1$ symmetry in the symmetry
group of the molecule plus surface, while the other vibrational modes
were kept in the ground state. The potential-energy surface is
characterised by an elongation of the C--H bonds when the molecule
approaches the surfaces, and a molecule-surface part appropriate for
Ni(111). We find that initial vibrational excitations enhance the
transfer of translational energy towards vibrational energy and increase
the accessibility of the entrance channel for dissociation, which means
an increase of inelastic scattering. The largest effects are observed in
the orientation with three bonds pointing to the surface.  Our
simulations predict that initial vibrational excitations of the
asymmetrical stretch ($\nu_3$) and especially the symmetrical stretch
($\nu_1$) modes will give the highest enhancement of the dissociation
probability of methane. 

A recent four-dimensional wave packet simulations of the O(${}^3$P) +
CH${}_4$ $\rightarrow$ OH + CH${}_3$ reaction, also finds also that the
symmetric stretch vibration of CH${}_4$ is more efficient in promoting
the reaction than the asymmetric stretch \cite{palma00}. It shows also
that models that only consider a local C--H stretching of CH${}_4$ are
not appropriate to make comparisons in which the asymmetric stretch mode
is selectively excited.

%\cite{luis00}.

\section{Further research}

The simulations with our different model PESs show that the internal
vibrations play an important role in the dissociation mechanism.
Excitation probabilities when the molecule hits the surface show how the
translational energy is converted into vibrational energy and it is
distributed over the internal modes. These probabilities vary strongly
with the PES. As only few internal vibrations contribute to the
dissociation, it is important to obtain more information on the real PES
for this system.  One way of obtaining an more accurate potential is by
doing electronic structure calculations. Since we have to deal with the
interaction with a metallic surface, only DFT calculation seem to be a
reasonable option.  It will be still very hard to calculate a full 15
dimensional potential. Therefore a reduced dimensional PES should be
constructed.

Our wave packet simulations give an indications that the isotope effect
in the methane dissociation is caused mostly by the difference in the
scattering behaviour of the molecule in the orientation with three bonds
pointing towards the surface. At least multiple vibrational stretch
modes should be included for a reasonable description of isotope effect
in the methane dissociation reaction. Also our wave packet simulations
of vibrational excited methane suggests that both the asymmetrical
stretch ($\nu_3$) and the symmetrical stretch ($\nu_1$) modes should be
included. Our classical trajectory simulations shows that parallel
translational is relatively unimportant, and that rotational energy
plays an important role for at least the scattering channel. It also
shows that it is possible that multiple bonds interact with the surface.

Very important for our understanding of the methane dissociation on
transition metal surfaces will be the progress made by molecular beam
experiments with state selective excitations with lasers. It can be
interesting if this technique is combined with the use of different
isotopomers of methane. This will give us also an idea, which reduced
dimensional models for our simulations should be tried. I think
especially the CHD${}_3$ molecule will be a good isotopomer to study
further, because we can approximate that the dissociation occurs only
along the C--H bond in our dynamics simulations while we include
multiple vibrational stretch modes.

%\bibliography{Methane}
%\bibliographystyle{prstyfull}

%% file: summary.tex
\chapter{Summary}

The dissociation of methane on transition metals is an important
reaction in catalysis. It is the rate limiting step in steam reforming
to produce syngas. Molecular beam experiments have shown that the energy
in the internal vibrations are about as effective as the translational
energy in inducing dissociation.

The published wave packet simulations on the methane dissociation
reaction on transition metals have treated the methane molecule always
as a diatomic up to now.  Besides the C--H bond and molecule surface
distance, a combination of other coordinates were included, like
(multiple) rotations and some lattice motion. None of them have looked
at the role of the internal vibrations. We were not able yet to simulate
the dissociation including all internal vibrations. Instead we simulated
the scattering of methane in fixed orientations, for which all internal
vibrations can be included, and used the results to deduce consequences
for the dissociation.

We have been using the multi-configurational time-dependent Hartree
(MCTDH) method for our wave packet simulation, because it can deal with a
large number of degrees of freedom and with large grids. We have started
with a study on different model potential energy surfaces (PESs) that
have been developed with Ni(111) in mind. We found that the scattering
of CH${}_4$ is almost completely elastic for all model PESs. Vibrational
excitations when the molecule hits the surface and the corresponding
deformation show that for methane to dissociate the interaction of the
molecule with the surface should lead to an elongated equilibrium C--H
bond length close to the surface.

We studied the isotope effects with CD${}_4$ in the same way, and found
an elastic scattering somewhat less than for CH${}_4$. Energy
distribution analysis at the surface of the expectation values of the
kinetic energy operators and terms potential energy terms gives enhanced
insight in the scattering process. Our simulations give an indications
that the isotope effect in the methane dissociation is caused mostly by
the difference in the scattering behaviour of the molecule in the
orientation with three bonds pointing towards the surface.

Next we looked at the role of single vibrational excitations at the
different orientations.  A high increase of vibrational kinetic energy
results in higher inelastic scattering. The highest increase of
vibrational kinetic energy and of the accessibility of the entrance
channel for dissociation are found for the $\nu_3$ asymmetrical stretch
mode, and especially for the $\nu_1$ symmetrical stretch mode. This
indicates that the $\nu_1$ will give the highest enhancement of the
dissociation probability.

We ended with classical trajectory calculations of the rotational
vibrational scattering of a non-rigid methane molecule from a Ni(111)
surface. Energy dissipation and scattering angles have been studied as a
function of the translational kinetic energy, the incident angle, the
(rotational) nozzle temperature, and the surface temperature.
Scattering angles are somewhat towards the surface for the incidence
angles of 30${}^{\circ}$, 45${}^{\circ}$, and 60${}^{\circ}$ at a
translational energy of 96 kJ/mol. Energy loss is primarily from the
normal component of the translational energy and transfered for somewhat
more than half to the surface and the rest mostly to rotational motion.

%% file: samenvatting.tex
\chapter{Samenvatting}

De dissociatie van methaan is een katalytisch interessante reactie.
Het is de snelheidsbepalende stap voor de vorming van synthese-gas.
Experimenten met moleculaire bundels op \'e\'enkristallen van
overgangsmetalen hebben aangetoond dat het een directe dissociatie
betreft, die zowel door translationele als door vibrationele energie
bevorderd wordt.

In het verleden heeft men geprobeerd de dynamica van de dissociatie te
simuleren met golfpakketmethoden door methaan te beschouwen als een
twee-atomig molecuul. Hoewel hiermee enkele tendensen beschreven kunnen
worden, kunnen met deze simulaties de rol van andere vibraties dan de
asymmetrische strekvibratie in de reactiedynamica niet bestudeerd
worden.  Het is echter nog steeds lastig om de dissociatiereactie in
hogere dimensies te beschrijven met behulp van golfpakketmethoden. In
plaats daarvan is gekeken naar de verstrooiing van methaan met een
specifieke ori\"entatie, waarbij wel alle vibraties kunnen worden
meegenomen, en hieruit informatie te winnen over de dissocatiereactie.

Voor de golfpakket-simulaties is gebruik gemaakt van de
\textsl{\foreignlanguage{english}{multi-configu\-rational time-dependent
    Hartree (MCTDH)}} methode, omdat deze methode zeer geschikt is voor
systemen met een hoge dimensie en grote roosters. Begonnen werd met de
bestudering van verschillende modelpotentialen voor de verstrooiing aan
Ni(111). Hierbij werd als eerste gekeken naar de
verstrooiingswaarschijnlijkheden van CH${}_4$. Het bleek dat de
verstrooiing bijna volledig elastisch is voor de bestudeerde
potentialen. Daarom is ook gekeken naar de excitatiewaarschijnlijkheden
en de verwachtingswaarden van de structuurdeformaties bij het oppervlak.
De potentiaal met een verlenging van de evenwichtsbindingslengte bij het
oppervlak bleek als enige geschikt om de ingang van het reactiekanaal te
beschrijven.

Om het isotoop-effect te bestuderen is vervolgens weer gekeken naar de
excitaties en deformaties bij het oppervlak voor CD${}_4$. De
verstrooiing was weer bijna volledig elastisch, maar minder dan voor
CH${}_4$. Om beter inzicht te verkrijgen in de verschillen tussen de
verstrooiings-processen van beide isotopomeren werd ook de
energiedistributie bij het oppervlak bestudeerd door de
verwachtingswaarden van de kinetische operatoren en de belangrijkste
termen van de potentiaal te berekenen. Het is gebleken dat bij een
ori\"entatie met drie bindingen gericht naar het oppervlak een groter
isotoop-effect mag worden verwacht dan bij andere ori\"entaties. Dit
ver\-klaart ook waarom de simulaties met een twee-atomig methaanmodel een
kleiner isotoop-effect geven dan experimenteel gevonden is.

Vervolgens is er gekeken naar het effect van een enkelvoudige excitatie
van de verschillende vibrationele modes bij verschillende ori\"entaties.
Waargenomen werd dat de inelastische verstrooiing hierdoor
toeneemt en meer overdracht van translationele energie naar
vibrationele energie plaatsvindt. Ook blijkt het eenvoudiger te worden
om het ingangskanaal van dissociatie binnen te gaan. Beide effecten zijn
het grootst voor de $\nu_3$ asymmetrische en vooral de $\nu_1$
symmetrische strekmodes in de ori\"entatie met drie bindingen naar het
oppervlak gericht. Excitatie van de $\nu_1$ symmetrische strekmode brengt
hierdoor waarschijnlijk een grotere toename van
dissociatiewaarschijnlijkheid teweeg dan de overige modes.

Afsluitend zijn er klassieke dynamica simulaties gedaan aan de
verstrooiing van methaan op een Ni(111) oppervlak, waarbij de
dimensionaliteit van het model kan worden verhoogd. Hierbij is gekeken
naar de effecten van translationele energie, de rotationele
\textsl{\foreignlanguage{english}{nozzle}} temperatuur, en de
temperatuur van het oppervlak voor het energie-overdrachtprocessen en de
verstrooiingshoeken. De verstrooiingshoeken zijn enkele graden gericht
naar het oppervlak voor de inkomende hoeken van 30${}^\circ$,
45${}^\circ$, en 60${}^\circ$ bij een translationele energie van 96
kJ/mol.  Energie-verlies is er voornamelijk in de normaal component van
de translationele energie en wordt overgedragen voor iets meer dan de
helft aan het oppervlak en voor de rest aan rotatiebeweging.

%% file: listofpub.tex
\chapter{List of publications}

\begin{itemize}

\item R.\ Milot and A.~P.~J.\ Jansen; {\sl Ten-dimensional wave packet
  simulations of methane scattering}, J.\ Chem.\ Phys.\ {\bf 109},
  1966 (1998). 

\item R.\ Milot and A.~P.~J.\ Jansen; {\sl Energy distribution
    analysis of the wave packet simulations of CH${}_4$ and CD${}_4$
    scattering}, Surf.\ Sci.\ {\bf 452}, 179 (2000).

\item R.\ Milot and A.~P.~J.\ Jansen; {\sl Bond breaking in
    vibrationally excited methane on transition-metal catalysts},
    Phys.\ Rev.\ B {\bf 61}, 15657 (2000).
    
\item R.\ Milot, A.~W.\ Kleyn, and A.~P.~J.\ Jansen; {\sl Energy
      dissipation and scattering angle distribution analysis of the
      classical trajectory calculations of methane scattering from a
      Ni(111) surface}, preprint available at
      \texttt{http://arXiv.org/abs/physics/0103053}.

%J.\ Chem.\ Phys.\ {\bf }, (submitted).

%\item R.\ Milot, H.\ Kooijman, J.\ Kroon, and E.\ Grech; {\sl
%    1,8-Bis(dimethylamino)\-naphthalene 2-hydroxy-1,4-naphthalenedione},
%  Acta Crystallogr.\ C {\bf }, (to be submitted).

\end{itemize}

%% file: dank.tex
\chapter{Dankwoord}

Graag wil ik van de gelegenheid gebruik maken om een paar mensen te
bedanken die hun bijdragen hebben geleverd aan de totstandkoming van
dit proefschrift. Om te beginnen mijn begeleider Tonek Jansen, die mij
vijf jaar geleden het bos heeft ingestuurd en wanneer nodig heeft
geholpen om er weer uit te komen. Aart Kleyn wil ik bedanken voor
zijn bijdrage aan beschrijving van de klassieke dynamica simulaties.

Een promovendus kan niet zonder de steun van zijn soortgenoten. Ik wil
daarom alle mensen van de theoriegroep, met wie ik heb mogen samen
werken op vloer 10 en in de rekenkamer, bedanken voor het draaiend
houden van de machine.  In het bijzonder wil ik nog mijn ex-kamergenoot
Eric Meijer bedanken voor zijn aangename gezelschap. Ook onze
gezamenlijke passie voor de dingetjes sprak mij zeer aan. Daarnaast heb
ik veel plezier beleefd aan zijn raad en daad aangaande binaire
aangelegenheden als het ware. Bij wijze van spreken is zijn onzichtbare
hand dan terug te vinden in de broncodes van mijn programma tot in de
povray figuren van dit proefschrift. Chr\' etien, Danny, Paul, Vili en
Willy bedankt voor jullie gezelschap tijdens de lunch en de
bijeenkomsten in Leuteren. Ik zal jullie missen, en ook de overige leden
van SKA met wie ik altijd gezellig koffie heb mogen drinken. Alle
achterblijvers wil ik succes en sterkte wensen met het afronden van hun
proefschrift binnen vier jaar.

De afgelopen jaren heb ik natuurlijk meer gedaan dan dit
promotie-onderzoek. De tijd die ik met familie, vrienden en teamgenoten
heb mogen doorbrengen was mij het meest dierbaar. Bedankt voor jullie
liefde, steun en vertrouwen. Hierbij is het jullie vergeven dat jullie
steeds weer vroegen wanneer het nu eens af is.

Het gaat jullie goed!

%% file: cv.tex
\chapter{Curriculum vitae}

Op 6 oktober 1972 werd Robin Milot geboren in Utrecht. Nadat hij in 1991
het (ongedeeld) VWO-diploma behaalde aan de Thorbecke Scholengemeenschap
te Utrecht, begon hij aan de Universiteit Utrecht met de studie
Scheikunde. In 1995 werd het doctoraalexamen behaald met als hoofdvak
Kristal- en Structuurchemie. Als keuzepakketten werden Fysische Chemie en
Fysica \& Chemie van Materialen gevolgd.
Op 1 april 1996 trad hij in dienst van NWO om zijn promotie-onderzoek,
zoals beschreven in dit proefschrift, uit te voeren bij het Schuit
Katalyse Instituut van de Technische Universiteit Eindhoven.

\newpage

\vspace*{13cm}

\vfill

Omslag: De foto op de omslag is door mij genomen van de \textsl{Canyon
  de Chelly} in Arizona (VS). De canyon staat symbool voor een
potentiaal-oppervlak met daarin de twee rotspieken als golfpakket.
\copyright \ 1999 Robin Milot.